\documentstyle{article}
\newcommand{\be}{\begin{equation}}
\newcommand{\ee}{\end{equation}}

\title{VARIOUS ASPECTS OF WHITHAM TIMES}
\author{Robert Carroll\\University of Illinois\\ email: 
rcarroll@math.uiuc.edu}

\date{May, 1999}

\begin{document}
\bibliographystyle{plain}
\maketitle
\begin{abstract}  
We sketch some of the different roles played by Whitham times
in connection with averaging, adiabatic invariants, soliton theory,
Hamiltonian structures, topological field theory (TFT), 
Seiberg-Witten (SW) theory, isomonodromy problems, Hitchin systems,
WDVV and Picard-Fuchs equations,
renormalization, soft supersymmetry breaking, etc.
\end{abstract}

\tableofcontents

\section{INTRODUCTION}
\renewcommand{\theequation}{1.\arabic{equation}}
\setcounter{equation}{0}

Whitham theory arises in various contexts and the corresponding
Whitham times play roles such as deformation parameters of moduli,
slow modulation times, coupling constants, etc.
When trying to relate these roles
some unifying perspective regarding Whitham theory
is needed and we try here to provide some steps in this spirit.  
The idea is to display Whitham theory in many (not perhaps all) of its
various aspects (with some sort of compatible notation)
and thus to exhibit (if not always explain)
the connections between various 
manifestations of Whitham times.
Some background work in this direction appears in
\cite{aa,bb,cc,ch,ci,cb,cf,cu,cs,cm,ct,ca,cd,de,da,dc,db,dz,dw,
en,ec,ed,ea,fa,fc,fb,ga,gc,gf,gi,gk,gd,ge,gg,gx,gy,gj,hb,ha,ib,
ia,ke,kf,kk,ki,ka,kc,ko,la,le,lw,lg,
lb,lc,ld,me,mo,mr,ms,ma,ml,mw,mp,mb,mz,mc,md,mj,na,nb,oe,oa,ob,ta,tc,te,tf,
tg,th,ti,tj,va,wa,ya}.  Symbols such as $(\bullet)$, $(\clubsuit)$, {\bf (A)},
{\bf (XIII)}, etc. may be repeated in different sections but references to them
are intrasectional.

\section{BACKGROUND}
\renewcommand{\theequation}{2.\arabic{equation}}
\setcounter{equation}{0}

Perhaps the proper beginning would be Whitham's book \cite{wa}
or the paper \cite{fa} where one deals with modulated wavetrains,
adiabatic invariants, etc.  followed by multiphase
averaging, Hamiltonian systems, and weakly deformed soliton
lattices as in \cite{dz,dw,ki}.  The next stages bring one into
contact with some current topics in theoretical physics such as
topological field theory
(TFT), strings, and SW theory to which some
references are already given above.  In Section 2 we begin
with the averaging method following \cite{cc,fb,ki} 
plus dispersionless theory as in \cite{ch,ci,ct,kk,ta}
(which is related to TFT); then we
develop
relations to isomonodromy and to Hitchin systems following
\cite{la,le,lw,oa,ob,te,tg,th,ti,tj} in Sections 3-6, while connections to
Seiberg-Witten (SW) theory, the renormalization group (RG)
and soft supersymmetry
(susy) breaking appear in Sections 7-9.  In Section 10 we add
a few remarks about the structure of contact terms in twisted
${\cal N}=2$ susy gauge theory on a 4-manifold following
\cite{en,ec,lg,me,mo,mg,tc}.
\\[3mm]\indent
The averaging method for soliton equations goes back to \cite{fa}
and early Russian work (cf. \cite{dz,dw} for references) but
the most powerful techniques appear first in \cite{ki} (with
refinements and clarification in \cite{cc,fb}).  For background on
KdV, KP, and RS see e.g. \cite{bu,cc,ce,df,kj,tl}.

\subsection{Riemann surfaces and BA functions}

We
take an arbitrary Riemann surface $\Sigma$
of genus $g$, pick a point $Q$ and a local variable $1/k$ near $Q$
such that $k(Q) = \infty$, and, for illustration, take $q(k)=kx+k^2y+k^3t$. 
Let $D = P_1 + \cdots + P_g$ be a non-special
divisor of degree $g$ and write $\psi$ for the (unique up to a 
multiplier by virtue of the Riemann-Roch theorem)
Baker-Akhiezer (BA) function characterized by the properties
({\bf A}) $\psi$ is meromorphic on $\Sigma$ except for $Q$ where $\psi
(P)exp(-q(k))$ is analytic 
and (*) after normalization
$\psi\sim exp(q(k))[1 + \sum_1^{\infty}(\xi_j/
k^j)]$ near $Q$.  ({\bf B}) On $\Sigma/Q,\,\,\psi$ has only a finite
number of poles (at the $P_i$).
In fact $\psi$ can be taken in the form ($P\in\Sigma,\,\,
P_0\not= Q$)
\be
\psi(x,y,t,P) =  exp[\int^P_{P_0}(xd\Omega^1 + yd\Omega^2 + td\Omega^3)]
\cdot\frac{\Theta({\cal A}(P) + xU + yV + tW + z_0)}{\Theta({\cal A}
(P) + z_0)}
\label{psi}
\ee
where $d\Omega^1 = dk + \cdots,\,\,d\Omega^2 = d(k^2) + \cdots,\,\,
d\Omega^3 = d(k^3) + \cdots, U_j = \int_{B_j}d\Omega^1,\,\,V_j = \int_
{B_j}d\Omega^2,\,\,W_j = \int_{B_j}d\Omega^3\,\,(j = 1,\cdots,g),\,\,z_0
= -{\cal A}(D) - K$, and $\Theta$ is the Riemann theta function.
The symbol $\sim$ will be used generally to mean ``corresponds
to" or ``is associated with"; occasionally it also denotes asymptotic
behavior; this should be clear from the context.
Here the
$d\Omega_j$ are meromorphic differentials of second kind normalized via
$\int_{A_k}d\Omega_j = 0\,\,(A_j,\,B_j$ are canonical homology cycles)
and we note that $xd\Omega^1 + yd\Omega^2 + td\Omega^3\sim
dq(k)$ normalized;
${\cal A}\sim$ Abel-Jacobi map ${\cal A}(P) = (\int^P_{P_0}d\omega_k)$
where
the $d\omega_k$ are normalized holomorphic differentials, $k = 1,\cdots,g,
\,\,\int_{A_j}d\omega_k = \delta_{jk}$, and $K = (K_j)\sim$ Riemann
constants ($2K = -{\cal A}(K_{\Sigma})$ where $K_{\Sigma}$ is the
canonical class of $\Sigma\sim$ equivalence class of meromorphic 
differentials).  Thus $\Theta({\cal A}(P) + z_0)$ has exactly $g$ zeros
(or vanishes identically.  The paths of integration are to be the
same in computing $\int_{P_0}^Pd\Omega^i$ or ${\cal A}(P)$ and it is
shown in \cite{bu,cc,df} that $\psi$ is well defined (i.e. path independent).  
Then the $\xi_j$ in (*) can be computed
formally and one determines Lax operators $L$ and $A$ such that
$\partial_y\psi = L\psi$ with $\partial_t\psi = A\psi$.  Indeed, given
the $\xi_j$ write $u = -2\partial_x\xi_1$ with $w = 3\xi_1\partial_x\xi_1
-3\partial^2_x\xi_1 - 3\partial_x\xi_2$.  Then formally, near $Q$, one
has $(-\partial_y + \partial_x^2 + u)\psi = O(1/k)exp(q)$ and 
$(-\partial_t + \partial^3_x + (3/2)u\partial_x + w)\psi = O(1/k)exp(q)$
(i.e. this choice of $u,\,w$ makes the coefficients of $k^nexp(q)$ vanish
for $n = 0,1,2,3$).  Now define $L = \partial_x^2 + u$ and $A = \partial^3_x
+ (3/2)u\partial_x + w$ so $\partial_y\psi = L\psi$ and $\partial_t\psi
= A\psi$.  This follows from the uniqueness of BA functions with the same
essential singularity and pole divisors (Riemann-Roch). 
Then we have, via compatibility
$L_t - A_y = [A,L]$, a KP equation $(3/4)u_{yy} = \partial_x[u_t
-(1/4)(6uu_x + u_{xxx})]$ and therefore such KP equations are parametrized
by nonspecial divisors or equivalently by points in general position
on the Jacobian variety $J(\Sigma)$.
The flow variables $x,y,t$ are put in by hand in ({\bf A}) via
$q(k)$ and then miraculously reappear in the theta function via
$xU+yV+tW$; thus the Riemann surface itself contributes to establish
these as linear flow variables on the Jacobian.
The pole positions
$P_i$ do not vary with $x,y,t$ and $(\dagger)\,\,
u = 2\partial^2_x log\Theta(xU + yV  + tW + z_0) + c$ 
exhibits $\Theta$ as a tau function.
\\[3mm]\indent
We recall also that
a divisor $D^{*}$ of degree $g$ is dual to $D$ (relative to $Q$) if
$D + D^{*}$ is the null divisor of a meromorphic differential $d\hat{\Omega}
= dk + (\beta/k^2)dk + \cdots$ with a double pole at $Q$ (look at
$\zeta = 1/k$ to recognize the double pole).  Thus 
$D + D^{*} -2Q\sim K_{\Sigma}$ so ${\cal A}(D^{*}) - {\cal A}(Q) + K =
-[{\cal A}(D) - {\cal A}(Q) + K]$.  One can define then a function
$\psi^{*}(x,y,t,P) = exp(-kx-k^2y-k^3t)[1 + \xi_1^{*}/k) + \cdots]$
based on $D^{*}$ (dual BA function)
and a differential $d\hat{\Omega}$ with zero divisor $D+D^*$, such that
$\phi = \psi\psi^{*}d\hat{\Omega}$ is
meromorphic, having for poles only
a double pole at $Q$ (the zeros of $d\hat{\Omega}$ cancel
the poles of $\psi\psi^{*}$).  Thus $\psi\psi^*d\hat{\Omega}\sim \psi\psi^*(1+
(\beta/k^2+\cdots)dk$ is meromorphic with a second order pole at $\infty$,
and no other poles.  
For $L^{*} = L$ and $A^{*} = -A + 2w
-(3/2)u_x$ one has then $(\partial_y + L^{*})\psi^{*} = 0$ and
$(\partial_t + A^{*})\psi^{*} = 0$.  Note that 
the prescription above seems to specify for $\psi^*$
($\vec{U}=xU+yV+tW,\,\,z_0^* =
-{\cal A}(D^*)-K$)
\be
\psi^*\sim e^{-\int^P_{P_o}(xd\Omega^1+yd\Omega^2+td\Omega^3)}
\cdot\frac{\Theta({\cal A}(P)-\vec{U}+z_0^*)}
{\Theta({\cal A}(P)+z_0^*)}
\label{star}
\ee
\indent
In any event the message here is that for any Riemann surface 
$\Sigma$ one can
produce a BA function $\psi$ with assigned flow variables $x,y,t,\cdots$
and this $\psi$ gives rise to a 
(nonlinear) KP equation with solution $u$ linearized
on the Jacobian $J(\Sigma)$.
For averaging with KP (cf. \cite{cc,fb,ki}) 
we can use formulas (cf. (\ref{psi}) and
(\ref{star}))
\be
\psi = e^{px+Ey+\Omega t}\cdot\phi(Ux+Vy
+Wt,P)
\label{YDD}
\ee
\be
\psi^* = e^{-px-Ey-\Omega t}\cdot\phi^*
(-Ux-Vy-Wt,P)
\label{YEE}
\ee
with $\phi,\,\phi^*$ periodic,
to isolate the quantities of interest in averaging
(here $p=p(P),\,\,E = E(P),\,\,\Omega =
\Omega(P),$ etc.)
We think here of a general Riemann surface $\Sigma_g$ with holomorphic
differentials $d\omega_k$ and quasi-momenta and quasi-energies
of the form $dp=d\Omega^1,\,\,dE=d\Omega^2,\,\,d\Omega=d\Omega^3,\cdots
\,\,(p=\int_{P_0}^Pd\Omega^1$ etc.) where the $d\Omega^j=d\Omega_j=
d(\lambda^j+O(\lambda^{-1}))$ are meromorphic differentials of the second
kind.  Following \cite{ki,ka} one could
normalize now via $\Re\int_{A_j}d\Omega^k=
\Re\int_{B_j}d\Omega^k=0$ so that e.g.
$U_k=(1/2\pi i)\oint_{A_k}dp$ and $U_{k+g}=-(1/2\pi i)\oint_{B_k}dp\,\,
(k=1,\cdots,g)$ with similar stipulations for $V_k\sim\oint d\Omega^2,\,\,
W_k\sim\oint d\Omega^3,$ etc.  This leads to real $2g$ period vectors
and evidently one could also normalize via $\oint_{A_m}d\Omega^k=0$ 
(which we generally adopt in later sections) or
$\Im\oint_{A_m}d\Omega^k=\Im\oint_{B_m}d\Omega^k=0$ (further we set
$B_{jk}=\oint_{B_k}d\omega_j$).

\subsection{General remarks on averaging}

Averaging can be rather mysterious at first due partly to some hasty treatments
and bad choices of notation - plus many inherent difficulties.  Some
of the clearest exposition seems to be in \cite{bb,fa,fb,mt}
for which we go to hyperelliptic curves (cf. also \cite{bu,cc,ce}).
For 
hyperelliptic Riemann surfaces one can pick
any $2g+2$ points $\lambda_j\in {\bf P}^1$ and there will be a unique
hyperelliptic curve $\Sigma_g$ with a 2-fold map $f:\,\Sigma_g\to{\bf P}^1$ 
having branch locus $B=\{\lambda_j\}$.  Since any 3 points $\lambda_i,\,
\lambda_j,\,\lambda_k$ can be sent to $0,\,1,\,\infty$ by an automorphism
of ${\bf P}^1$ the general hyperelliptic surface of genus $g$ can be
described by $(2g+2)-3=2g-1$ points on ${\bf P}^1$.  Since $f$ is unique
up to an automorphism of ${\bf P}^1$ any hyperelliptic $\Sigma_g$ corresponds
to only finitely many such collections of $2g-1$ points so locally there
are $2g-1$ (moduli) parameters.  Since the moduli space of algebraic
curves has dimension $3g-3$ one sees that for $g\geq 3$ the generic 
Riemann surface is nonhyperelliptic whereas for $g=2$ all Riemann
surfaces are hyperelliptic (with 3 moduli).  For $g=1$ we have tori
or elliptic curves with one modulus $\tau$ and $g=0$ corresponds to
${\bf P}^1$.  In many papers on soliton mathematics and integrable systems
one takes real distinct branch points $\lambda_j,\,\,1\leq j\leq 2g+1$, and 
$\infty$, with $\lambda_1<\lambda_2<\cdots<\lambda_{2g+1}<\infty$ and
\be
\mu^2=\prod_1^{2g+1}(\lambda-\lambda_j)=P_{2g+1}(\lambda,\lambda_j)
\label{mu}
\ee
as the defining equation for $\Sigma_g$.  Evidently one could choose
$\lambda_1=0,\,\,\lambda_2=1$ in addition so for $g=1$ we could use
$0<1<u<\infty$ for a familiar parametrization with elliptic integrals,
etc.  One can take $d\lambda/\mu,\,\,\lambda d\lambda/\mu,\cdots,
\lambda^{g-1} d\lambda/\mu$ as a basis of holomorphic differentials on
$\Sigma_g$ but usually one takes linear combinations of these denoted
by $d\omega_j,\,\,1\leq j\leq g$, normalized via $\oint_{A_i}d\omega_j=
\delta_{ij}$, with period matrix defined via $\oint_{B_i}d\omega_j=
\Pi_{ij}$.  The matrix $\Pi=(\Pi_{ij})$ is symmetric with $\Im\Pi>0$ and
it determines the curve.  Frequently in situations arising from KdV
(Korteweg-deVries equation) for example one regards the intervals
$[\lambda_1,\lambda_2],\cdots,[\lambda_{2g+1},\infty)$ as spectral
bands and intervals $(\lambda_2,\lambda_3),\cdots,(\lambda_{2g},\lambda_
{2g+1})$ as gaps with the $a_i$ cycles around $(\lambda_{2i},\lambda_
{2i+1})\,\,(i=1,\cdots,g)$.
One will also want to consider another representation of hyperelliptic
curves of genus $g$ via
\be
\mu^2=\prod_0^{2g+1}(\lambda-\lambda_j)=P_{2g+2}(\lambda,\lambda_j)
\label{muu}
\ee
where $\infty$ is now not a branch point and in fact there are two points
$\mu_{\pm}$ corresponding to $\lambda=\infty$.
\\[3mm]\indent
We recall next some of the results and techniques of \cite{fa,mt}
where one can see explicitly the nature of things.
The presentation here follows \cite{cc}.
First from \cite{mt}, in a slightly different notation,
write $q_t = 6qq_x - q_{xxx}$ with Lax pair
$L = -\partial^2_x + q,\,\,B = -4\partial^3_x + 3(q\partial_x + \partial_x q),
\,\,L_t = [B,L],\,\,L\psi = \lambda\psi,$ and $\psi_t = B\psi$.
Let $\psi$ and $\phi$ be two solutions of the Lax pair equations and
set $\Psi = \psi\phi$; these are the very important ``square eigenfunctions"
which arise in many ways with interesting and varied meanings 
(cf. \cite{ch,ci,ct,cj,ce}).  Evidently
$\Psi$ satisfies
\be
[-\partial^3_x + 2(q\partial_x+\partial_x q)]\Psi = 4\lambda\partial_x\Psi;
\,\,\partial_t\Psi = -2q_x\Psi + 2(q+2\lambda)\partial_x\Psi
\label{AF}
\ee
From (\ref{AF}) one finds immediately the conservation law ({\bf C}):
$\partial_t[\Psi] + \partial_x[6(q-2\lambda)\Psi - 2\partial^2_x\Psi] = 0$.
If one looks for solutions of (\ref{AF}) of the form $\Psi(x,t,\lambda) =
1 + \sum_1^{\infty}[\Psi_j(x,t)]\lambda^{-j}$ as $\lambda\to\infty$ then
one obtains a recursion relation for polynomial densities
\be
\partial_x\Psi_{j+1} = [-\frac{1}{2}\partial^3_x + (q\partial_x + 
\partial_x q)]\Psi_j\,\,(j = 1,2,\cdots);\,\,\Psi_0 = 1
\label{AG}
\ee
Now consider the operator $L = -\partial^2_x + q$ in $L^2(-\infty,\infty)$
with spectrum consisting of closed intervals separated by exactly $N$ gaps
in the spectrum.  The $2N+1$ endpoints $\lambda_k$
of these spectral bands are
denoted by $-\infty<\lambda_0<\lambda_1<\cdots<\lambda_{2N}<\infty$ 
and called the
simple spectrum of $L$.  They can be viewed as constants of motion for
KdV when $L$ has this form.  We are dealing here with the hyperelliptic
Riemann surface determined via $R^2(\lambda)=\prod_0^{2N}(\lambda - \lambda_k)$
($\infty$ is a branch point)
and one can think of a manifold ${\cal M}$ of $N$-phase waves with
fixed simple spectrum as an $N$-torus based on $\theta_j\in [0,2\pi)$.
Hamiltonians in the KdV hierarchy generate flows on this torus and one
writes $q = q_N(\theta_1,\cdots,\theta_N)$ (the $\theta$ variables
arise if we use theta functions
for the integration - cf. also below).  
Now there is no $y$ variable so let us write
$\theta_j = x\kappa_j+tw_j$ (we will continue to use $d\omega_j$ for
normalized holomorphic differentials).
For details concerning the Riemann surface we refer to
\cite{bu,ce,du,df,fa,nd} and
will summarize here as follows.  For any $q_N$ as indicated one can find
functions $\mu_j(x,t)$ via $\Psi(x,t,\lambda) = \prod_1^N(\lambda -
\mu_j(x,t))$ where $\mu_j(x,t)\in [\lambda_{2j-1},\lambda_{2j}]$ and
satisfies
\be
\partial_x\mu_j = -2i(R(\mu_j)/\prod_{i\not= j}(\mu_j-\mu_i));
\label{AH}
\ee
$$\partial_t\mu_j = -2i[2(\sum_0^{2N}\lambda_k - 2\sum_{i\not= j}\mu_i)]
\cdot (R(\mu_j)/\prod_{i\not= j}(\mu_j - \mu_i)$$
In fact the $\mu_j$ live on the Riemann surface of $R(\lambda)$
in the spectral gaps
and as $x$ increases $\mu_j$ travels from $\lambda_{2j-1}$ to $\lambda_{2j}$
on one sheet and then returns to $\lambda_{2j-1}$ on the other sheet; this
path will be called the $j^{th}\,\mu$-cycle ($\sim A_j$).  
In the present context we
will write the theta function used for integration purposes as
$\Theta({\bf z},\tau) = \sum_{m\in {\bf Z}^N}exp[\pi i(2({\bf m},{\bf z})
+ ({\bf m},\tau{\bf m}))]$ where ${\bf z}\in {\bf Z}^N$ and $\tau$ 
denotes the $N\times N$ period matrix ($\tau$ is symmetric with $\Im\tau
> 0$).  We take canonical cuts $A_i,\,B_i\,\,(i = 1,\cdots,N)$ and
let $d\omega_j$ be holomorphic diffentials normalized via
$\int_{A_j} d\omega_k = \delta_{jk}$ (the cycle $A_j$ corresponds to
a loop around the cut $A_j$).  Then $q_N$ can be represented in the form
\be
q_N(x,t) = \Lambda + \Gamma -2\partial^2_xlog\Theta({\bf z}(x,t);\tau);\,\,
\Lambda = \sum_0^{2N}\lambda_j;
\label{AI}
\ee
$$\tau = (\tau_{ij}) = (\oint_{B_i}d\omega_j);\,\,\tau^{*}_{ij} = -\tau_{ij};
\,\,\Gamma = -2\sum_1^N\oint_{A
_j}\lambda d\omega_j$$
and ${\bf z}(x,t) = -2i[{\bf c}^N(x-x_0) + 2(\Lambda{\bf c}^N +
2{\bf c}^{N-1})t] + {\bf d}$ where $({\bf c}^N)_i = c_{iN}$ arises 
from the representation $d\omega_i = 
(\sum_1^N c_{ij}\lambda^{j-1})[d\lambda/R(\lambda)]$ (${\bf d}$ is a 
constant whose value is not important here).  Then the wave number
and frequency vectors can be defined via $\vec{\kappa} = -4i\pi
\tau^{-1}{\bf c}^N$ and $\vec{w} = -8i\pi\tau^{-1}[\Lambda
{\bf c}^N + 2{\bf c}^{N-1}]$ with $\theta_j(x,t) = \kappa_j x
+ w_j t + \theta_j^0$ (where the $\theta_j^0$ represent initial phases).
\\[3mm]\indent
To model the modulated wave now one writes
now $q = q_N(\theta_1,\cdots,\theta_N;\vec{\lambda})$ where $\lambda_j
\sim\lambda_j(X,T)$ and $\vec{\lambda}\sim(\lambda_j)$.
Then consider the first $2N+1$ polynomial
conservation laws arising from (\ref{AF}) - (\ref{AG})
and ${\bf C}$ for example (cf. below for KP)
and write these as $\partial_t{\cal T}_j(q) 
+ \partial_x{\cal X}_j(q) = 0$ (explicit formulas are given in 
\cite{fa} roughly as follows).
We note that the adjoint linear KdV equation (governing
the evolution of conserved densities) 
is $\partial_t\gamma_j + \partial^3_x\gamma_j - 6q\partial_x
\gamma_j = 0\,\,(\gamma_j\sim\nabla H_j)$ and (\ref{AG}) has the form
$\partial\gamma_{j+1} = (-(1/2)\partial^3 + q\partial +\partial q)\gamma_j$.
One then rewrites this to show that $6q\partial_x\gamma_j = \partial_x
[6\gamma_{j+1}-6q\gamma_j+3\partial^2\gamma_j]$ so that the adjoint
equation becomes
\be
\partial_t\gamma_j +\partial[-2\partial^2\gamma_j + 6q\gamma_j
-6\gamma_{j+1}]=0
\label{NF}
\ee
which leads to (\ref{AJ}) and (\ref{AN}) below (after simplification
of (\ref{NF})).  
Then for
the averaging step, write $\partial_t=\epsilon\partial_T$, etc.,
and average over the fast variable $x$ to obtain
\be
\partial_T<{\cal T}_j(q_N)> + \partial_X<{\cal X}_j(q_N)> = 0
\label{AJ}
\ee
((\ref{AJ}) makes the first order term in $\epsilon$ vanish).
The procedure involves averages
\be
<{\cal T}_j(q_N)> = lim_{L\to\infty}\frac{1}{2L}
\int_{-L}^L{\cal T}_j(q_N)dx
\label{AK}
\ee
for example (with a similar expression for $<{\cal X}_j(q_N)>$) and an
argument based on ergodicity is used.  Thus if the wave numbers
$\kappa_j$ are incommensurate the trajectory $\{q_N(x,t);\,\,x\in (-\infty,
\infty)\}$ will densely cover the torus ${\cal M}$.  Hence we can replace
$x$ averages with 
\be
<{\cal T}_j(q_N)> = \frac{1}{(2\pi)^N}\int_0^{2\pi}\cdots\int_0^{2\pi}
{\cal T}_j(q_N(\vec{\theta}))\prod_1^Nd\theta_j
\label{AL}
\ee
For computational purposes one can change the $\theta$ integrals to
$\mu$ integrals and obtain simpler calculations. By this procedure
one obtains a system of $2N+1$ first order partial differential equations
for the $2N+1$ points $\lambda_j(X,T)$, or equivalently for the
physical characteristics $(\vec{\kappa}(X,T),\vec{w}(X,T))$ (plus
$<q_N>$).
One can think of freezing the slow variables in the averaging and it is
$\underline{assumed}$ that (\ref{AJ}) is the correct first order description of
the modulated wave. 
\\[3mm]\indent
The above argument may or may not
have sounded convincing but it was in any case rather loose.
Let us be more precise following \cite{fa}.  One looks at the KdV
Hamiltonians beginning with $H = H(q) = lim_{L\to\infty}(1/2L)\int^L_
{-L}(q^2 + (1/2)q_x^2)dx$ (this form is appropriate for quasi-periodic
situations).  Then $\{f,g\} = lim_{L\to\infty}
(1/2L)\int^L_{-L}(\delta f/\delta q)\partial_x(\delta g/\delta q)dx$ 
(averaged Gardner bracket) and  
$q_t = \{q,H\}$.
The other Hamiltonians are found via
\be
\partial\frac{\delta H_{m+1}}{\delta q} = (q\partial + \partial q -
\frac{1}{2}\partial^3)\frac{\delta H_m}{\delta q}\,\,(m\geq 0);
\,\,\frac{\delta H_0}{\delta q} = 1
\label{AM}
\ee
where $\gamma_j\sim\nabla H_j\sim(\delta H_j/\delta q)$
(cf. here \cite{ce,fa}).  It is a general situation in the study of 
symmetries and conserved gradients (cf. \cite{cj}) that symmetries
will satisfy the linearized KdV equation $(\partial_t -6\partial_x q +
\partial^3_x)Q = 0$ and conserved gradients will satisfy the adjoint
linearized KdV equation $(\partial_t - 6q\partial_x + \partial^3_x)Q^{\dagger}
=0$; the important thing to notice here is that one is linearizing about
a solution $q$ of KdV.  Thus in our averaging processes the function
$q$, presumed known, is inserted in the integrals.  This leads then to
\be
{\cal T}_j(q) = \frac{\delta H_j}{\delta q};\,\,{\cal X}_j(q) =
-2\partial^2_x\frac{\delta H_j}{\delta q} - 6\frac{\delta H_{j+1}}{\delta q}
+ 6q\frac{\delta H_j}{\delta q}
\label{AN}
\ee
with (\ref{AJ}) holding, where $<\partial^2\phi>=0$ implies
\be
<{\cal X}_j> = lim_{L\to\infty}\frac{1}{2L}\int^L_{-L}(-6\frac
{\delta H_{j+1}}{\delta q_N} + 6q_N\frac{\delta H_j}{\delta q_N})dx
\label{AO}
\ee
In \cite{fa} the integrals are then simplified in terms of $\mu$ integrals
and expressed in terms of abelian differentials.  This is a beautiful
and important procedure linking the averaging process to the Riemann 
surface and is summarized in
\cite{mt} as follows.
One defines differentials 
\be
\hat{\Omega}_1 = -\frac{1}{2}[\lambda^N - \sum_1^Nc_j\lambda^{j-1}]\frac
{d\lambda}{R(\lambda)}
\label{AP}
\ee
$$
\hat{\Omega}_2 = [-\frac{1}{2}\lambda^{N+1} + \frac{1}{4}(\sum\lambda_j)
\lambda^N + \sum_1^N E_j\lambda^{j-1}]\frac{d\lambda}{R(\lambda)}
$$
where the $c_j,\,\,E_j$ are determined via 
$\oint_{b_i}\hat{\Omega}_1 = 0=\oint_{b_i}\hat{\Omega}_2\,\,(i = 1,
2,\cdots,N)$.  Then it can be shown that
\be
<\Psi>\sim
<{\cal T}>\sim\sum_0^{\infty}\frac{<{\cal T}_j>}{(2\mu)^j};\,\,
<{\cal X}>\sim\sum_0^{\infty}\frac{<{\cal X}_j>}{(2\mu)^j}
\label{NZ}
\ee
with
$\hat{\Omega}_1\sim<{\cal T}>(d\xi/\xi^2)$ and $<{\cal X}>(d\xi/\xi^2)\sim
12[(d\xi/\xi^4)-\hat{\Omega}_2]$ where $\mu=\xi^{-2}\to\infty\,\,(\mu\sim
(1/\sqrt{\xi})^{1/2}$) so $d\mu = -2\xi^{-3}d\xi\Rightarrow(d\xi/\xi^2)
\sim -(\xi/2)d\mu\sim-(d\mu/2\sqrt{\mu})$.  Since $\hat{\Omega}_1 =
O(\mu^N/\mu^{N+(1/2)})d\mu = O(\mu^{-(1/2)}d\mu,\,\,
\hat{\Omega}_2 =
O(\mu^{1/2})d\mu$ (with lead term $-(1/2)$) we obtain 
$<\Psi>\sim <{\cal T}>=O(1)$ and $<{\cal X}> = O(1)$.
Thus
(\ref{AF}), (\ref{AG}), ({\bf C}) generate all conservation laws
simultaneously with $<{\cal T}_j>$ (resp. $<{\cal X}_j>$) giving rise to
$\hat{\Omega}_1$ (resp. $\hat{\Omega}_2$).
It is then proved that all of the modulational
equations are determined via the equation 
\be
\partial_T\hat{\Omega}_1 = 12\partial_X\hat{\Omega}_2
\label{AR}
\ee
where the Riemann surface is thought of as depending on $X,T$ through
the points $\lambda_j(X,T)$.
In particular if the first $2N+1$ averaged conservation laws are satisfied
then so are all higher averaged conservation laws.  These equations
can also be written directly in terms of the $\lambda_j$ as Riemann
invariants via $\partial_T\lambda_j = S_j\partial_X\lambda_j$ for
$j = 0,1,\cdots,2N$ where $S_j$ is a computable characteristic speed
(cf. (\ref{FRR})-(\ref{FSS})).
Thus we have displayed the prototypical model for the Whitham or
modulational equations.
\\[3mm]\indent
Another way of looking at some of this goes as follows.
We will consider surfaces defined via $R(\Lambda)=
\prod_1^{2g+1}(\Lambda-\Lambda_i)$.
For convenience take the branch points $\Lambda_i$ real with $\Lambda_1
<\cdots<\Lambda_{2g+1}<\infty$.  This corresponds to spectral bands
$[\Lambda_1,\Lambda_2],\cdots,[\Lambda_{2g+1},\infty)$ and gaps 
$(\Lambda_2,\Lambda_3),\cdots,(\Lambda_{2g},\Lambda_{2g+1})$ with the
$A_i$ cycles around the gaps (i.e. $a_i\sim (\Lambda_{2i},\Lambda_{2i+1}),
\,\,i=1,\cdots,g$).
The notation is equivalent to what preceeds with a shift of index.
For this kind of situation one usually defines
the period matrix via $iB_{jk}=\oint_{B_k}\omega_j$ and sets
\be
\oint_{A_k}d\omega_j = 2\pi\delta_{jk}\,\,(j,k=1,\cdots,g);\,\,
d\omega_j = \sum_1^g\frac{c_{jq}\Lambda^{q-1}d\Lambda}{\sqrt{R(\Lambda)}}
\label{FHH}
\ee
(cf. \cite{bu,cc,ce}).
The $B_i$ cycles can be drawn from a common vertex ($P_0$ say) passing
through the gaps $(\Lambda_{2i},\Lambda_{2i+1})$.  One chooses now e.g.
$p=\int dp$ and $\Omega = \int d\Omega$ in the form
\be
p(\Lambda) = \int dp(\Lambda) = \int \frac{{\cal P}(\Lambda)d\Lambda}
{2\sqrt{R(\Lambda)}};\,\,{\cal P} = \Lambda^g+\sum_1^ga_j\Lambda^{g-j};
\label{FFF}
\ee
$$\Omega(\Lambda) = \int d\Omega(\Lambda) = \int\frac{6\Lambda^{g+1} 
+{\cal O}(\Lambda)}{\sqrt{R(\Lambda}}d\Lambda;\,\,{\cal O} = 
\sum_0^gb_j\Lambda^{g-j};\,\,b_0 = -3\sum_1^{2g+1}\Lambda_i$$
with normalizations
$\int^{\Lambda_{2i+1}}_{\Lambda_{2i}}dp(\Lambda) = \int^{\Lambda_{2i+1}}_
{\Lambda_{2i}}d\Omega(\Lambda) = 0;\,\,i=1,\cdots,g$.
We note that one is thinking here of (*) $\psi=exp[ipx+i\Omega t]\,\,\cdot$
theta functions (cf. (\ref{psi}) with e.g. 
$p(\Lambda) = -i\overline{(log\psi)_x};\,\,\Omega(\Lambda) = 
-i\overline{(log\psi)_t}$.
Recall that the notation $<\,,\,>_x$ simply
means $x$-averaging (or ergodic averaging)
and $\overline{(\log\psi)_x}\sim <(log\psi)_x>_x\not= 0$ here since e.g.
$(log\psi)_x$ is not bounded.
Observe that
(*) applies to any finite zone quasi-periodic situation.  
The KdV equation here arises from
$L\psi=\Lambda\psi,\,\,L=-\partial^2+q,\,\,\partial_t\psi=A\psi,\,\,
A=4\partial^3-6q\partial-3q_x$ and there is no need to put this
in a more canonical form since this material is only illustrative.
\\[3mm]\indent
In this context one has also the Kruskal integrals
$I_0,\cdots,I_{2g}$ which arise via a generating function
\be
p(\Lambda) = -i<(log\psi)_x>_x = \sqrt{\Lambda} + \sum_0^{\infty}
\frac{I_s}{(2\sqrt{\Lambda})^{2s+1}}
\label{FLL}
\ee
where $I_s = <P_s>_x = \overline{P_s}
\,\,(s = 0,1,\cdots$) with $-i(log\psi)_x =
\sqrt{\Lambda} + \sum_0^{\infty}[P_s/(2\sqrt{\Lambda})^{2s+1}]$.
Similarly 
\be
-i\overline{(log\psi)_t} = -i\frac{A\psi}{\psi} = 4(\sqrt{\Lambda})^3 +
\sum_0^{\infty}\frac{\Omega_s}{(2\sqrt{\Lambda})^{2s+1}}
\label{FMM}
\ee
and one knows $\partial_tP_s = \partial_x\Omega_s$ since 
$(\spadesuit)\,\, [(log\psi)_x]_t =
[(log\psi)_t]_x$.  The expansions are standard (cf. \cite{cc,ch,ci,ct}).
Now consider a ``weakly deformed" soliton lattice of the form 
\be
\theta(\tau|B)=\sum_{-\infty<n_1<\cdots<n_m<\infty}exp(-\frac{1}{2}
\sum_{j,k}B_{jk}n_jn_k + i\sum_jn_j\tau_j)
\label{FII}
\ee
with the $\Lambda_i\,\,(i=1,\cdots,2g+1)$ (or equivalently the parameters
$u^i = I_i;\,\,i=0,\cdots,2g$) slowly varying functions of $x,t$ (e.g.
$u^i = u^i(X,T),\,\,X = \epsilon x,\,T=\epsilon t,\,\,i = 0,1,\cdots,2m$).
Now one wants to obtain a version of (\ref{AR}) directly via
$(\spadesuit)$.  Thus insert the slow variables in $(\spadesuit)$ and average,
using $\epsilon\partial_X$ or $\epsilon\partial_T$ in the external derivatives,
to obtain
\be
\partial_T\overline{(log\psi)_x} = \partial_X\overline{(log\psi)_t}
\label{FOO}
\ee
or $\partial_Tp(\Lambda) = \partial_XE(\Lambda)$.  Then from (\ref{FFF})
differentiating in $\Lambda$ one gets
\be
\partial_Tdp = \partial_Xd\Omega
\label{FPP}
\ee
Now (recall $\partial_tP_s = \partial_x
\Omega_s$) expanding (\ref{FPP}) in powers of $(\sqrt{\Lambda})^{-1}$ one
obtains the slow modulation equations in the form
\be
\partial_T u^s = \partial_X\bar{\Omega}_s\,\;\,(s=0,\cdots,2g)
\label{FQQ}
\ee
where $\overline{\Omega_s}$ is a function of the $u^i\,\,(0\leq i\leq 2g)$.
This leads to equations
\be
\partial_T\Lambda_i = v_i(\Lambda_1,\cdots,\Lambda_{2g+1})\partial_X
\Lambda_i\,\;\,(i=1,\cdots,2g+1)
\label{FRR}
\ee
for the branch points $\Lambda_k$ as Riemann invariants.  The characteristic
``velocities" have the form
\be
v_i = \left.\frac{d\Omega}{dp}\right|_{\Lambda=\Lambda_i} =
2\frac{6\Lambda_i^{g+1} +
\Omega(\Lambda_i)}{p(\Lambda_i)}\,\;\,(1\leq i\leq 2g+1)
\label{FSS}
\ee
To see this simply multiply (\ref{FPP}) by $(\Lambda-\Lambda_i)^{3/2}$ and
pass to limits as $\Lambda\to\Lambda_i$.

\subsection{Averaging with $\psi^*\psi$}

Let us look now at \cite{ki} but in the spirit of \cite{cc,fb}.
We will only sketch this here and refer to \cite{cc} for more detail.
Thus
consider KP in a standard 
form $3\sigma^2u_{yy} + \partial_x(4u_t -
6uu_x + u_{xxx}) = 0$ via compatibility $[\partial_y -L,\partial_t - A]
=0$ where $L = \sigma^{-1}(\partial^2-u)$ and $A = \partial^3 - (3/2)
u\partial + w\,\, (\sigma^2 = 1$ is used in \cite{fb} which we follow for
convenience but the procedure should work in general with minor
modifications - note $\partial$ means $\partial_x$ and $u\to -u$ in the
development of (\ref{psi})).  We have then
$(\partial_y - L)\psi = 0$ with $(\partial_t -A)\psi = 0$ and for the 
adjoint or dual wave function 
$\psi^{*}$ one writes in \cite{ka} $\psi^{*}L = -\partial_y\psi^{*}$
with $\psi^{*}A = \partial_t\psi^{*}$ where $\psi^{*}(f\partial^j)\equiv
(-\partial)^j(\psi^{*}f)$.  We modify the formulas used in \cite{ki}
(and \cite{fb}) in taking (cf. \cite{cc} - $\phi$ and $\phi^*$ are periodic)
\be
\psi = e^{px+Ey+\Omega t}\cdot\phi(Ux+Vy
+Wt,P);\,\,
\psi^* = e^{-px-Ey-\Omega t}\cdot\phi^*
(-Ux-Vy-Wt,P)
\label{YE}
\ee
where $p=p(P),\,\,E = E(P),\,\,\Omega =
\Omega(P),$ etc. (cf. (\ref{YDD}) - (\ref{YEE})), which isolate
the quantities needed in averaging. 
The arguments to follow are essentially the same for this
choice of notation, or that in \cite{fb} or \cite{ki}.
Now one sees immediately that 
\be
(\psi^{*}L)\psi = \psi^{*}L\psi + \partial_x(\psi^{*}L^1\psi) +
\partial^2_x(\psi^{*}L^2\psi) + \cdots
\label{AS}
\ee
where e.g. $L^r = ((-1)^r/r!)(d^r L/d(\partial)^r)$.  In particular
$L^1 = -2\partial$ and $L^2 = 1$ while
$A^1 = -3\partial^2+(3/2)u,\,\,A^2 = 3\partial,$ and $A^3 =
-1$.
We think of a general Riemann surface $\Sigma_g$.  Here one picks
holomorphic differentials 
$d\omega_k$ as before and quasi-momenta, quasi-energies, etc. via
$dp\sim d\Omega_1,\,\,dE\sim d\Omega_2,\,\,d\Omega\sim d\Omega_3,\cdots$
where $\lambda\sim k,\,\,p = \int^P_{P_0}d\Omega_1,$ etc.
Normalize
the $d\Omega_k$ so that $\Re\int_{A_i}d\Omega_k = 0 = \Re\int_{B_j}
d\Omega_k$ (cf. remarks after (\ref{YEE})); then
$U,\,V,\,W,\cdots$ are real $2g$ period vectors.
and one has BA functions $\psi(x,y,t,P)$ as
in (\ref{YE}).  As before we look for approximations 
based on
$u_0(xU+yV+tW|I) = u_0(\theta_j,I_k)$.  For averaging, $\theta_j \sim
xU_j + yV_j + tW_j,\,\,1\leq j\leq 2g$,
with period $2\pi$ in the
$\theta_j$ seems natural (but note
$\theta_j,\,\theta_{g+j}\sim U_j,$ etc.).  
Then again by ergodicity
$<\phi>_x = lim_{L\to\infty}(1/2L)\int_{-L}^L\phi dx$ becomes
$<\phi> = (1/(2\pi)^{2g}\int\cdots\int\phi d^{2g}\theta$ and one
notes that $<\partial_x\phi> = 0$ automatically
for $\phi$ bounded.  In \cite{fb} one thinks
of $\phi(xU+\cdots)$ with $\phi_x = \sum U_i(\partial\phi/\partial\theta_i)$
and $\int\cdots\int(\partial
\phi/\partial\theta_i)d^{2g}\theta = 0$. 
\\[3mm]\indent
Now for averaging we think of $u_0\sim u_0
(\frac{1}{\epsilon}\hat{S}|I)$ 
with $(\hat{S},I)\sim (\hat{S},I)(X,Y,T),
\partial_X \hat{S} = U,\,\,\partial_Y
\hat{S} = V,$ and $\partial_T\hat{S} = W$.  We think of expanding
about $u_0$ with $\partial_x\to\partial_x + \epsilon\partial_X$.
This step will cover both $x$ and $X$ dependence for subsequent averaging.  
Then look at the compatibility condition $(\clubsuit):\,\,\partial_tL
-\partial_y A + [L,A] = 0$.  As before we will want the term of first
order in $\epsilon$ upon writing e.g. $L = L_0 + \epsilon L_1 + \cdots$
and $A = A_0 + \epsilon A_1 + \cdots$ where slow variables appear only
in the $L_0,\,A_0$ terms.
Details are indicated in \cite{cc} and after some calculation
$(\clubsuit)$ becomes
\be
\partial_tL_0 -\partial_yA_0 +[L_0,A_0] +\epsilon\{\partial_tL_1
-\partial_yA_1 +\partial_TL-\partial_YA +[L_0,\hat{A}_1] +[\hat{L}_1,A_0]\}
+O(\epsilon^2)
\label{YK}
\ee
Now the $\partial_t
L_0-\partial_yA_0+[L_0,A_0]$ term vanishes and further calculation shows
that to make the coefficient of $\epsilon$ vanish one wants
\be
\partial_t L_1 - \partial_y A_1 + [L_0,A_1] + [L_1,A_0] + F = 0;\,\,
F=\partial_TL - \partial_YA - (L^1\partial_XA - A^1\partial_XL)
\label{AT}
\ee
However via ergodicity in $x,y$, or $t$ flows, averaging of derivatives in
$x,y$, or $t$ gives zero, so from (\ref{AT}) and further 
computation we obtain
the Whitham equations in the form $<\psi^*F\psi> = 0$.
In order to spell
this out in \cite{fb} one imagines $X,Y,T$ as a parameter $\xi$ and 
considers $L(\xi),\,\,A(\xi)$, etc. and $\psi\sim\psi(\xi)$ (with
$I\sim (I_k)$ being any additional factors dependent on the slow times) so that
\be
\psi(\xi) = e^{p(\xi)x+E(\xi)y+\Omega(\xi)t}\cdot
\phi(U(\xi)x + V(\xi)y 
+ W(\xi)t|I(\xi))
\label{AV}
\ee
but one keeps $\psi^* = exp(-px-Ey-\Omega t)\phi^*(-Ux-Vy-Wt|I)$
fixed in $\xi$ 
(i.e. assume $p,E,\Omega,U,V,W,I$ fixed in $\psi^*$).
We recall that
one expects $\lambda_k = \lambda_k(X,Y,T)$ etc. so the Riemann surface
varies with $\xi$.  The procedure here is somewhat contrived but seems
appropriate for heuristic purposes at least.
Also recall that
$x,y,t$ and $X,Y,T$ can be considered as independent variables.  
Note also from (\ref{AV}) for $P$ fixed
($\theta\sim xU+yV+yW$)
\be
\partial_{\xi}\psi^*\psi(\xi)|_{\xi=0}= (\dot{p}x+\dot{E}y+\dot{\Omega}t)
\psi^*\psi + 
\label{AX}
\ee
$$ + (\dot{U}x+\dot{V}y +\dot{W}t)\cdot\psi^*\partial_{\theta}\psi +
\dot{I}\cdot\psi^*\partial_I\psi$$
where $\dot{f}\sim\partial f/\partial\xi$.  In \cite{fb} one assumes
that it is also permitted to vary $\xi$ and hold
e.g. the $I_k$ constant while allowing say the $P$ to vary.
Now one computes terms like
$\partial_{\xi}[\partial_t(\psi^*\psi(\xi))]|_{\xi=0}$,
averages, and subsequently inserts $X,Y,T$ for $\xi$.
Taking into account compatibility relations
\be
\partial_YU = \partial_XV;\,\,\partial_TU = \partial_XW;\,\,\partial_TV
= \partial_YW
\label{BH}
\ee
after considerable calculation one arrives at the Whitham
equations in the form
\be
p_T = \Omega_X;\,\,p_Y = E_X;\,\,E_T = \Omega_Y
\label{BM}
\ee
(cf. Section 7 for general forms).
We feel that this derivation from \cite{cc}, based on \cite{fb,ki}, 
is important since it
again exhibits again the role of square eigenfunctions (now in the form
$\psi^*\psi$) in dealing with averaging processes (a derivation
based on (\ref{FOO}) also seems important). 
In view of the
geometrical nature of such square eigenfunctions (cf. \cite{cj} for
example) one might look for underlying geometrical objects related to
the results of averaging.  Another (new) direction 
involves the Cauchy kernels expressed via $\psi^*\psi$
and their dispersionless limits (cf. \cite{cf,cb}).   It is also proved in
\cite{cc} that
$dp = <\psi\psi^*>d\hat{\Omega},\,\,
dE\sim -<\psi^*L^1\psi>d\hat{\Omega}$, and
$d\Omega \sim -<\psi^*A^1\psi>d\hat{\Omega}$.
This shows in particular how
the quantity $\psi\psi^*$ determines the Whitham 
differentials $dp,\,\,dE,\,\,d\Omega$, etc.
\\[3mm]\indent {\bf REMARK 2.1}.$\,\,$  Since $D+D^*-2\infty\sim K_{\Sigma},
\,\,\psi\psi^*$ is determined by a section of $K_{\Sigma}$ (global
point of view) but it is relations based on $<\psi^*L^1\psi>,\,\,
<\psi^*A^1\psi>,$ etc. (based on the Krichever averaging process)
which reveal the ``guts" of $\psi\psi^*$ needed for averaging and the
expression of differentials.

\subsection{Dispersionless theory}

We give next a brief sketch of some ideas regarding dispersionless KP
(dKP) following mainly \cite{ch,ci,ct,kk,ta} to which we refer for 
philosophy.  We will make various notational adjustments as we go along.  One
can think of fast and slow variables with $\epsilon x=X$ and $\epsilon t_n=
T_n$ so that $\partial_n\to\epsilon\partial/\partial T_n$ and $u(x,t_n)
\to\tilde{u}(X,T_n)$ to obtain from the KP equation $(1/4)u_{xxx}+3uu_x
+(3/4)\partial^{-1}\partial^2_2u=0$ the equation $\partial_T\tilde{u}=3
\tilde{u}\partial_X\tilde{u}+(3/4)\partial^{-1}(\partial^2\tilde{u}/
\partial T_2^2)$ when $\epsilon\to 0$ ($\partial^{-1}\to(1/\epsilon)
\partial^{-1}$).  In terms of hierarchies the theory can be built around the
pair $(L,M)$ in the spirit of \cite{ci,cj,ta}.  Thus writing $(t_n)$ for
$(x,t_n)$ (i.e. $x\sim t_1$ here) consider
\be
L_{\epsilon}=\epsilon\partial+\sum_1^{\infty} u_{n+1}(\epsilon,T)
(\epsilon\partial)^{-n};\,\,M_{\epsilon}=\sum_1^{\infty}nT_nL^{n-1}_{\epsilon}
+\sum_1^{\infty}v_{n+1}(\epsilon,T)L_{\epsilon}^{-n-1}
\label{YAA}
\ee
Here $L$ is the Lax operator $L=\partial+\sum_1^{\infty}u_{n+1}\partial^{-n}$
and $M$ is the Orlov-Schulman operator defined via $\psi_{\lambda}=M\psi$.
Now one assumes $u_n(\epsilon,T)=U_n(T)+O(\epsilon)$, etc. and 
set (recall $L\psi=\lambda\psi$)
$$
\psi=\left[1+O\left(\frac{1}{\lambda}\right)\right]exp
\left(\sum_1^{\infty}\frac{T_n}
{\epsilon}\lambda^n\right)=exp\left(\frac{1}
{\epsilon}S(T,\lambda)+O(1)\right);$$
\be
\tau=exp\left(\frac{1}{\epsilon^2}F(T)+O\left(\frac{1}{\epsilon}\right)
\right)
\label{YBB}
\ee
We recall that $\partial_nL=[B_n,L],\,\,B_n=L^n_{+},\,\,\partial_nM
=[B_n,M],\,\,[L,M]=1,\,\,L\psi=\lambda\psi,\,\,\partial_{\lambda}\psi
=M\psi,$ and $\psi=\tau(T-(1/n\lambda^n))exp[\sum_1^{\infty}T_n\lambda^n]/
\tau(T)$.  Putting in the $\epsilon$ and using $\partial_n$ for
$\partial/\partial T_n$ now, with $P=S_X$, one obtains
\be
\lambda=P+\sum_1^{\infty}U_{n+1}P^{-n};\,\,
P=\lambda-\sum_1^{\infty}P_i\lambda^{-1};
\label{YCC}
\ee
$${\cal
M}=\sum_1^{\infty}nT_n\lambda^{n-1}+\sum_1^{\infty}V_{n+1}\lambda^{-n-1};
\,\,\partial_nS={\cal B}_n(P)\Rightarrow \partial_nP=\hat{\partial}
{\cal B}_n(P)$$
where $\hat{\partial}\sim \partial_X+(\partial P/\partial X)\partial_P$
and $M\to {\cal M}$.
Note that one assumes also $v_{i+1}(\epsilon,T)=V_{i+1}(T)+O(\epsilon)$; 
further
for $B_n=\sum_0^nb_{nm}\partial^m$ one has ${\cal B}_n=\sum_0^nb_{nm}
P^m$ (note also $B_n=L^n+\sum_1^{\infty}\sigma_j^nL^{-j}$).  
We list a few additional formulas which are 
easily obtained (cf. \cite{ci}); thus, writing $\{A,B\}=\partial_PA\partial A
-\partial A\partial_PB$ one has
\be
\partial_n\lambda=\{{\cal B}_n,\lambda\};\,\,\partial_n{\cal M}
=\{{\cal B}_n,{\cal M}\};\,\,\{\lambda,{\cal M}\}=1
\label{YDDD}
\ee
Now we can write $S=\sum_1^{\infty}T_n\lambda^n+\sum_1^{\infty}S_{j+1}
\lambda^{-j}$ with $\partial_mS_{j+1}=\tilde{\sigma}_j^m,\,\,V_{n+1}=-nS_{n+1}$,
and $\partial_{\lambda}S={\cal M}\,\, (
\sigma_j^m\to\tilde{\sigma}_
j^m$).  Further 
\be
{\cal B}_n=\lambda^n+\sum_1^{\infty}\partial_nS_{j+1}\lambda^{-j};\,\,
\partial S_{n+1}\sim -P_n\sim -\frac{\partial V_{n+1}}{n}\sim
-\frac{\partial\partial_n F}{n}
\label{YEEE}
\ee
\indent
We sketch next a few formulas from \cite{kk}.  First
it will be important to rescale the $T_n$ variables and write 
$t'=nt_n,\,\,T_n'=nT_n,\,\,
\partial_n=n\partial'_n=n(\partial/\partial T'_n)$.  Then
\be
\partial'_nS=\frac{\lambda^n_{+}}{n};\,\,\partial'_n\lambda=\{{\cal Q}_n,
\lambda\}\,\,({\cal Q}_n=\frac{{\cal B}_n}{n});
\label{YFF}
\ee
$$\partial'_nP=\hat{\partial}{\cal Q}_n=\partial{\cal Q}_n+\partial_P
{\cal Q}_n\partial P;\,\,\partial'_n{\cal Q}_m-\partial'_m{\cal Q}_n=
\{{\cal Q}_n,{\cal Q}_m\}$$
Think of $(P,X,T'_n),\,\,n\geq 2,$ as basic Hamiltonian variables
with $P=P(X,T'_n)$.  Then $-{\cal Q}_n(P,X,T'_n)$ will serve as a
Hamiltonian via
\be
\dot{P}'_n=\frac{dP'}{dT'_n}=\partial{\cal Q}_n;\,\,\dot{X}'_n=\frac
{dX}{dT'_n}=-\partial_P{\cal Q}_n
\label{YGG}
\ee
(recall the classical theory for variables $(q,p)$ involves $\dot{q}=
\partial H/\partial p$ and $\dot{p}=-\partial H/\partial q$).  The function
$S(\lambda,X,T_n)$ plays the role of part of a generating function 
$\tilde{S}$ for the Hamilton-Jacobi theory with action angle variables 
$(\lambda,-\xi)$ where
\be
PdX+{\cal Q}_ndT'_n=-\xi d\lambda-K_ndT'_n+d\tilde{S};\,\,K_n=-R_n=-
\frac{\lambda^n}{n};
\label{YHH}
\ee
$$\frac{d\lambda}{dT'_n}=\dot{\lambda}'_n=\partial_{\xi}R_n=0;\,\,\frac
{d\xi}{dT'_n}=\dot{\xi}'_n=-\partial_{\lambda}R_n=-\lambda^{n-1}$$
(note that $\dot{\lambda}'_n=0\sim\partial'_n\lambda=\{{\cal Q}_n,
\lambda\}$).  To see how all this fits together we write
\be
\frac{dP}{dT'_n}=\partial'_nP+\frac{\partial P}{\partial X}\frac{dX}
{dT'_n}=\hat{\partial}{\cal Q}_n+\frac{\partial P}{\partial X}\dot
{X_n}'=\partial{\cal Q}_n+\partial P\partial_P{\cal Q}_n+\partial P
\dot{X}'_n
\label{YI}
\ee
This is compatible with (\ref{YGG}) and Hamiltonians $-{\cal Q}_n$.  Furthermore
one wants
\be
\tilde{S}_{\lambda}=\xi;\,\,\tilde{S}_X=P;\,\,\partial'_n\tilde{S}=
{\cal Q}_n-R_n
\label{YJJ}
\ee
and from (\ref{YHH}) one has
\be
PdX+{\cal Q}_ndT'_n=-\xi d\lambda+R_ndT'_n+\tilde{S}_XdX+\tilde{S}_
{\lambda}d\lambda+\partial'_n\tilde{S}dT'_n
\label{YKK}
\ee
which checks.  We note that $\partial'_nS={\cal Q}_n={\cal B}_n/n$ and
$S_X=P$ by constructions and definitions.  Consider $\tilde{S}=S-\sum_2^{\infty}
\lambda^nT'_n/n$.  Then $\tilde{S}_X=S_X=P$ and $\tilde{S}_n'=S_n'-R_n=
{\cal Q}_n-R_n$ as desired with $\xi=\tilde{S}_{\lambda}=S_{\lambda}-
\sum_2^{\infty}T'_n\lambda^{n-1}$.  It follows that
$\xi\sim{\cal M}-\sum_2^{\infty}T'_n\lambda^{n-1}=X+\sum_1^{\infty}V_{i+1}
\lambda^{-i-1}$.  If $W$ is the gauge operator such that $L=W\partial W^{-1}$
one sees easily that
\be
M\psi
=W\left(\sum_1^{\infty}kx_k\partial^{k-1}\right)W^{-1}\psi=\left(G+
\sum_2^{\infty}kx_k\lambda^{k-1}\right)\psi
\label{YLL}
\ee
from which follows that $G=WxW^{-1}\to\xi$.  This shows that $G$ is
a very fundamental object and this is encountered in various places
in the general theory (cf. \cite{ci,cj,ya}).
\\[3mm]\indent {\bf REMARK 2.2.}$\,\,$ We refer here also to \cite{ch,ct}
for a complete characterization of dKP and the solution of the dispersionless
Hirota equations.
and will sketch this here.  Thus we follow
\cite{ch} (cf. also \cite{kk,ta}) and begin
with two pseudodifferential operators ($\partial = 
\partial/\partial x$),
\be
L = \partial + \sum_1^{\infty}u_{n+1}\partial^{-n};\, \ \,W = 1 + 
\sum_1^ {\infty}w_n\partial^{-n} 
\label{ZQ}
\ee
called the Lax operator and gauge operator respectively, where the 
generalized Leibnitz rule with $\partial^{-1} \partial = \partial 
\partial^{-1} = 1$ applies \be
\partial^if = \sum_{j=0}^{\infty}{i \choose j}(\partial^j f) 
\partial^ {i-j} 
\label{AAB}
\ee
for any $i \in {\bf Z}$, and $L = W\partial\,W^{-1}$. The KP hierarchy 
then is determined by
the Lax equations ($\partial_n = \partial/\partial t_n$), \be \partial_n L = 
[B_n,L] = B_n L - L B_n 
\label{AAC} \ee 
where $B_n = L^n_{+} $ is the 
differential part of $L^n = L^n_{+} + L^n_{-} = \sum_0^{\infty}
\ell_i^n\partial^i + \sum_{-\infty}^ {-1}\ell_i^n\partial^i$. One can also 
express this via the Sato equation, \be \partial_n W\,W^{-1} = -L^n_{-} 
\label{AAD}
\ee
which is particularly well adapted to the dKP theory. Now define the wave 
function
via
\be
\psi = W\,e^{\xi} = w(t,\lambda)e^{\xi};\, \ \,\xi = \sum_1^{\infty}t_n 
\lambda^n;
\, \ \,w(t,\lambda) = 1 + \sum_1^{\infty}w_n(t)\lambda^{-n} 
\label{AAE} 
\ee 
where $t_1 = x$. There is also an adjoint wave function $\psi^{*} = 
W^{*-1} \exp(-\xi) = w^{*}(t,\lambda)\exp(-\xi),\\w^{*}(t,\lambda) = 1 + 
\sum_1^ {\infty}w_i^{*}(t)\lambda^{-i}$, and one has equations 
\be L\psi = 
\lambda\psi;\, \ \,\partial_n\psi = B_n\psi;\, \ \,L^{*}\psi^{*} = 
\lambda\psi^{*};\, \ \,\partial_n\psi^{*} = -B_n^{*}\psi^{*} 
\label{AAF} 
\ee
Note that the KP hierarchy (\ref{AAC}) is then given by the compatibility 
conditions among these equations, treating $\lambda$ as a constant.
Next one 
has the fundamental tau function $\tau(t)$ and vertex operators ${\bf X},
\,\,{\bf X}^{*}$ satisfying
\be
\psi(t,\lambda) = \frac{{\bf X}(\lambda)\tau (t)}{\tau (t)} = \frac{e^{\xi} 
G_{-}(\lambda)\tau (t)}{\tau (t)} = \frac{e^{\xi}\tau(t-[\lambda^{-1}])} 
{\tau (t)};
\label{AAG}
\ee
$$\psi^{*}(t,\lambda) = \frac{{\bf X}^{*}(\lambda)\tau (t)} {\tau (t)} = 
\frac{e^{-\xi}
G_{+}(\lambda)\tau (t)}{\tau (t)} = \frac{e^{-\xi}\tau(t+[\lambda^{-1}])} 
{\tau (t)} $$
where $G_{\pm}(\lambda) = \exp(\pm\xi(\tilde{\partial},\lambda^{-1}))$ with 
$\tilde{\partial} = (\partial_1,(1/2)\partial_2,(1/3)\partial_3, \cdots)$ 
and $t\pm[\lambda^{-1}]= (t_1\pm \lambda^{-1},t_2 \pm (1/2) \lambda^{-2}, 
\cdots)$.
One writes also
\be
e^{\xi} = \exp \left({\sum_1^{\infty}t_n\lambda^n}\right) = \sum_0^ 
{\infty}\chi_j(t_1, t_2, \cdots ,t_j) \lambda^j  
\label{AAH} \ee 
where 
the $\chi_j$ are the elementary Schur polynomials, which arise in many 
important formulas (cf. below). \\[3mm]\indent We mention now the famous 
bilinear identity which generates the entire KP hierarchy. This has the 
form 
\be
\oint_{\infty}\psi(t,\lambda)\psi^{*}(t',\lambda)d\lambda = 0 
\label{AAI} 
\ee 
where $\oint_{\infty}(\cdot)d\lambda$ is the residue integral about 
$\infty$, which we also denote $Res_{\lambda}[(\cdot)d\lambda]$. Using 
(\ref{AAG}) this can also be written in terms of tau functions as 
\be 
\oint_{\infty}\tau(t-[\lambda^{-1}])\tau(t'+[\lambda^{-1}]) 
e^{\xi(t,\lambda)-\xi(t',\lambda)}d\lambda = 0  
\label{AAJ} 
\ee This leads to the characterization of the tau function in 
bilinear form expressed via ($t\to t-y,\,\,t'\to t+y$) 
\be 
\left(\sum_0^{\infty}\chi_n(-2y)\chi_{n+1}(\tilde{\partial})
e^{\sum_1^{\infty}
y_i \partial_i}\right)\tau\,\cdot\,\tau = 0  
\label{AAK} 
\ee 
where $\partial^m_j a\,\cdot\,b = (\partial^m/
\partial s_j^m) a(t_j+s_j)b(t_j-s_j)|_{s=0}$ and $\tilde{\partial} = 
(\partial_1,(1/2) \partial_2,(1/3)\partial_3,\cdots)$. 
In particular, we have from the 
coefficients of $y_n$ in (\ref{AAK}), 
\be \label{hirota}
\partial_1\partial_n\tau \cdot \tau = 2 \chi_{n+1} (\tilde{\partial}) 
\tau \cdot \tau  
\ee 
which are called the Hirota bilinear equations.  One has also 
the Fay identity via (cf.
\cite{af,ci,ct,ce} - c.p. means cyclic permutations) 
\be \sum_{c.p.}
(s_0-s_1)(s_2-s_3)\tau(t+[s_0]+[s_1])\tau(t+[s_2]+[s_3]) = 0  
\label{ZR}
\ee
which can be derived from the bilinear identity (\ref{AAJ}). 
Differentiating this in $s_0$, then setting $s_0 = s_3 = 0$, then 
dividing by $s_1 s_2$, and finally shifting $t\to t-[s_2]$, leads to 
the differential Fay identity,
\begin{eqnarray}
\nonumber
& &\tau(t)\partial\tau(t+[s_1]-[s_2]) - \tau(t+[s_1] -[s_2])\partial 
\tau(t) \\ & &= (s_1^{-1}-s_2^{-1}) \left[\tau(t+[s_1]-[s_2]) \tau(t) - 
\tau(t+[s_1])\tau(t-[s_2])\right]  
\label{AAL} \end{eqnarray} 
The 
Hirota equations
(\ref{hirota}) can be also derived from (\ref{AAL}) by taking the 
limit $s_1 \to s_2$. The identity (\ref{AAL}) will play an important 
role later. \\[3mm]\indent
Now for the dispersionless theory (dKP) one can think of fast and slow 
variables, etc., or averaging procedures, but simply one takes 
$t_n\to\epsilon t_n = T_n\,\,(t_1 = x\to \epsilon x = X)$ in the 
KP equation $u_t = (1/4)u_{xxx} + 3uu_x + (3/4)\partial^{-1}u_{yy},
\,\, (y=t_2,\,\,t=t_3)$, with $\partial_n\to \epsilon\partial/\partial 
T_n$ and $u(t_n)\to U(T_n)$ to obtain $\partial_T U = 3UU_X + 
(3/4)\partial^ {-1}U_{YY}$ when $\epsilon\to 0\,\,(\partial =
\partial/\partial X$ now). Thus the dispersion term $u_{xxx}$ is
removed. In terms of hierarchies we write 
\be
L_{\epsilon} = \epsilon\partial + \sum_1^{\infty}u_{n+1}(T/\epsilon) 
(\epsilon\partial)^{-n} 
\label{AAM}
\ee
and think of $u_n( T/\epsilon)= U_n(T) + O(\epsilon)$, etc. One takes 
then a WKB form for the wave function with the action $S$  
\be \psi = \exp \left[\frac{1}{\epsilon}S(T,\lambda) \right] \label{AAN} 
\ee 
Replacing now $\partial_n$ by $\epsilon\partial_n$, where $\partial_n = 
\partial/\partial T_n$ now, we define $P = \partial S = S_X$. Then 
$\epsilon^i\partial^i\psi\to P^i\psi$ as $\epsilon\to 0$ and the equation 
$L\psi = \lambda\psi$ becomes
\be
\lambda = P + \sum_1^{\infty}U_{n+1}P^{-n};\, \ \,P = \lambda - 
\sum_1^{\infty}P_{i+1}\lambda^{-i}  
\label{AAO} \ee 
where the second 
equation is simply the inversion of the first. We also note from 
$\partial_n\psi =
B_n\psi = \sum_0^nb_{nm}(\epsilon\partial)^m\psi$ that one obtains 
$\partial_n S = {\cal B}_n(P) = \lambda^n_{+}$ where the subscript (+) 
refers now to powers of $P$ (note $\epsilon\partial_n\psi/\psi \to 
\partial_n S$). Thus $B_n = L^n_{+}\to {\cal B}_n(P) = \lambda^n_{+} = 
\sum_0^nb_{nm}P^m$ and the KP hierarchy goes to 
\be \partial_n P = 
\partial {\cal B}_n 
\label{YYCC}
\ee
which is the dKP hierarchy
(note $\partial_n S = {\cal B}_n\Rightarrow \partial_n P = 
\partial{\cal B}_n$).
The action $S$ in (\ref{AAN}) can be computed from (\ref{AAG}) in the 
limit $\epsilon \to 0$ as
\be
\label{action}
S = \sum_{1}^{\infty} T_n \lambda^n - \sum_{1}^{\infty} {\partial_mF 
\over m} \lambda^{-m}
\ee
where the function $F=F(T)$ (free energy) is defined by  
\be 
\label{tau}
\tau = \exp \left[ {1 \over \epsilon^2} F(T) \right] \ee The formula 
(\ref{action}) then solves the dKP hierarchy (\ref{YCC}), i.e. $P={\cal B}_1 = 
\partial S$ and
\be
\label{B}
{\cal B}_n = \partial_n S =
\lambda^n - \sum_{1}^{\infty} {F_{nm} \over m} \lambda^{-m}  \ee 
where $F_{nm} = \partial_n\partial_m F$ which play an important role in 
the theory of dKP.
\\[3mm]\indent
Now following \cite{ta} one writes the differential Fay identity (\ref{AAL}) 
with $\epsilon\partial_n$ replacing $\partial_n$, looks at logarithms, 
and passes $\epsilon\to 0$ (using (\ref{tau})).  Then
only the second order derivatives survive, and one 
gets the dispersionless differential Fay identity 
\be \sum_{m,n=1}^{\infty}\mu^{-m}\lambda^{-n}\frac{F_{mn}}
{mn} = \log \left(1- \sum_1^{\infty}\frac{\mu^{-n}-\lambda^{-n}}{\mu-\lambda} 
\frac{F_{1n}}{n} \right)  
\label{AAT}
\ee
Although (\ref{AAT}) only uses a subset of the Pl\"ucker relations defining 
the KP hierarchy it was shown in \cite{ta} that this subset is sufficient to 
determine KP; hence (\ref{AAT}) characterizes the function $F$ for dKP. 
Following \cite{ch,ct}, we now derive a dispersionless limit of the Hirota 
bilinear equations (\ref{hirota}), which we call the dispersionless 
Hirota equations. We first note from (\ref{action}) and (\ref{AAO}) 
that $F_{1n} = nP_{n+1}$
so
\be
\sum_1^{\infty}\lambda^{-n}\frac{F_{1n}}{n} = \sum_1^{\infty}P_{n+1} 
\lambda^{-n} = \lambda - P(\lambda) 
\label{AAU}
\ee
Consequently the right side of (\ref{AAT}) becomes $\log[\frac{P(\mu) - 
P(\lambda)}{\mu-\lambda}]$ and for $\mu\to \lambda$ with $\dot{P} = 
\partial_{\lambda}P$ we have
\be
\log\dot{P}(\lambda) = \sum_{m,n=1}^{\infty}\lambda^{-m-n}\frac{F_{mn}} 
{mn} = \sum_{j=1}^{\infty} \left(\sum_{n+m=j} {F_{mn} \over mn} \right) 
\lambda^{-j} 
\label{AAV}
\ee
Then using the elementary Schur polynomial defined in (\ref{AAH}) and 
(\ref{AAO}), we obtain
$$
\dot{P}(\lambda) = \sum_0^{\infty} \chi_j(Z_2, \cdots,Z_j) \lambda^{-j} 
= 1 + \sum_1^{\infty}F_{1j} 
\lambda^{-j-1};$$ 
\be
Z_i = \sum_{m+n=i} {F_{mn} \over mn}\,\,\,\,(Z_1 = 0) 
\label{AAW}
\ee 
Thus we obtain the dispersionless Hirota 
equations, \be \label{F}
F_{1j} = \chi_{j+1}(Z_1=0,Z_2, \cdots,Z_{j+1})  
\ee These can 
be also derived directly from (\ref{hirota}) with (\ref{tau}) in the limit 
$\epsilon \to 0$ or by expanding (\ref{AAV}) in powers of $\lambda^{-n}$
as in  \cite{ch,ct}).  
The equations (\ref{F}) then characterize dKP.
\\[3mm]\indent
It is also interesting to note that the dispersionless Hirota equations 
(\ref{F})  can be regarded as algebraic equations for 
``symbols" $F_{mn}$, which are defined via (\ref{B}), i.e. \be
{\cal B}_n := \lambda^n_+= \lambda^n - \sum_1^{\infty}\frac{F_{nm}}{m} 
\lambda^{-m}
\label{AAY}
\ee
and in fact
\be
F_{nm} = F_{mn} = Res_P[\lambda^m d \lambda^n_+]  \label{AAZ} \ee 
Thus 
for $\lambda,\,\,P$ given algebraically as in (\ref{AAO}), 
with no a priori connection to dKP, and for ${\cal B}_n$ defined as in
(\ref{AAY}) via a formal collection of symbols with two
indices $F_{mn}$, it follows that the dispersionless Hirota equations 
(\ref{F}) are nothing but polynomial identities among 
$F_{mn}$.  In particular one has from \cite{cd}
\\[3mm]\indent {\bf THEOREM 2.3}.$\,\,$
(\ref{AAZ}) with (\ref{F}) completely characterizes and solves the
dKP hierarchy.
\\[3mm]\indent
Now one very natural way of developing dKP begins with (\ref{AAO}) and 
(\ref{YCC}) since
eventually the $P_{j+1}$ can serve as universal coordinates (cf. here 
\cite{aa} for a discussion of this in connection with topological field 
theory = TFT). This point of view is also natural in terms of developing 
a Hamilton-Jacobi theory involving ideas from the hodograph $-$ Riemann 
invariant approach (cf. \cite{ci,cs,ga,kk} and in 
connecting
NKdV ideas to TFT, strings, and quantum gravity.
It is natural here to work with $Q_n := (1/n){\cal B}_n$ and note that 
$\partial_n S = {\cal B}_n$ corresponds to $\partial_n P = \partial 
{\cal B}_n = n\partial Q_n$.
In this connection one often uses different time variables, say $T'_n = 
nT_n$, so that $\partial'_nP = \partial Q_n$, and $G_{mn} = F_{mn}/mn$ is 
used in place of $F_{mn}$. Here however we will retain the $T_n$ notation 
with $\partial_n S = nQ_n$ and $\partial_n P = n\partial Q_n$ since one 
will be connecting a number of formulas to standard KP notation. Now given 
(\ref{AAO}) and (\ref{YYCC}) the equation
$\partial_n P = n\partial Q_n$ corresponds to Benney's moment equations 
and is equivalent to a system of Hamiltonian equations defining the dKP 
hierarchy (cf. \cite{ga,kk});
the Hamilton-Jacobi equations are 
$\partial_n S = nQ_n$ with Hamiltonians $nQ_n(X, P=\partial S)$). 
There is now an important formula involving the functions $Q_n$ 
(cf. \cite{ch,cf,kk}), namely
the generating function of $\partial_P Q_n(\lambda)$ is given by 
\be \frac{1}{P(\mu) - P(\lambda)} = \sum_1^{\infty}\partial_P Q_n(\lambda) 
\mu^{-n}  \label{ABFF}
\ee
In particular one notes 
\be \oint_{\infty} {\mu^n \over 
P(\mu) - P(\lambda)} d\mu = \partial_P Q_{n+1}(\lambda) \ , \label{canonicalT}
\ee
which gives a key formula in the Hamilton-Jacobi method for the 
dKP \cite{kk}. 
Also note here that the function 
$P(\lambda)$ alone provides all the information necessary for the dKP theory. 
It is proved in \cite{ch} that 
\\[3mm]\indent {\bf THEOREM 2.4}.$\,\,$
The kernel formula (\ref{ABFF}) is
equivalent to the dispersionless differential Fay identity (\ref{AAT}).
\\[3mm]\indent
The proof uses
\be \label{ShP}
\partial_P Q_n = \chi_{n-1}(Q_1,\cdots,Q_{n-1}) 
\ee 
where $\chi_n(Q_1,\cdots,Q_n)$ can be expressed as a polynomial in $Q_1 = P$
with the coefficients given by polynomials in the $P_{j+1}$.
Indeed
\be
\chi_n = det
\left[
\begin{array}{ccccccc}
P & -1 & 0 & 0 & 0 & \cdots & 0\\
P_2 & P & -1 & 0 & 0 & \cdots & 0\\
P_3 & P_2 & P & -1 & 0 & \cdots & 0\\
\vdots & \vdots & \vdots & \vdots & \vdots & \ddots & \vdots\\ P_n & P_{n-1} 
& \cdots & P_4 & P_3 & P_2 & P 
\end{array} 
\right] = \partial_P Q_{n+1} 
\label{ABMM}
\ee
and this leads to the observation that the $F_{mn}$ can be 
expressed
as polynomials in $P_{j+1} = F_{1j}/j$.
Thus
the dHirota
equations can be solved totally algebraically via $F_{mn} = 
\Phi_{mn}(P_2,P_3,\cdots,P_{m+n})$ where $\Phi_{mn}$ is a polynomial in 
the $P_{j+1}$ so the $F_{1n} = nP_{n+1}$ are generating elements for the 
$F_{mn}$, and serve as universal coordinates.  Indeed
formulas such as (\ref{ABMM}) and (\ref{ShP})
indicate that in fact dKP theory can be characterized 
using only elementary Schur polynomials since these provide all the 
information necessary for the kernel (\ref{ABFF}) or equivalently for the 
dispersionless differential Fay identity. This amounts
also to observing that in the passage from KP to dKP only certain Schur 
polynomials survive the limiting process $\epsilon\to 0$. Such terms 
involve second derivatives of $F$ and these may be characterized in terms of 
Young diagrams with only vertical or horizontal boxes. This is also related 
to the explicit form of the hodograph transformation where one needs only 
$\partial_P Q_n = \chi_{n-1}(Q_1,\cdots,Q_{n-1})$ and the $P_{j+1}$ in the 
expansion of $P$ (cf. \cite{ch}).
Given KP and dKP theory we can now discuss nKdV or dnKdV easily although
many special aspects of nKdV for example are not visible in KP.  In
particular for the $F_{ij}$ one will have $F_{nj} = F_{jn} = 0$
for dnKdV.
We note also (cf. \cite{cf,kk}) that from (\ref{ShP}) one has
a formula of Kodama
\be
\frac{1}{P(\mu)-P(\lambda)} = \sum_1^{\infty}\partial_PQ_n\mu^{-n} =
\sum_0^{\infty}\chi_n(Q)\mu^{-n}=exp(\sum_1^{\infty}Q_m\mu^{-m})
\label{ABNN}
\ee

\section{ISOMONODROMY PROBLEMS}
\renewcommand{\theequation}{3.\arabic{equation}}
\setcounter{equation}{0}

We begin with \cite{tg} where the goal is to exhibit the relations 
between SW theory and Whitham dynamics via isomonodromy
deformations.  One considers algebraic curves $C\to C_0$
(spectral covering) over the quantum moduli space of $z\in
C_0\sim$ algebraic curve of genus $0$ or $1$ via ${\bf (3A)}\,\,
det(w-L(z))=0$.  For the $N=2,\,\,SU(s)$ Yang-Mills (YM) theory 
without matter $C_0={\bf CP^1}$ and $z=log(h)$ for $h\in{\bf
CP^1}/\{0,\infty\}$ with Lax matrix
\be
L(z)=\left(
\begin{array}{ccccc}
b_1 & 1 & \cdots & 0 & c_sh^{-1}\\
c_1 & b_2 & \cdots & 0 & 0\\
0 & 0 & \cdots & 0& 0\\
0 & 0 & \cdots & b_{s-1} & 1\\
h & 0 & \cdots & c_{s-1} & b_s
\end{array}\right)
\label{1}
\ee
Here $b_j=\dot{q}_j=dq_j/dt$ with $c_j=exp(q_j-q_{j+1})$ and 
one compares with \cite{ia} where $q_{s+1}=q_1$ and $q_j\to-q_j
\sim c_j\to c_j^{-1}$.  The eigenvalue equation {\bf (3A)} becomes
($\Lambda^{2s}=\prod_1^sc_j=exp(q_1-q_{s+1}$)
\be
h+\Lambda^{2s}h^{-1}=P(w);\,\,y^2=P^2(w)-\Lambda^{2s};\,\,
h=\frac{1}{2}(y+P(w));\,\,P(w)=w^s+\sum_2^su_jw^{s-j}
\label{2}
\ee
(where $\Lambda^{2s}=1$ in \cite{ia}).  Thus one has hyperelliptic
curves as in the case of a finite periodic Toda chain where such curves
arise via a commutative set of isospectral flows in Lax representation
\be
\partial_nL(z)=[P_n(z),L(z)];\,\,[\partial_n-P_n,\partial_m-P_m]=0
\label{3}
\ee
The $P_n$ are suitable matrix valued meromorphic functions on $C_0$.
The associated linear problem
\be
w\psi=L\psi;\,\,\partial_n\psi=P_n\psi
\label{4}
\ee
then determines a vector (or matrix) Baker-Akhiezer (BA) function.
Now in \cite{tg} one replaces such an isospectral problem by an
isomonodromy problem
\be
\epsilon\frac{\partial\Psi}{\partial z}=Q\Psi;\,\,\partial_n\Psi=
P_n\Psi
\label{5}
\ee
The idea is that the $t$ flows leave the monodromy data of the
$z$ equation intact if and only if
\be
[\partial_n-P_n,\epsilon\partial_z-Q]=0;\,\, [\partial_n-P_n,\partial_m
-P_m]=0
\label{6}
\ee
An idea from \cite{fc} is to write $\Psi$ in a WKB form ($\epsilon
\to 0$)
\be
\Psi=\left(\phi+\sum_1^{\infty}\epsilon^n\phi_n\right)
\left(\frac{\partial^2 S}{\partial z^2}\right)^{-1/2}exp[\epsilon^{-1}
S(z)]
\label{7}
\ee
where $dS=wdz$ corresponds to the SW differential.  Thus in
the leading order term for $\epsilon\partial_z\Psi=Q\Psi$ one
finds ${\bf (3B)}\,\,(\partial_zS)\phi=Q\phi$ and if we identify
${\bf (3C)}\,\,w=S_z\equiv dS=wdz$ then the algebraic formulation
gives essentially the same eigenvalue problem as {\bf (3A)} (where
the relation between $\phi$ and $\psi$ is clarified below).  The idea
from \cite{fc} is that isomonodromic deformations in 
WKB approximation look like modulation of isospectral 
deformations.  The passage isospectral $\to$ isomonodromy is
achieved via ${\bf (3D)}\,\,w\to\epsilon\partial_z$ which is a kind
of quantization with $\epsilon\sim\hbar$ and $\Psi\sim$
a quantum mechanical wave function.
\\[3mm]\indent
Now one separates the isomonodromy problem into a combination
of fast (isospectral - $t_n$) and slow (Whitham -$T_n$) dynamics
via multiscale analysis.  The variables are connected via $T_n=\epsilon t_n$ 
and one assumes all fields $u_{\alpha}=u_{\alpha}(t,T)$.  Then writing
$\partial_n\sim\partial/\partial t_n$ we have ${\bf (SE)}\,\,\partial_n
u_{\alpha}(t,\epsilon t)=\partial_n u_{\alpha}(t,T)+\epsilon (\partial
u_{\alpha}(t,T)/\partial T_n)|_{T=\epsilon t}$.  Further take $P_n$
and  $Q$ as functions of $(t,T,z)$ and look for a wave function
$\Psi$ as in (\ref{7}) with $\phi_n=\phi_n(t,T,z)$ (say $\phi_0=\phi$)
and $S=S(T,z)$.  The leading order terms in (\ref{4}) become, for
${\bf (3F)}\,\, w=\partial_zS$ and $\psi=\phi(z)exp(\sum t_n(\partial S/
\partial T_n)$,
\be
w\psi=Q\psi;\,\,\partial_n\psi=P_n\psi
\label{8}
\ee
which is an isospectral problem with associated curve $C$ defined by
$det(w-Q(t,T,z))=0$ which is typically $t$ independent but now
depends on the $T_n$ as adabatic parameters.  One can then produce
a standard BA function $\tilde{\psi}=\tilde{\phi}exp(\sum t_n\Omega_
n)$ (cf. \cite{ce,df} and Section 2) where $\Omega_n=
\int^zd\Omega_n$.  The amplitude $\tilde{\phi}$ is then
composed of theta functions of the form $\theta(\sum t_n\sigma_n+
\cdots)$ where ${\bf (3G)}\,\,\sigma_n=(\sigma_{n,1},\cdots,
\sigma_{n,g})^T$with $\sigma_{n,j}=(1/2\pi i)\oint_{B_j}d\Omega_n$
(we assume a suitable homology basis $(A_j,B_j)$ has been chosen).
The theta functions provide the quasi-periodic (fast) dynamics which is 
eventually averaged out and does not contribute to the Whitham
(slow) dynamics.  The main contribution to the Whitham dynamics arises
by matching $\tilde{\psi}\sim\psi$ and $\tilde{\phi}\sim\phi$ with 
${\bf (3H)}\,\,\partial S/\partial T_n=\Omega_n(z)$ which in fact
provides a definition of a Whitham system via ${\bf (3I)}\,\,
\partial d\Omega_n/\partial T_m=\partial d\Omega_m/\partial T_n$
representing a dynamical system on the moduli space of spectral
curves.  Thus one starts with monodromy and the WKB leading
order term is made isospectral (via {\bf (3F)}) which leads to a
$t$ independent algebraic curve.  Then the BA
function corresponding to this curve is subject to averaging and matching
to $\psi$.  We ignore here (as in \cite{tg}) all complications due to
Stokes multipliers etc.  The object now is to relate $Q$ and $L$ and this is
done with the $N=2, \,\,SU(s)$ YM example.  The details are spelled
out in \cite{tg} (cf. also \cite{na}).
\\[3mm]\indent
We go next
to \cite{ti} where isospectrality and isomonodromy are considered
in the context of Schlesinger equations.  Thus in the small 
$\epsilon$ limit solutions of the isomonodromy problem are expected
to behave as slowly modulated finite gap solutions of an isospectral
problem.  The modulation is caused by slow deformation of the
spectral curve of the finite gap solution.  Now from \cite{ti} let
$gl(r,{\bf C})^N\sim\oplus_1^Ngl(r,{\bf C})\sim N$-tuples $(A_1,
\cdots,A_N)$ of $r\times r$ matrices.  There is a $GL(2,{\bf C})$ 
coadjoint action $A_i\to gA_ig^{-1}$ and the Schlesinger equation
{\bf (SE)} is
\be
\frac{\partial A_i}{\partial t_j}=\left[A_i,(1-\delta_{ij})\frac{A_j}
{t_i-t_j}-\delta_{ij}\sum_{k\ne i}\frac{A_k}{t_i-t_k}\right]
\label{9}
\ee
and each coadjoint orbit ${\cal O}_i$ is invariant under the $t$ flows.
Thus {\bf (SE)} is a collection of non-autonomous dynamical 
systems on $\prod_1^N{\cal O}_i$.  We consider only semisimple (ss)
orbits labeled by the eigenvalues $\theta_{i\alpha}\,\,(\alpha=1,\cdots,r)$ 
of $A_i$.  Thus these eiqenvalues (and in general the Jordan canonical
form) are invariants of {\bf (SE)}.  There are also extra invariants
in the form of the matrix elements of $A_{\infty}=-\sum_1^NA_i$,
invariant via $\partial A_{\infty}/\partial t_i=0$.  Assuming 
$A_{\infty}$ is ss it can be diagonalized in advance by a constant
gauge transformation $A_i\to CA_iC^{-1}$ and then only the 
eigenvalues $\theta_{\infty,\alpha}\,\,(\alpha=1,\cdots,r)$ of $A_{\infty}$
are nontrivial invariants.  One can introduce a Poisson structure
on $gl(r,{\bf C})^N$ via 
$${\bf (3J)}\,\,
\{A_{i,\alpha\beta},A_{j,\rho\sigma}\}=\delta_{ij}\left(-\delta_{\beta
\rho}A_{i,\alpha\sigma}+\delta_{\sigma\alpha}A_{i,\rho\beta}\right)$$
which in each component of the direct sum is the ordinary
Kostant-Kirillov bracket.  The {\bf (SE)} can then be written as
\be
\frac{\partial A_j}{\partial t_i}=\{A_j,H_i\};\,\,H_i=Res_{\lambda=
t_i}\frac{1}{2}Tr\,M(\lambda)^2=\sum_{j\ne i}Tr\left(\frac{A_iA_j}
{t_i-t_j}\right)
\label{10}
\ee
with $\{H_i,H_j\}=0$.
\\[3mm]\indent
Now {\bf (SE)} gives isomonodromy deformations of ${\bf (3K)}\,\,
dY/d\lambda=M(\lambda)Y$ with $M(\lambda)=
\sum_1^N[A_i/(\lambda-t_i)]$.  One can assume for simplicity
that {\bf (3L)}  The $A_i$ and $A_{\infty}$ are diagonalizable
and $\theta_{i\alpha}-\theta_{i\beta}\not\in {\bf Z}$ if $\alpha
\ne\beta$.  This means that local solutions at the singular points
$\lambda=t_1,\cdots, t_N,\infty$ do not develop logarithmic terms.
The isomonodromy deformations are generated by ${\bf (3M)}\,\,
\partial Y/\partial t_i=-[A_i/(\lambda- t_i)]Y$ and the Frobenius
integrability conditions are
\be
\left[\frac{\partial}{\partial t_j}+\frac{A_j}{\lambda-t_j},M(\lambda)
-\frac{\partial}{\partial \lambda}\right]=0;\,\,\left[\frac{\partial}{\partial
t_i}+\frac{A_i}{\lambda-t_i},\frac{\partial}{\partial t_j}+\frac{A_j}
{\lambda-t_j}\right]=0
\label{11}
\ee
and these are equivalent to {\bf (SE)}.  Next, since $\lambda=t_1,
\cdots,t_N,\infty$ are regular singular points of {\bf (3K)} there will
be local solutions
\be
Y_i=\hat{Y}_i\cdot(\lambda-t_i)^{\Theta_i};\,\,\hat{Y}_i=
\sum_0^{\infty}Y_{in}(\lambda-t_i)^n;
\label{12}
\ee
$$Y_{\infty}=\hat{Y}_{\infty}\cdot\lambda^{-\Theta_{\infty}};\,\,\hat
{Y}_{\infty}=\sum_0^{\infty}Y_{\infty,n}\lambda^{-n}$$
where the $Y_{in}$ are $r\times r$ matrices, $Y_{i0}$ and $Y_{\infty,0}$ 
are invertible, and $\Theta_i,\,\Theta_{\infty}$ are diagonal matrices
of local monodromy exponents.  This leads to ${\bf (3N)}\,\,
A_i=Y_{i0}\Theta_iY^{-1}_{i0}$ for $i=1,\cdots,N,\infty$ where
$\Theta_i=diag(\theta_{i1},\cdots,\theta_{ir})$.  The tau function
for {\bf (SE)} can be defined in two equivalent ways ${\bf (3O)}\,\,
d\,log\tau=\sum_1^NH_idt_i$ or $\partial\,log\tau/\partial t_i=Tr(\Theta_i
Y^{-1}_{i0}Y_{i1}$ and the integrability condition $\partial H_i/\partial
t_j=\partial H_j/\partial t_i$ is ensured by {\bf (SE)} (see \cite{ti}
for details).  The spectral curve is now ${\bf (3P)}\,\,det(M(\lambda)-\mu
I)=0$ and this generally varies under isomonodromic deformations
(see \cite{ti}).
\\[3mm]\indent
Consider now Garnier's autonomous analogue of {\bf (SE)}.  This is
given by
\be
\left[\frac{\partial}{\partial t_i}+
\frac{A_i}{\lambda-c_i},M(\lambda)\right]=0;
\,\,\left[\frac{\partial}{\partial t_i}+\frac{A_i}{\lambda-c_i},
\frac{\partial}{\partial t_j}+\frac{A_j}{\lambda-c_j}\right]=0
\label{13}
\ee
and this is an isospectral problem with $det(M(\lambda)-\mu I)$ 
independent of $t$.  An auxiliary linear problem is given by
\be
w\psi=M(\lambda)\psi;\,\,\frac{\partial \psi}{\partial t_i}=-\frac
{A_i}{\lambda-c_i}\psi
\label{14}
\ee
where $\psi$ is a column vector.  This kind of problem can be mapped
to linear flows on the Jacobian variety of the spectral curve and the
case of $r=2$ is particularly interesting since here the Painlev\'e
VI (and Garnier's multivariable version) emerge.
\\[3mm]\indent
For the geometry one goes now to \cite{ab,ac,ba,hc,hd}.  Let $C_0$
be the spectral curve ${\bf (3Q)}\,\,F(\lambda,\mu)=
det(M(\lambda)-\mu I)=0$ which can be thought of as a ramified
cover of the punctured Riemann sphere $\pi:\,\,C_0\to {\bf CP^1}/
\{c_1,\cdots,c_N,\infty\}$ where $\pi(\lambda,\mu)=\lambda$.  Then
generically $\pi^{-1}(\lambda)\sim\{(\lambda,\mu_{\alpha}),\,\,
\alpha=1,\cdots,r\}$ and the $\mu_{\alpha}$ are eigenvalues of 
$M(\lambda)$.  Near $\lambda=c_i$ one has $\mu_{\alpha}=
\theta_{i\alpha}/(\lambda-c_i)+$ nonsingular terms, where the $\theta_
{i\alpha}$ are the eigenvalues of $A_i$.  Similarly near $\lambda=
\infty$ one has $\mu_{\alpha}=-\theta_{\infty,\alpha}\lambda^{-1}
+O(\lambda^{-2})$ where the $\theta_{\infty,\alpha}$ are eigenvalues
of $A_{\infty}$.  One can compactify $C_0$ by adding points over
the punctures.  Thus near $c_i$ replace $\mu$ by $\tilde{\mu}=
f(\lambda)\mu$ where $f(\lambda)=\prod_1^N(\lambda-c_i)$.
Then the spectral curve equation becomes ${\bf (3R)}\,\,
\tilde{F}(\lambda,\tilde{\mu})=det(f(\lambda)M(\lambda)-
\tilde{\mu}I)=0$ and $\pi^{-1}(\lambda)=\{(\lambda,\tilde{\mu}_
{\alpha}),\,\,\alpha=1,\cdots,r\}$ where ${\bf (3S)}\,\,
\tilde{\mu}_{\alpha}=f'(c_i)\theta_{i\alpha}+O(\lambda-c_i)$.  Since
the $\theta_{i\alpha},\,\,(\alpha=1,\cdots,r)$ are pairwise distinct
one is adding to $C_0$ at $c_i$, $r$ extra points $(\lambda,
\tilde{\mu}_{\alpha})=(c_i,f'(c_i)\theta_{i\alpha})$ to fill the holes
above $c_i$.  At $\infty$ one uses $\tilde{\mu}=\lambda\mu$ with
$\pi^{-1}(\infty)=\{(\infty,-\theta_{\infty,\alpha})\}$.  Thus by adding
$rN+r$ points to $C_0$ one gets a compactification $C$ of $C_0$ with a 
unique extension of the covering map $\pi$ to a ramified $\pi:\,\,
C\to {\bf CP^1}$.  The genus of $C$ will be $g=(1/2)(r-1)(rN-r-2)$.
Taking ${\bf (3T)}\,\,M^0(\lambda)=\sum_1^N[A^0_i/(\lambda-c_i)],
\,\,A^0_i=diag(\theta_{i1},\cdots,\theta_{ir})$ as a reference point
on the coadjoint orbit of $M(\lambda)$ one compares the
characteristic polynomials via ${\bf (3U)}\,\,\tilde{F}(\lambda,
\tilde{\mu})-\tilde{F}^0(\lambda,\tilde{\mu})=f(\lambda)\sum_
{\ell=2}^rp_{\ell}(\lambda)\tilde{\mu}^{r-\ell}$ where
$p_{\ell}(\lambda)=\sum_0^{\delta_{\ell}}h_{m\ell}\lambda^m$
with $\delta_{\ell}=(N-1)\ell-N$.  This leads to
\be
\tilde{F}(\lambda,\tilde{\mu})=\tilde{F}^0(\lambda,\tilde{\mu})+
\sum_{\ell=2}^r\sum_0^{\delta_{\ell}}h_{m\ell}\lambda^m\tilde{\mu}^
{r-\ell}
\label{15}
\ee
The spectral curve has therefore three sets of parameters ${\bf (3V)}\,\,
{\bf (1)}$
Pole positions $c_i\,\,(i=1,\cdots,N) \,
\,{\bf (2)}$ Coadjoint orbit invariants
$\theta_{i\alpha}\,\,(i=1,\cdots,N,\infty)$, and 
${\bf (3)}$ Isospectral invariants
$h_{m\ell}\,\,(\ell=2,\cdots,r;\,\,m=0,\cdots,\delta_{\ell}-1)$.
\\[3mm]\indent
Now one develops the Whitham ideas as follows.
First reformulate the {\bf (SE)} equation via (cf. \cite{va})
\be
\frac{\partial}{\partial t_i}\to\epsilon\frac{\partial}{\partial T_i};\,\,
\frac{\partial}{\partial\lambda}\to\epsilon\frac{\partial}{\partial\Lambda};
\,\,\frac{A_i}{\lambda- t_i}\to\frac{{\cal A}_i}{\Lambda-T_i}
\label{16}
\ee
Note this corresponds to $\epsilon t_i=T_i$ with $\partial_i=\epsilon
\partial/\partial T_i$ and $\epsilon\lambda=\Lambda$ 
with $\partial_{\lambda}=\epsilon\partial_{\Lambda}$ and
$A_i/(\lambda-t_i)\to {\cal A}_i/(\Lambda-T_i)$ so ${\cal A}_i\sim
\epsilon A_i$.  Then (\ref{9}) becomes
\be
\epsilon\frac{\partial{\cal A}_i}{\partial T_j}=(1-\delta_{ij})
\frac{[{\cal A}_i,{\cal A}_j]}{T_i-T_j}-\delta_{ij}\frac{[{\cal A}_i,
{\cal A}_j]}{T_i-T_j}
\label{17}
\ee
(the notation in \cite{ti} is confusing here since the step $A_j\to{\cal A}_j=
\epsilon A_j$ was not clarified).  The auxiliary linear problem is
(cf. {\bf (3K)} and {\bf (3M)})
\be
\epsilon\frac{\partial{\cal Y}}{\partial\Lambda}={\cal M};\,\,
\epsilon\frac{\partial{\cal Y}}{\partial T_i}=-\frac{{\cal A}_i}
{\Lambda-T_i}{\cal Y}
\label{18}
\ee
(thus $M\to {\cal M}$ and $Y\to {\cal Y}$).  Then the $\epsilon$ 
dependent {\bf (SE)} equation (\ref{17}) can be reproduced from the
Frobenius integrability condition
\be
\left[\epsilon\frac{\partial}{\partial T_j}+\frac{{\cal A}_i}{\Lambda-T_i},
{\cal M}(\Lambda)-\epsilon\frac{\partial}{\partial\Lambda}\right]=0;
\label{19}
\ee
$$\left[\epsilon\frac{\partial}{\partial T_i}+\frac{{\cal A}_i}
{\Lambda-T_i},\epsilon\frac{\partial}{\partial T_j}+\frac{{\cal A}_j}
{\Lambda-T_j}\right]=0$$
Now start with (\ref{17})-(\ref{19}) and introduce first
variables $t_i=\epsilon^{-1}T_i$ so that (\ref{17}) can be written as
\be
\frac{\partial{\cal A}_i}{\partial t_j}=(1-\delta_{ij})\frac{[{\cal A}_i,
{\cal A}_j]}{T_i-T_j}-\delta_{ij}\sum_{k\ne i}\frac{[{\cal A}_i,
{\cal A}_k]}{T_i-T_k}
\label{20}
\ee
In the scale of the fast variables the $T_i$ can be regarded as 
approximately constant and hence (\ref{20}) looks approximately
like Garnier's autonomous isospectral problem.  Of course the
$T_i$ vary slowly as do the $h_{m\ell}$ (but not the $\theta_{ik}$).
\\[3mm]\indent
For multiscale analysis following \cite{dg} one 
writes now ${\cal A}_i={\cal A}_i
(t,T,\epsilon)$ and $\epsilon\partial/\partial T_i\to\partial/\partial t_i+
\epsilon\partial/\partial T_i$ in (\ref{20}) (note $t_i$ and $T_i$ are 
considered to be independent now with $t_i=\epsilon^{-1}T_i$ 
imposed only in the left side of the differential equation (\ref{20})).
Now assume ${\bf (3W)}\,\,{\cal A}_i={\cal A}^0_i(t,T)+\epsilon
{\cal A}^1_i(t,T)+\cdots$.  The lowest order term in $\epsilon$ gives
\be
\frac{\partial{\cal A}_i^0}{\partial t_j}=(1-\delta_{ij})\frac
{[{\cal A}^0_i,{\cal A}^0_j]}{T_i-T_j}-\delta_{ij}\sum_{k\ne i}
\frac{[{\cal A}_i^0,{\cal A}_k^0]}{T_i-T_k}
\label{21}
\ee
For $T_i=c_i$ this is Garnier's autonomous system.  The next order
equation is
\be
\frac{\partial {\cal A}_i^1}{\partial t_j}=(1-\delta_{ij}\frac
{[{\cal A}_i^0,{\cal A}_j^1]+[{\cal A}_i^1,{\cal A}_j^0]}{T_i-T_j}-
\label{22}
\ee
$$-\delta_{ij}\sum_{k\ne i}\frac{[{\cal A}_i^0,{\cal A}_k^1]+
[{\cal A}_i^1,{\cal A}_k^0]}{T_i-T_k}+\Xi$$
where $\Xi$ refers to terms in ${\cal A}^0_m$ and their $T$ 
derivatives.
\\[3mm]\indent
A standard procedure in multiscale analysis is to eliminate the $t$
derivatives of ${\cal A}_i^1$ by averaging over the $t$ space but this
involves some technical difficulties.  The spectral curve for {\bf (SE)} is
hyperelliptic only if $r=2$ and even in that situation there are problems.
So another approach is adopted.  In addition to the multiscale expression
for the ${\cal A}_i$ one assumes ${\bf (3X)}\,\,{\cal Y}=[\phi^0(t,T,
\Lambda)+\phi^1(t,T,\Lambda)\epsilon+\cdots]exp[\epsilon^{-1}
S(T,\Lambda)]$ (${\cal Y}$ and the $\phi^k$ are vector valued
functions with $S$ a scalar).  Then one writes the auxiliary linear
problem (\ref{18}) in the multiscale form
\be
\epsilon\frac{\partial {\cal Y}}{\partial\Lambda}={\cal M}(\Lambda)
{\cal Y};\,\,\left(\frac{\partial}{\partial t_i}+\epsilon\frac{\partial}
{\partial T_i}\right){\cal Y}=-\frac{{\cal A}_i}{\Lambda-T_i}{\cal Y}
\label{23}
\ee
The leading order term reproduces the auxiliar linear problem of the
isospectral problem
\be
\frac{\partial S}{\partial\Lambda}\phi^0={\cal M}^0\phi^0;\,\,\frac
{\partial\phi^0}{\partial t_i}+\frac{\partial S}{\partial T_i}\phi^0
=-\frac{{\cal A}^0_i}{\Lambda-T_i}\phi^0
\label{24}
\ee
where ${\cal M}^0=\sum_1^N[{\cal A}_i^0/(\Lambda-T_i)$.  Define
now ${\bf (3Y)}\,\,\mu=\partial S/\partial\Lambda$ and $\psi=
\phi^0exp[\sum t_i(\partial S/\partial T_i)]$ so (\ref{24}) becomes
(\ref{14}) in the form
\be
\mu\psi={\cal M}^0\psi;\,\,\frac{\partial\psi}{\partial t_i}=-\frac
{{\cal A}_i^0}{\Lambda-T_i}\psi
\label{25}
\ee
Next one compares this $\psi$ with the ordinary BA function $\psi=\phi
exp[\sum t_i\Omega_i]$ where $\Omega_i=\int^{\Lambda,\mu}
d\Omega_i$ and the vector function $\phi$ involves $\theta$ functions
as before.  For matching we want now ${\bf (3Z)}\,\,\phi^0=\phi$
(or more generally $\phi^0=h(T,\Lambda)\phi$) and $\partial S/
\partial T_i=\Omega_i$.  Thus one proposes $(\bullet)\,\,\mu=
\partial S/\partial\Lambda$ and $\partial S/\partial T_i=\Omega_i$
(and consequently $\partial\Omega_i/\partial T_j=\partial\Omega_j/
\partial T_i$) as the modulation equations governing the slow 
dynamics of the spectral curve $(\bullet\bullet)\,\,det
({\cal M}^0(\Lambda)-\mu I)=0$.  Details in this direction are given in 
Section 6 of \cite{ti} where it is determined that
\be
d\Omega_i=\frac{\theta_{i\alpha}}{(\Lambda-T_i)^2}d\Lambda+
\,\,nonsingular\,\, terms
\label{26}
\ee
near $P_{\alpha}(\Lambda)\in\pi^{-1}(\Lambda)\,\,(\alpha=
1,\cdots,r)$.  Normalization is achieved via $(\bullet\bullet\bullet)\,\,
\oint_{A_j}d\Omega_i=0$ for $j=1,\cdots,g$.
\\[3mm]\indent
Next consider the $h_I$ for $I=1,\cdots,g$ as the set of isospectral
invariants $h_{m\ell}$ (cf. (\ref{15})) and let $S=S(T,h,\Lambda)$ with
$dS=\mu(T,h,\Lambda)d\Lambda$.  One defines
\be
d\tilde{\omega}_I=\frac{\partial dS}{\partial h_I}=\frac{\partial
\mu(T,h,\Lambda)}{\partial h_I}d\Lambda=\frac{1}{f(\Lambda)}
\frac{\partial\tilde{\mu}}{\partial h_I}d\Lambda
\label{27}
\ee
Then $(\clubsuit)\,\,[\partial\tilde{F}(\Lambda,\tilde{\mu})/\partial
h_{m\ell}]=f(\Lambda)\Lambda^{\ell}\tilde{\mu}^{r-m}$ and there
results 
$$(\spadesuit)\,\,d\tilde{\omega}_{m\ell}=-[\Lambda^{\ell}
\tilde{\mu}^{r-m}/(\partial\tilde{F}/\partial\tilde{\mu})]d\Lambda$$
These differentials $d\tilde{\omega}_I$ form a basis of holomorphic
differentials on the spectral curve and one shows that
\be
\frac{\partial a_J}{\partial h_I}=\oint_{A_J}\frac{\partial dS}{\partial
h_I}=\oint_{A_J}d\tilde{\omega})I=A_{IJ}
\label{28}
\ee
is an invertible matrix ($d\tilde{\omega}_I=\sum A_{IJ}d\omega_J$
where the $d\omega_J$ are determined via $\oint_{A_I}d\omega_J=
\delta_{IJ}$).  Thus one has an invertible map $h\to a$ where the
$a_J$ are the standard action variables of period integrals
$a_J=\oint_{A_J}dS$.  One can also express $dS$ now as 
$dS(T,a)$ with $\partial dS/\partial T_i=d\Omega_i$ and
$\partial dS/\partial a_I=d\omega_I$ since
\be
\frac{\partial dS}{\partial a_I}=\sum_J\frac{\partial h_J}{\partial a_I}
\frac{\partial dS}{\partial h_J}=\sum_J(A_{IJ})^{-1}d\tilde{\omega}_J
=d\omega_I
\label{29}
\ee
One can also introduce a prepotential as in SW theory with
\be
\frac{\partial {\cal F}}{\partial a_I}=b_I=\oint_{B_I}dS;\,\,
\frac{\partial {\cal F}}{\partial T_i}=H_i
\label{30}
\ee
(cf. (\ref{30}) for $H_i$).

\section{JMMS EQUATIONS}
\renewcommand{\theequation}{4.\arabic{equation}}
\setcounter{equation}{0}

We go now to \cite{th}, which is partly a rehash of \cite{tg} for
the Jimbo-Miwa-Mori-Sato (JMMS) equations, giving isomonodromic
deformations of the matrix system ($r\times r$)
\be
\frac{dY}{d\lambda}=M(\lambda)Y;\,\,M(\lambda)=u+\sum_1^N
\frac{A_i}{\lambda-t_i};\,\,u=\sum_1^ru_{\alpha}E_{\alpha}
\label{31}
\ee
where $E_{\alpha}\sim\delta_{\alpha\beta}\delta_{\alpha\gamma}$
in the $(\beta,\gamma)$ position so $u=diag(u_1,\cdots,u_r)$ where
the $u_{\alpha}\sim$ time variables of isomonodromic deformations.
Evidently $\lambda=t_i$ are regular singular points and $\lambda
=\infty$ is an irregular singular point of Poincar\'e rank $1$.  In
order to avoid logarithmic terms one assumes the eigenvalues
$\theta_{i\alpha}$ of $A_i$ have no integer difference and that the
$u_i$ are pairwise distinct.  Then isomonodromic deformations
are generated by
\be
\frac{\partial Y}{\partial t_i}=-\frac{A_i}
{\lambda-t_i}Y;\,\,\frac{\partial Y}
{\partial u_{\alpha}}=(\lambda E_{\alpha}+B_{\alpha})Y;
\label{32}
\ee
$$B_{\alpha}=-\sum_{\beta\ne\alpha}\frac{E_{\alpha}A_{\infty}
E_{\beta}+E_{\beta}A_{\infty}E_{\alpha}}{u_{\alpha}-u_{\beta}};\,\,
A_{\infty}=-\sum_1^NA_i$$
with Frobenius integrability conditions
\be
\left[\frac{\partial}{\partial t_i}+\frac{A_i}{\lambda-t_i},\frac
{\partial}{\partial\lambda}-M(\lambda)\right]=0;
\label{33}
\ee
$$\left[\frac{\partial }{\partial u_{\alpha}}-
\lambda E_{\alpha}-B_{\alpha},
\frac{\partial}{\partial\lambda}-M(\lambda)\right]=0$$
This leads to
\be
\frac{\partial A_j}{\partial u_{\alpha}}=[t_jE_{\alpha}+
B_{\alpha},A_j];
\label{34}
\ee
$$\frac{\partial A_j}{\partial t_i}=(1-\delta_{ij})\frac{[A_i,A_j]}
{t_i-t_j}+\delta_{ij}\left[u+\sum_{k\ne i}\frac{A_k}{t_i-t_k},
A_j\right]$$
There are two sets of invariants {\bf (4A)}$\,\,$ Eigenvalues
$\theta_{j\alpha}$ of $A_j:\,\,\,(\partial \theta_{j\beta}/\partial t_i)=
(\partial \theta_{j\beta}/\partial u_{\alpha})=0$ and {\bf (4B)}$\,\,$
Diagonal elements of $A_{\infty}:\,\,\,(\partial A_{\infty,\beta\beta}/
\partial t_i)=(\partial A_{\infty,\beta\beta}/\partial u_{\alpha})=0$.
The JMMS equation is a nonautonomous dynamical system on the
direct product $\prod {\cal O}_i$ of coadjoint orbits and can be
written in the Hamiltonian form
$$
\frac{\partial A_j}{\partial t_i}=\{A_j,H_i\};\,\,\frac{\partial A_j}
{\partial u_{\alpha}}=\{A_j,K_{\alpha}\};\,\,H_i=Tr\left(uA_i+\sum_
{j\ne i}\frac{A_iA_j}{t_i-t_j}\right);$$
\be
K_{\alpha}=Tr\left(E_{\alpha}\sum_1^Nt_iA_i+\sum_
{\beta\ne\alpha}\frac{E_{\alpha}A_{\infty}E_{\beta}A_{\infty}}
{u_{\alpha}-u_{\beta}}\right)
\label{35}
\ee
\indent
The dual isomonodromic problem can be formulated in a general
setting where we suppose rank $A_i=\ell_i$ (which is also a
coadjoint orbit invariant and constant under the JMMS equation
in the generalized sense).  Thus write first with $r\geq \ell_i$ (cf.
\cite{hc,hd,he} for more details and expanded frameworks)
\be
M(\lambda)=u-G^T(\lambda I-T)^{-1}F
\label{36}
\ee
where $F=(F_{a\alpha})$ and $G=(G_{a\alpha})$ are $\ell\times r$
matrices, $\ell=\sum\ell_i$, and $T$ is a diagonal matrix of the form
${\bf (4C)}\,\,T=\sum_1^Nt_iD_i$ with $D_i=E_{\ell_1+\cdots+\ell_
{i-1}+1}+\cdots+E_{\ell_1+\cdots+\ell_i}$ 
(so $D_1=E_1+\cdots+E_{\ell}$
with $D_2=E_{\ell_1}+\cdots+E_{\ell}$, etc.)
In particular $A_i$
can be written as $A_i=-G^TD_iF$
and $F_{a\alpha},\,\,G_{a\alpha}$ may be understood as canonical
coordinates with Poisson bracket
\be
\{F_{a\alpha},F_{b\beta}\}=\{G_{a\alpha},G_{b\beta}\}=0;\,\,
\{F_{a\alpha},G_{b\beta}\}=\delta_{ab}\delta_{\alpha\beta}
\label{37}
\ee
As an example consider $\ell_1=3$ and $\ell_2=2$ which implies
$\ell=5$ and take $r=3$ so $F,G$ are $5\times 3$ matrices with $D_1=
E_1+\cdots+E_5=I_5$ and $D_2=E_4+E_5$.  Then $A_1=-G^TI_5F$
is $3\times 3$ of rank 3 and $A_2=-G^TD_2F$ is $3\times 3$ of rank 2.
The map $(F,G)\to (A_1,\cdots,A_N)$ then becomes a Poisson map
and the JMMS equation can be derived from a Hamiltonian
system in the $(F,G)$ space with the same Hamiltonians $H_i$ and
$K_{\alpha}$.  The dual problem is formulated in terms of the 
rational $\ell\times\ell$ matrix ${\bf (4D)}\,\,L(\mu)=T-
F(\mu I-u)^{-1}G^T$ where $L(\mu)=
T+\sum[P_{\alpha}/(\mu-u_{\alpha})]$
with $P_{\alpha}=-FE_{\alpha}G^T$ of rank $1$ (but this restriction
can also be relaxed).  The map $(F,G)\to (P_1,P_2,\cdots)$ is Poisson for
the Kostant-Kirillov bracket ${\bf (4E)}\,\,\{P_{\alpha,ab},P_{\beta,
cd}\}=\delta_{\alpha\beta}(-\delta_{bc}P_{\beta,ad}+\delta_{da}
P_{\beta,cb})$ and one can write the dual isomonodromy problem 
in the form
$$
\frac{dZ}{d\lambda}=-L(\mu)Z;\,\,\frac{\partial Z}{\partial u_{\alpha}}=
\frac{P_{\alpha}}{\mu-u_{\alpha}}Z;\,\,\frac{\partial Z}
{\partial t_i}=-(\mu D_i+Q_i)Z$$
\be
Q_i=-\sum_{j\ne i}\frac{D_iP_{\infty}D_j+D_jP_{\infty}D_i}{t_i-t_j};
\,\,P_{\infty}=-\sum P_{\alpha}
\label{38}
\ee
\indent
Now introduce a small parameter $\epsilon$ via (cf. \cite{va})
\be
\frac{\partial}{\partial\lambda}\to\epsilon\frac{\partial}
{\partial\Lambda};\,\,\frac{\partial}{\partial t_i}\to\epsilon
\frac{\partial}{\partial T_i};\,\,\frac{\partial}{\partial u_{\alpha}}\to
\epsilon\frac{\partial}{\partial U_{\alpha}}
\label{39}
\ee
with $\sum u_{\alpha}\to\sum U_{\alpha}E_{\alpha}$ and the
JMMS equatins take the form
\be
\epsilon\frac{\partial{\cal A}}{\partial T_i}=(1-\delta_{ij})\frac
{[{\cal A}_i,{\cal A}_j]}{T_i-T_j}+\delta_{ij}\left[U+\sum_{k\ne i}
\frac{{\cal A}_k}{T_i-T_k},{\cal A}_j\right];
\label{40}
\ee
$$\epsilon\frac{\partial{\cal A}_j}{\partial U_{\alpha}}=[T_jE_{\alpha}
+{\cal B}_{\alpha},{\cal A}_j];\,\,{\cal B}_{\alpha}=
-\sum_{\beta\ne\alpha}\frac{E_{\alpha}{\cal A}_{\infty}E_{\beta}
+E_{\beta}{\cal A}_{\infty}E_{\alpha}}{U_{\alpha}-U_{\beta}}$$
(where ${\cal A}_i=\epsilon A_i,\,\,T_i=\epsilon t_i,\,\,
U_{\alpha}=\epsilon u_{\alpha},$ and ${\cal B}_{\alpha}=
\epsilon B_{\alpha}$).
Working as before we assume ${\cal A}_j={\cal A}_j(t,u,T,U)$ and
then, using $t_i=\epsilon^{-1}T_i$ and $u_{\alpha}=\epsilon^{-1}
Y_{\alpha}$, one has e.g.
\be
\epsilon\frac{\partial {\cal A}_j}{\partial T_i}\to\frac{\partial{\cal A}_j}
{\partial t_i}+\epsilon\frac{\partial{\cal A}_j}{\partial T_i}
\label{41}
\ee
as in (\ref{20})-(\ref{21}), leading to $({\cal A}_j={\cal A}_j^0
+{\cal A}_j^1\epsilon+\cdots$)
\be
\frac{\partial {\cal A}_j^0}{\partial U_{\alpha}}=\left[T_jE_{\alpha}+
{\cal B}_{\alpha}^0,{\cal A}_j^0\right];
\label{42}
\ee
$$\frac{\partial {\cal A}_j^0}{\partial t_i}=(1-\delta_{ij})\frac
{[{\cal A}_i^0,{\cal A}_j^0]}{T_i-T_j}+\delta_{ij}\left[U+\sum_{k\ne i}
\frac{{\cal A}_k^0}{T_i-T_k},{\cal A}_j^0\right]$$
where ${\cal B}_{\alpha}^0$ is given by the same formula as
${\cal B}_{\alpha}$ but with ${\cal A}_{\infty}$ replaced by
${\cal A}_{\infty}^0=-\sum_1^N{\cal A}_i^0$.  The slow variables
are parameters at the lowest order and to determine the slow dynamics
one goes to the next order and one uses here the
following approach. 
\\[3mm]\indent
The lowest order equations are again an isospectral problem with
Lax representation
$$
\left[\frac{\partial}{\partial t_i}+\frac{{\cal A}_j^0}{\Lambda-T_i},
{\cal M}^0(\Lambda)\right]=0;\,\,{\cal M}^0(\Lambda)=U+\sum_1^N
\frac{{\cal A}_i^0}{\Lambda-T_i};$$
\be
\left[\frac{\partial}{\partial u_{\alpha}}
-\Lambda E_{\alpha}-{\cal B}_{\alpha}^0,{\cal M}^0\right]=0
\label{43}
\ee
and $det(\mu I-{\cal M}^0(\Lambda))=\Xi$ is constant under
$(t,u)$ flows.  $\Xi$ depends on $(T,U)$ however and the slow
dynamics may be described as slow deformations of the characteristic
polynomial of the spectral curve, namely $\Xi=0$.  This spectral
curve $C_0$ on the $(\lambda,\mu)$ plane can be compactified
to a nonsingular curve $C$; the projection $\pi(\Lambda,\mu)\to
\Lambda:\,\,C_0\to {\bf CP^1}/\{T_1,\cdots,T_N,\infty\}$ extends
to $C$ to give an $r$ fold ramified covering of $C$ over ${\bf CP^1}$.
One adds points to the holes over $\Lambda=(T_i,\infty)$ as before.
There will be $r$ points $(\infty,U_{\alpha})\,\,(\alpha=1,\cdots,r)$
over $\Lambda=\infty$.  For the $T_i$ write ${\bf (4F)}\,\,\tilde{\mu}
=f(\Lambda)\mu$ with $f(\Lambda)=\prod_1^N(\Lambda-T_i)$ so 
$\Xi$ becomes ${\bf (4G)}\,\,F(\Lambda,\tilde{\mu})=det
(\tilde{\mu} I-f(\Lambda){\cal M}^0(\Lambda))=0$.  Add then the
$r$ points $(\Lambda,\tilde{\mu})=(T_i,f'(T_i)\theta_{i\alpha})\,\,(\alpha
=1,\cdots,r)$ over $T_i$ to obtain a curve of genus $g=(1/2)(r-1)(rN-2)$.
To describe the slow dynamics one wants a suitable system of
moduli in the space of permissible curves and the slow dynamics
involves differential equations for such moduli.  Now one can write
\be
F(\Lambda,\tilde{\mu})=F^0(\Lambda,\tilde{\mu})+f(\Lambda)
\sum_{s=2}^r\sum_{m=0}^{\delta_s-1}h_{ms}\Lambda^m\tilde{\mu}^s
\label{44}
\ee
where ${\bf (4H)}\,\,\delta_s=(N-1)s-N$.
$F^0(\Lambda,\tilde{\mu})$ is a polynomial whose coefficients are
determined by the isomonodromy invariants and $(T,U)$.  The
coefficients $h_{\delta_s-1,s}\,\,(2\leq s\leq r)$ are also determined
by these quantities and the remaining coefficients $h_{m,s}$ for
$2\leq s\leq r,\,0\leq m\leq\delta_s-2$ give the appropriate moduli
of number $\sum_{s=2}^r(\delta_s-1)=g$.
\\[3mm]\indent
Thus the ${\cal A}_i^0$ are solutions of the isospectral problem with
slowly varrying spectral invariants $h_{m,s}(T,U)$ and to derive
the modulation equations one writes
\be
Y=\left(\phi^0(t,u,T,U,\Lambda) + \phi^1\epsilon+\cdots\right)
\times exp[\epsilon^{-1}S(T,U,\Lambda)]
\label{45}
\ee
The lowest order equations are
$$
\frac{\partial S}{\partial\Lambda}\phi^0={\cal M}^0\phi^0;\,\,
\frac{\partial\phi^0}{\partial t_i}+\frac{\partial S}{\partial T_i}\phi^0=
-\frac{\partial{\cal A}^0_i}{\Lambda-T_i}\phi^0;$$
\be
\frac{\partial\phi^0}{\partial u_{\alpha}}+\frac{\partial S}{\partial
U_{\alpha}}\phi^0=(\Lambda E_{\alpha}+{\cal B}^0_{\alpha})
\phi^0
\label{46}
\ee
which determines $\phi^0$ up to a multiplier $\phi^0\to \phi^0
h(T,U)$.  These equations can be rewritten into the isospectral
linear problem
\be
\mu\psi={\cal M}^0\psi;\,\,\frac{\partial\psi}{\partial t_i}=-\frac
{{\cal A}_i^0}{\Lambda-T_i}\psi;\,\,\frac{\partial\psi}{\partial u_
{\alpha}}=(\Lambda E_{\alpha}+{\cal B}^0_{\alpha})\psi
\label{47}
\ee
$$\mu=\frac{\partial S}{\partial\Lambda};\,\,\psi=\phi^0exp\left(\sum t_i
\frac{\partial S}{\partial T_i}+\sum u_{\alpha}\frac{\partial S}{\partial
U_{\alpha}}\right)$$
In particular the characteristic equation $det(\mu I-{\cal M}^0(\Lambda))
=0$ is satisfied by $\mu=\partial_{\Lambda}S$.  One now identifies
this $\psi$ with a corresponding algebro-geometric BA function via
\be
\psi=\phi exp\left(\sum t_i\Omega_{T_i}+\sum u_{\alpha}\Omega_
{U_{\alpha}}\right);\,\,\Omega_L=\int^{(\Lambda,\mu)}d\Omega_L
\label{48}
\ee
One must choose here a symplectic homology basis $A_I,\,B_I$ so that
the WKB solution gives a correct approximation to the isospectral
problem and we assume that this has been done (the problem is 
nontrivial).  The meromorphic differentials are characterized now by
the following conditions: ${\bf (4I)}\,\,d\Omega_{T_i}$ has poles at
the $r$ points in $\pi^{-1}(T_i)$ and is holomorphic outside of
$\pi^{-1}(T_i)$.  Its singular behavior at these points is such that
$d\Omega_{T_i}=-d\mu +$ nonsingular.  ${\bf (4J)}\,\,d\Omega_{U_
{\alpha}}$ has poles at the $r$ points in $\pi^{-1}(\infty)$ with
singular behavior $d\Omega_{U_{\alpha}}=d\Lambda+$ nonsingular.
${\bf (4K)}\,\,\oint_{A_I}d\Omega_{T_i}=\oint_{A_I}d\Omega_
{U_{\alpha}}=0$.  Matching the exponential parts in the two forms
of $\psi$ gives then the modulation equations
\be
\frac{\partial S}{\partial T_i}=\Omega_{T_i};\,\,\frac{\partial S}
{\partial U_{\alpha}}=\Omega_{U_{\alpha}}
\label{49}
\ee
These can be rewritten as
\be
\left.\frac{\partial dS}{\partial T_i}\right|_{\Lambda=c}=
d\Omega_{T_i};\,\,\left.\frac{\partial dS}{\partial 
U_{\alpha}}\right|_{\Lambda=c}=d\Omega_{U_{\alpha}};\,\,
dS=\mu d\Lambda
\label{499}
\ee
leading to ``classical" Whitham equations of the form
(evaluation at $\Lambda=c$)
\be
\frac{\partial d\Omega_{T_j}}{\partial T_i}=\frac{\partial d\Omega_
{T_i}}{\partial T_j};
\label{50}
\ee
$$\frac{\partial d\Omega_{U_{\beta}}}{\partial U_{\alpha}}=\frac
{\partial d\Omega_{U_{\alpha}}}{\partial U_{\beta}};\,\,
\frac{\partial d\Omega_{U_{\alpha}}}{\partial T_i}=\frac
{\partial d\Omega_{T_i}}{\partial U_{\alpha}}$$
\indent
A similar multiscale analysis applies to the dual isomonodromic
problem with WKB ansatz
\be
Z=\left(\chi^0(t,u,T,U,\mu)+\chi^1\epsilon+\cdots\right)\times
exp[\epsilon^{-1}\Sigma(T,U,\mu)]
\label{51}
\ee
This leads to slow dynamics of the dual isospectral problem based
on the dual expressions ${\bf (4L)}\,\,det(\Lambda I-L^0(\mu))=0$ of the 
same spectral curve (cf. \cite{ac}).  The modulation equations are
obtained in the dual form (evaluation at $\mu=c$)
\be
\frac{\partial d\Sigma}{\partial T_i}=d\Omega_{T_i};\,\,
\frac{\partial d\Sigma}{\partial U_{\alpha}}=d\Omega_{U_{\alpha}};\,\,
d\Sigma=-\Lambda d\mu
\label{52}
\ee
(cf. \cite{dc,ia}).
\\[3mm]\indent
As before one can relate all this to SW ideas as follows:  ${\bf (A)}\,\,$
One has a $g$ dimensional period map $h=(h_{ms})\to a=(a_I)$ where
$a_I=\oint_{A_I}dS$ for $I=1,\cdots,g$.  One can show that the
Jacobian of this map is nonvanishing and one may write $h=h(T,U,a)$
(inverse period map).  For $a$ fixed this gives a family of deformations 
of the spectral curve producing a solution of the modulation equations;
varying $a$ one has a general solution.  ${\bf (B)}\,\,$ One will have
(for suitable holomorphic differentials $d\omega_J$)
\be
\frac{\partial dS}{\partial a_I}=d\omega_I;\,\,\oint_{A_I}d\omega_J
=\delta_{IJ}
\label{53}
\ee
${\bf (C)}\,\,$ There is a prepotential ${\cal F}(T,U,a)$ satisfying
\be
\frac{\partial{\cal F}}{\partial a_I}=b_I=\oint_{B_I}dS;\,\,\frac
{\partial{\cal F}}{\partial T_i}=H_i;
\label{54}
\ee
$$\frac{\partial {\cal F}}{\partial U_{\alpha}}=K_{\alpha};\,\,
\frac{\partial^2{\cal F}}{\partial a_I\partial a_J}={\cal T}_{IJ}=
\oint_{B_J}d\omega_I$$
(where $H_i$ and $K_{\alpha}$ are as before).

\section{GAUDIN MODEL AND KZ EQUATIONS}
\renewcommand{\theequation}{5.\arabic{equation}}
\setcounter{equation}{0}

We go to \cite{tj} now in order to exhibit connections of isomonodromy
problems and the Knizhnik-Zamolodchikov (KZ) equations
(see e.g. \cite{bd,fd,hf,id,kb,kd,kg,la,ra,se,tk} and cf. 
also \cite{ep,ic,oc}).  Here
\cite{tj} follows \cite{kg} at first and later \cite{ic}; given the
background in Sections 3 and 4 it is appropriate to follow \cite{tj} here
instead of the Russian school (to which we go later
via \cite{la,le,oa,ob}).
First a generalized Gaudin model is based on a generalization of the
XYZ Gaudin model to an $SU(n)$ spin system following \cite{kg}.
Let $X$ be a torus with modulus $\tau$, i.e. $X={\bf C}/({\bf Z}
+{\bf Z}\tau)$.  For integer $(a,b)$ define
\be
\theta_{[ab]}(z)=\theta_{\frac{a}{n}-\frac{1}{2},\frac{1}{2}-
\frac{b}{n}}(z,\tau);
\label{55}
\ee
$$\theta_{kk'}(z,\tau)=\sum_{m\in {\bf Z}}exp[\pi in(m+k)^2+
2\pi i(m+k)(z+k')]$$
Define $n\times n$ matrices ${\bf (5A)}\,\,J_{ab}=g^ah^b$ and
$J^{ab}=(1/n)J^{-1}_{ab}$ where $g=diag(1,\omega,\cdots,
\omega^{n-1})$ with $\omega=exp(2\pi i/n)$ and $h=(\delta_{i-1,j})$;
one notes that $gh=\omega hg$.  The $J$ matrices give a basis of
$su(n)$ over {\bf R} and of $sl(n,{\bf C})$ over {\bf C} with
$Tr(J_{ab}J^{cd})=\delta_{ac}\delta_{bd}$.  As an R matrix define
\be
R(\lambda)=\sum_{(a,b)\in {\bf Z}_n\times {\bf Z}_n}W_{ab}(\lambda,
\eta)J_{ab}\otimes J^{ab};\,\,W_{ab}=\frac{\theta_{[ab]}(\lambda
+\eta)}{\theta_{[ab]}(\eta)}
\label{56}
\ee
These Boltzman weights are n-periodic so the summation is as
indicated.  The r-matrix is the leading part in the $\eta$ expansion
of $R$ at $\eta=0$, i.e.
\be
r(\lambda)=\sum_{(a,b)\ne (0,0)}w_{ab}J_{ab}\otimes J^{ab};\,\,
w_{ab}(z)=\frac{\theta_{[ab]}(\lambda)\theta'_{[00]}(0)}
{\theta_{[ab]}(0)\theta_{[00]}(\lambda)}
\label{57}
\ee
where $'\sim d/dz$.  The sum is now over $({\bf Z}_n\times
{\bf Z}_n)/\{(0,0)\}$ (alternatively one could take $w_{00}(z)=0$).
The $r$ matrix satisfies the classical Yang-Baxter (YB) equation
\be
[r^{(13)}(\lambda),r^{(23)}(\mu)]=-[r^{(12)}(\lambda-\mu),
r^{(13)}(\lambda)+r^{(23)}(\mu)]
\label{58}
\ee
where e.g. $r^{(12)}(\lambda)=\sum w_{ab}(\lambda)J_{ab}
\otimes J^{ab}\otimes I$.  The generalized Gaudin model is a limit
as $\eta\to 0$ of an inhomogeneous $SU(n)$ spin chain with $N$
lattice sites and the monodromy matrix of the spin chain is
${\bf (5B)}\,\,T(\lambda)=L_N(\lambda-t_N)\cdots L_1(\lambda-
t_1)$.  The $t_i$ are inhomogeneous parameters which eventually 
correspond to time variables in an isomonodromy problem and 
${\bf (5G)}\,\,L_i(\lambda)=\sum_{(ab)} W_{ab}(\lambda)
J_{ab}\otimes\rho_i(J^{ab})$ acting on ${\bf C^n}\otimes V_i$
where $V_i$ is the representation space of an irreducible representation
$(\rho_i,V_i)$ of $su(n)$.  The $L$ operators satisfy ${\bf (5D)}\,\,
RLL=LLR$ and the monodromy matrix acts nontrivially on
${\bf C^n}\otimes V$ for $V=V_1\otimes \cdots\otimes V_N$.  The
leading nontrivial part in the $\eta$-expansion of $T(\lambda)$ at
$\eta=0$ is ${\bf (5E)}\,\,{\cal T}(\lambda)=\sum_1^N\sum_{(ab)}
w_{ab}(\lambda-t_i)J_{ab}\otimes \rho_i(J^{ab})$ and  one has
the fundamental commutation relation ${\bf (5F)}\,\,[{\cal T}^1
(\lambda),{\cal T}^2(\mu)]=-[r(\lambda-\mu),{\cal T}^1(\lambda)+
{\cal T}^2(\mu)]$ where e.g. ${\cal T}^1(\lambda)=\sum_i\sum_
{(ab)}w_{ab}(\lambda-t_i)J_{ab}\otimes I\otimes\rho_i(J^{ab})$
(i.e. the monodromy acts nontrivially in the first ${\bf C^n}$ of
${\bf C^n}\otimes {\bf C^n}\otimes V$).  Mutually commuting
Hamiltonians $H_i\,\,(i=0,1,\cdots,N)$ of the generalized Gaudin
model are defined via
\be
\frac{1}{2}Tr_{n\times n}{\cal T}(\lambda)^2=\sum_1^NC_i
{\cal P}(\lambda-t_i)+\sum_1^NH_i\zeta(\lambda-t_i)+H_0
\label{59}
\ee
where ${\cal P}(z)$ and $\zeta(z)$ are Weierstrass functions with
modulus $\tau$.  The $C_i$ are quadratic Casimir
elements of the algebra generated by $\rho_i(J^{ab})$, i.e.
one has 
${\bf (5G)}\,\,C_i=(1/2)\sum_{(ab)}\rho_i(J_{ab})\rho_i(J^{ab})$ and
${\bf (5H)}\,\,H_i=\sum_{j\ne i}\sum_{(ab)}w_{ab}
(t_i-t_j)\rho_i(J_{ab} \rho_j(J^{ab})$
for $i>0$ with $\sum_1^NH_i=0$.
\\[3mm]\indent
One considers now
${\bf (5I)}\,\,
M(\lambda)=\sum_1^N\sum_{(ab)} w_{ab}(\lambda-t_i)J_{ab}
A_i^{ab}$ as a classical analogue of the ${\cal T}$ operator in
{\bf (5E)}; the $A_i^{ab}$ are scalar functions of $t_i$ and $\tau$
(for $i=1,\cdots,N$) and the isomonodromy problem will be
formulated as differential equations for these functions.  Note the
$A_i^{ab}$ depend on $t$ but the $\rho_i(J^{ab})$ do not;
this is related to the difference between Heisenberg and Schr\"odinger
pictures (as indicated below).  The commutation relations for 
$\rho_i(J^{ab})$ can be packed into
\be
\{M(\lambda)\stackrel{\otimes}{,}M(\mu)\}=-[r(\lambda-\mu),M(\lambda)
\otimes I+I\otimes M(\mu)]
\label{60}
\ee
Here the left side is an abbreviation for ${\bf (5J)}\,\,\sum\{M^{ab}
(\lambda),M^{cd}(\mu)\}J_{ab}\otimes J_{cd}$ where 
$M^{ab}$ are the coefficients in $M=\sum M^{ab}J_{ab}$.  In
terms of the residue matrix ${\bf (5K)}\,\,A_i=Res_{\lambda=t_i}M(\lambda)
=\sum_{(ab)}J_{ab}A_i^{ab}$ the Poisson structure is simply the
one induced from the Kirillov-Kostant Poisson structure on 
$sl(n,{\bf C})$; symplectic leaves are the direct product ${\cal O}_1
\times\cdots\times {\cal O}_N$ of coadjoint orbits ${\cal O}_i$ in
$sl(n,{\bf C})$ on which $A_i$ is living.  These leaves become the
phase spaces of the nonautonomous Hamilton system
described as follows:  Using $M(\lambda)$ for ${\cal T}$
in (\ref{59}) with $C_i$ Casimir elements in the Poisson algebra
above, one thinks of Hamiltonians $H_i$ for the classical Gaudin
system in the form ${\bf (5L)}\,\,H_i=\sum_{j\ne i}\sum_{(ab)}
w_{ab}(t_i-t_j)A_{i,ab}A^{ab}_i$.  $H_0$ is also a quadratic form
in the $A_i^{ab}$ but is more complicated.
The $t_i$ here are just parameters of the classical Gaudin system, which
is a Hitchin system on the punctured torus $X/\{t_1,\cdots,t_N\}$.
The nonautonomous Hamiltonian system will be a nonautonomous
analogue of this Hitchin system in the form
\be
\frac{\partial A_j^{ab}}{\partial t_i}=\{A_j^{ab},H_i\};\,\,
\frac{\partial A_j^{ab}}{\partial\tau}=\left\{A_j^{ab},
\frac{1}{2\pi i}\left(H_0-
\eta_1\sum_1^Nt_iH_i\right)\right\}
\label{61}
\ee
where $\eta_1$ arises via the known transformation law ${\bf (5M)}\,\,
\zeta(z+1)=\zeta (z)+\eta_1$.  This nonautonomous Hamiltonian
system is an isomonodromy problem which can be put in Lax form.
The key is a relation
\be
\left\{M(\lambda),\frac{1}{2}Tr M(\mu)^2\right\}=\left[\sum_1^N
\sum_{(ab)}w_{-a,-b}(\mu-\lambda)w_{ab}(\mu-t_i)J_{ab}
A_i^{ab},M(\mu)\right]
\label{62}
\ee
(not proved - one can use function theoretic methods as with the
YB equation).  Taking residues gives
\be
\{M(\lambda),H_i\}=-[A_i(\lambda),M(\lambda)];\,\,A_i(\lambda)=
\sum_{(ab)} w_{ab}(\lambda-t_i)J_{ab}A_i^{ab}
\label{63}
\ee
and after lengthy calculations
\be
\{M(\lambda),H_0\}=[4\pi iB(\lambda)-\eta_1\sum_1^Nt_iA_i(\lambda),
M(\lambda)];
\label{64}
\ee
$$B(\lambda)=\sum_1^N\sum_{(ab)}Z_{ab}(\lambda-t_i)J_{ab}
A_i^{ab};\,\,Z_{ab}=\frac{w_{ab}(\lambda)}{4\pi i}\left(\frac
{\theta'_{[ab]}(\lambda)}{\theta_{[ab]}(\lambda)}-\frac{\theta'_{[ab]}
(0)}{\theta_{[ab]}(0)}\right)$$
The first formula can be written in the form
\be
\{M(\lambda),H_0-\eta_1\sum_1^Nt_iH_i\}=[4\pi iB(\lambda),M(\lambda)]
\label{65}
\ee
and one obtains the Lax form of the nonautonomous Hamiltonian
system as
\be
\frac{\partial M(\lambda)}{\partial t_i}=-[A_i(\lambda),M(\lambda)]-
\frac{\partial A_i(\lambda)}{\partial\lambda};
\label{66}
\ee
$$\frac{\partial M(\lambda)}{\partial\tau}=[2B(\lambda),M(\lambda)]+2
\frac{\partial B(\lambda)}{\partial\lambda}$$
One notes that $\tau$ derivatives correspond to $\lambda$ derivatives
via 
$${\bf (5N)}\,\,(\partial\theta_{[ab]}(\lambda)/\partial\tau)=
(1/4\pi i)(\partial^2\theta_{[ab]}(\lambda)/\partial\lambda^2)$$
These equations are integrability conditions for a linear system
\be
\left(\frac{\partial}{\partial\lambda}-M(\lambda)\right)Y(\lambda)=0;
\label{67}
\ee
$$\left(\frac{\partial}{\partial t_i}+A_i(\lambda)\right)Y(\lambda)=0;\,\,
\left(\frac{\partial}{\partial\tau}-2B(\lambda)\right)Y(\lambda)=0$$
where the last two equations can be interpreted as isomonodromy
deformations of the first equation.
\\[3mm]\indent
Now the elliptic KZ equations of \cite{kg} can be written as
\be
\left(\kappa\frac{\partial}{\partial t_i}+\sum_{j\ne i}\sum_{(ab)}
w_{ab}(t_i-t_j)\rho_i(J_{ab})\rho_j(J^{ab})\right)F(t)=0;
\label{68}
\ee
$$\left(\kappa\frac{\partial}{\partial \tau}+\sum_{i,j=1}^N\sum_{(ab)}
Z_{ab}(t_i-t_j)\rho_i(J_{ab})\rho_j(J^{ab})\right)F(t)=0$$
where $\kappa=k+n$ with $k$ the level of a twisted WZW model.  
These equations characterize N-point conformal blocks with 
irreducible representations $\rho_i$ sitting at $t_i$.  Following
\cite{ra} one adds another marked point at $\lambda$ with fundamental
representation $({\bf C}^n,id)$ to obtain
\be
\left(\kappa\frac{\partial}{\partial\lambda}+\sum_1^N\sum_{(ab)}
w_{ab}(\lambda-t_i)J_{ab}\rho_i(J^{ab})\right)G(\lambda,t)=0;
\label{69}
\ee
$$\left(\kappa\frac{\partial}{\partial t_i}+\sum_{j\ne i}\sum_{(ab)}
w_{ab}(t_i-t_j)\rho_i(J_{ab})\rho_j(J^{ab})+\sum_{(ab)}w_{ab}
(t_i-\lambda)\rho_i(J_{ab})J^{ab}\right)G(\lambda,t)=0;$$
$$\left(\kappa\frac{\partial}{\partial\tau}+\sum_{i,j=1}^N\sum_{(ab)}
Z_{ab}(t_i-t_j)\rho_i(J_{ab})\rho_j(J^{ab})+\sum_1^N\sum_{(ab)}
Z_{ab}(t_i-\lambda)\rho_i(J_{ab})J^{ab}+\right.$$
$$\left.
+\sum_1^N\sum_{(ab)}Z_{ab}(\lambda-t_j)J_{ab}\rho_j(J^{ab})+
\sum_{(ab)}Z_{ab}(0)J_{ab}J^{ab}\right)G(\lambda,t)=0$$
Let $F=exp(S)$ be a fundamental solution of (\ref{68}) (cf. \cite{ra})
and consider 
the resulting equations for $X=F^{-1}G$ of the form
\be
\left(\kappa\frac{\partial}{\partial\lambda}+\sum_1^N\sum_{(ab)}
w_{ab}(\lambda-t_i)J_{ab}A_i^{ab}\right)X(\lambda,t)=0;
\label{70}
\ee
$$\left(\kappa\frac{\partial}{\partial t_i}-\sum_{(ab)}w_{ab}(\lambda
-t_i)J_{ab}A_i^{ab}\right)X(\lambda,t)=0;$$
$$\left(\kappa\frac{\partial}{\partial\tau}+2\sum_1^N\sum_{(ab)}Z_{ab}(\lambda
-t_i)J_{ab}A_i^{ab}+\sum_{(ab)}Z_{ab}(0)J_{ab}J^{ab}\right)X
(\lambda,t)=0$$
where ${\bf (5O)}\,\,A_i^{ab}=F(t)^{-1}\rho_i(J^{ab})F(t)$.  This
is almost the same as (\ref{67}) except that:  {\bf (A)}$\,\,$ The last
equation contains an extra term.  {\bf (B)}$\,\,$  The $A_i^{ab}$ are
not scalar functions but operators on $V=V_1\otimes\cdots\otimes V_N$.
{\bf (C)}$\,\,$ There is an arbitrary parameter $k$; the previous
monodromy (i.e. (\ref{67})) can be reproduced by setting $\kappa=-1$.
Now the first descrepency {\bf (A)} can be removed by gauging
it away via $X\to [exp(f(\tau))]X$.  As for {\bf (B)} one remarks
that the system (\ref{70}) is a quantization of the isomonodromy
problem (\ref{67}); this is parallel to the relation between quantized
and classical Hitchin systems.  The operators $A_i^{ab}$ depend
on $(t,\tau)$ but inherit the same commutativity as the $J$ matrices
from $\rho_i(J^{ab})$.  The passage from $\rho_i(J^{ab})$ to 
$A_i^{ab}$ amounts to a change from the Schr\"odinger picture
to the Heisenberg picture and classical limit now means replacing
these operators by functions on a phase space, thus giving the
isomonodromy problem.

\section{ISOMONODROMY AND HITCHIN SYSTEMS}
\renewcommand{\theequation}{6.\arabic{equation}}
\setcounter{equation}{0}

We will extract here from \cite{oa} which gives an excellent survey of
many matters (cf. also \cite{id,la,le,lw,ob}).  However first some remarks
about RS and VB are appropriate.
\\[3mm]\indent
{\bf REMARK 6.1.}$\,\,$  In a first version of this survey we made up a
number of remarks (seven) on connections, vector bundles (VB), and Hitchin
systems following \cite{ae,be,de,da,dy,dx,ed,fh,gz,hu,hg,hz,hx,hw,hh,ie,
kh,kz,lf,la,le,lw,lz,mv,mx,mu,oa,ob,pa,vb}.  The result was complete
enough but lacked coherence so we will try again using \cite{he} as a 
springboard.  We will first simply indicate, generally without proof, some facts
about RS, VB, and sheaves from \cite{hi} (for background information
cf. \cite{ce,du,fz,fy,gu,ie,mq,pa} - we will assume various definitions are
known).
First a holomorphic section of a line bundle $L$ over $\Sigma\,\, (\pi:\,\,
L\to\Sigma$) is a holomorphic map $s:\,\,\Sigma\to L$ such that $\pi\circ
s=id_{\Sigma}$ and the space of such sections $H^0(\Sigma,L)$ is finite
dimensional.  Given a point $p\in\Sigma$, with $U$ a neighborhood (nbh)
of $p,\,\,V=\Sigma/\{p\}$, and $z(p)=0$, then, using $z$ as a transition
function
on $U\cap V$, one produces a line bundle $L_p$ which will be useful later.
The canonical bundle $K$ is the bundle of holomorphic 1-forms and on
$\Sigma={\bf P}^1$ one defines ${\cal O}(n)$ as the line bundle with transition
function $z^n$ on $U\cap V$ where e.g. $0\in U$ and $V={\bf P}^1/\{0\}$);
the dimension of $H^0({\bf P}^1,{\cal O}(n))$ is $n+1$.  Given a line bundle
$L$ one forms $L^*=L^{-1}$ via transition functions $g_{ab}(L^*)=g^{-1}_
{ab}(L)$ and there are obvious relations $Hom(L,\hat{L})\simeq L^*
\otimes \hat{L}$, etc.  If $\Sigma$ is compact then $genus(\Sigma)=g=dim\,
H^0(\Sigma,K)$.  
\\[3mm]\indent
One defines sheaf cohomology in the standard manner
and we recall for a line bundle $L,\,\,H^1(\Sigma,L)\simeq H^0(\Sigma,
K\otimes L^*)$ (Serre duality).  The isomorphism classes of holomorphic
line bundles on $\Sigma$ are given by elements of $H^1(\Sigma,{\cal O}^*)$
(called the Picard group) and for given degree $d,\,\,J^d\sim$ line bundles
of degree $d$ is a complex torus isomorphic to $Jac(\Sigma)$  
(more on this below).  Note also for
${\cal S}={\cal O}(L)=$ sheaf of holomorphic sections of $L$, that
$H^p(\Sigma,{\cal S})=0$ for $p>1$; if ${\cal S}={\bf C}$ or ${\bf Z}$ then
$H^p(\Sigma,{\cal S})=0$ for $p>2$.  From the short exact sequence
$(\bullet)\,\,0\to {\bf Z}\to {\cal O}\stackrel{\phi}{\to}{\cal O}^*\to 1$
with $\phi\sim exp(2\pi if)$ one determines a long exact sequence leading to
\be
0\to\frac{H^1(\Sigma,{\cal O})}{H^1(\Sigma,{\bf Z})}\to H^1(\Sigma,
{\cal O}^*)\stackrel{\delta}{\to}H^2(\Sigma,{\bf Z})=0
\label{900}
\ee
where $H^2(\Sigma,Z)\simeq {\bf Z},\,\,H^1(\Sigma,{\cal O})\simeq 
{\bf C}^g$, and $H^1(\Sigma,{\bf Z})\simeq {\bf Z}^{2g}$.  The 
coboundary operator $\delta$ in (\ref{900}) yields (since $H^1(\Sigma,
{\cal O}^*)\simeq$ equivalence classes of line bundles) $\delta([L])=deg(L)
=c_1(L)\in{\bf Z}$ (first Chern class).  Thus for $L$ such that $\delta([L])=
d$ one obtains $J^d(\Sigma)\simeq Jac(\Sigma)$  One shows next that if
a section $s\in H^0(\Sigma,L)$ vanishes at $p_i\in\Sigma$ with multiplicities
$m_i$ then $deg(L)=\sum m_i$ so if $deg(L)<0$ then $L$ has no nontrivial
holomorphic sections.  For vector bundles $V$ one 
defines $deg(V)=deg(det(V))=c_1(V)$ 
where $det(V)\sim\wedge^mV$ for $rank(V)=m$ has transition functions
$det(g_{UW})$ in an obvious notation.  We recall also the Riemann-Roch
(RR) theorem which states
\be
dim\,H^0(\Sigma,V)-dim\,H^1(\Sigma,V)=deg(V)+rank(V)(1-g)
\label{901}
\ee
Many nice results can be obtained from this.  Another classical theorem
states that any rank $m$ holomorphic VB
$V\to{\bf P}^1$ has the form $V\simeq {\cal O}(a_1)\oplus\cdots
\oplus {\cal O}(a_m)$ for $a_i\in {\bf Z}$.  One defines direct image sheaves
via $f_*{\cal S}(U)={\cal S}(f^{-1}(U))$ for $f:\,\,\tilde{\Sigma}\to\Sigma$
and for ${\cal S}={\cal O}(L)$ there results ${\bf (A)}\,\,H^0(\Sigma,
f_*{\cal O}(L))\simeq H^0(\tilde{\Sigma},{\cal O}(L))$ with 
${\bf (B)}\,\,f_*{\cal O}(L)={\cal O}(E)$ for $E$ a rank $m$ holomorphic
VB (where $m=deg(f)=deg(f^*L_p)=\#[f^{-1}(p)]$) and {\bf (C)}$\,\,$
If $V$ is a holomorphic VB on $\Sigma$ then $f_*{\cal O}(L\otimes
f^*V)\simeq{\cal O}(E\otimes V)$ (where $E$ is given in {\bf (B)}).
One obtains then for $L$ and $E$ as indicated $deg(E)=deg(L)+
(1-\tilde{g})-deg(f)(1-g)$ (note here a useful fact in proving theorems is that
for $V$ a VB on $\Sigma$
$H^0(\Sigma,VL_p^{-n})=0$ for $n$ sufficiently large or
equivalently, on ${\bf P}^1$,
$V(n)=V\otimes {\cal O}(n)$ will have holomorphic sections
for $n$ sufficiently large).
Next note that if $deg(E)=0$ above then $E$ is trivial if and 
only if $L(-1)=L\otimes
f^*{\cal O}(-1)$ has no nontrivial sections.  Concerning $Jac(\Sigma)$ one 
usually defines $J^{g-1}(\Sigma)=Jac(\Sigma)$ and the 
the image of the holomorphic map $(p_1,\cdots,p_{g-1})\to L_{p_1}\cdots
L_{p_{g-1}}$ (tensor product) from $\Sigma\times\cdots\times\Sigma\to
J^{g-1}(\Sigma)$ is called the $\Theta$ divisor; it is of codimension 1
in $Jac(\Sigma)$.
\\[3mm]\indent
Now going to Lax equations and integrable systems, take a section
$w\in H^0(\Sigma,f^*{\cal O}(n))$ where $f:\,\Sigma\to{\bf P}^1$ and
for $U\subset{\bf P}^1$ multiplication by $w$ defines a linear map
$w:\,H^0(f^{-1}(U),L)\to H^0(f^{-1}(U),L(n))$ where $L(n)=L\otimes
f^*{\cal O}(n)$.  One takes here $L$ such that $L(-1)\in J^{g-1}/\Theta$ and
$f_*{\cal O}(L)={\cal O}(E)$ with $E$ trivial (thus $deg(E)=0$ and $L(-1)=
L\otimes f^*{\cal O}(-1)$ has no nontrivial holomorphic sections).  By 
definition of $E$ this $w$ determines a homomorphism $W:\,H^0(U,E)\to
H^0(U,E(n))$ and thence
\be
W:\,\,H^0({\bf P}^1,E)\simeq {\bf C}^m\to H^0({\bf P}^1,E(n))\simeq {\bf C}^m
\otimes H^0({\bf P}^1,{\cal O}(n))
\label{902}
\ee
Thus one has an $m\times m$ matrix valued holomorphic section of 
${\cal O}(n)$ so $W=A(z)=A_0+A_1z+\cdots+A_nz^n$.  This gives a
construction $L\to A(z)$ and from $A(z)s_i=ws_i$ on sections (via local
constructions) one is led to the algebraic curve 
$(\bullet\bullet)\,\,S:\,det(wI-A(z))=w^m+
a_1(z)w^{m-1}+\cdots+a_m(z)=0$ which is an m-fold covering of ${\bf P}^1$.
One can think (rather crudely)
of $w$ embedding $\Sigma$ as a one dimensional submanifold
$S$ of the total space of ${\cal O}(n)\,\,(\subset {\bf C}^2$ locally)
and eigenspaces of $A(z)$ correspond to line bundles on $\Sigma$ or $S$.
This is not clearly stated and the matter should be clarified below.
As for Lax pair equations $dA/dt=[A,B]$ with $A(z),\,B(z)$ matrix
polynomials one sees that the spectrum of $A$ is preserved in such 
evolutions so the eigenspace bundle moves in a complex torus and if a line
bundle follows a linear motion there is a basis for which $A(z)$ evolves
as a Lax pair for a specific $B$ depending on $A$ (cf. \cite{hi} for details).  
Finally one comes to completely integrable Hamiltonian systems (CIHS).
We recall that symplectic manifolds $M^{2N}=M$ have a nondegenerate
closed 2-form $\omega$ (and we assume $\omega$ is holomorphic for
complex manifolds $M$).  This provides an isomorphism $TM\leftrightarrow
T^*M$ so that for a given function $H:\,M\to {\bf C}$ (Hamiltonian) one
has a vector field $X_H$ characterized by $\omega(X_H,Y)=Y(H)$ for vector
fields $Y$.  The Poisson bracked is defined via $\{f,g\}=X_f(g)=-X_g(f)$ and 
$X_{\{f,g\}}=[X_f,X_g]$.  One speaks of coadjoint orbits of a Lie group
$G$ as follows.  If $\xi\in\tilde{g}^*$ then write $\omega_{\xi}(X,Y)=\xi([X,
Y])$ where $X,Y\in\tilde{g}$ define tangent vectors to the orbit at $\xi$.
Further for $f,g\in C^{\infty}(\tilde{g}^*)$ one can express $df(\xi)\in
\tilde{g}$ as a linear form on $T_{\xi}\tilde{g}^*=\tilde{g}^*$ and write
$\{f,g\}(\xi)=<\xi,[df(\xi),dg(\xi)]>$.  One writes also $Ad_g^*:\,
\tilde{g}^*\to \tilde{g}^*$ via $<Ad_g^*(\xi),X>=<\xi,Ad_{g^{-1}}X>$ (and
similarly $<ad^*_X\xi,Y>=-<\xi,[X,Y]>$).
The momentum map is an equivariant map $\mu:\,M\to\tilde{g}^*$ and as an
example of all this let $G=SL(m,{\bf C})$ and use $Tr(AB)$ to identify
$\tilde{g}$ and $\tilde{g}^*$.  Take a product of coadjoint orbits ${\cal O}
\times\cdots\times {\cal O}_k$ of elements of $\tilde{g}\simeq\tilde{g}^*$
with distinct eigenvalues.  The moment map is 
$(R_1,\cdots,R_k)\to\sum R_i$ with symplectic quotient $M=\{(R_i)/\sum R_i=0
\}$ which has dimension $(\clubsuit)\,\,2N=k(m^2-m)-2(m^2-1)$.  Let
\be
A(z)=p(z)\sum_1^k\frac{R_i}{z-\alpha_i};\,\,p(z)=\prod_1^k(z-\alpha_i);
\label{903}
\ee
$$det(w-A(z))=w^m+a_2(z)w^{m-2}+\cdots+ a_m(z)=p(z,w)$$
The eigenvalues $\lambda_k$ of $A(\alpha_k)=R_k$ are fixed in advance and
$p(\alpha_j,w)=det(w-R_j)=0$ for $w=\lambda_j$ so the polynomial coefficients
$a_i(z)$ are not arbitrary.  It is shown in \cite{hi} how Lax pair equations
arise in this example upon fixing the $a_i$ subject to constraints (cf. also
\cite{ac}).  In any event one defines a CIHS to be a symplectic manifold
$M^{2N}$ with $N$ Hamiltonian functions $H_i$ such that $\{H_i,H_j\}=0$
and $\wedge_1^NdH_i\not\equiv 0$.  The bigger picture developed in 
particular in \cite{de,dy,dx,hg,hz,hx,hw,hi,la,le,lw,mv,oa,ob,sf} can be 
illustrated now by considering $R(z)=\sum[R_jdz/(z-\alpha_j)]$ with
$R_j$ the residues of a matrix valued differential on ${\bf P}^1$.  The
fact $\sum R_i=0$ corresponds to the vanishing of the moment map for
symplectic reduction.  Now replace $R(z)$ by a 1-form on an arbitrary 
compact RS with values in $End(V)$ for an arbitrary holomorphic VB
$V$.  The numerical miracle now involved consists in obtaining in suitable
circumstances precisely the right number $N$ of Hamiltonian functions.
Thus e.g. in the example above one requires the number of 
independent coefficients
$a_i$ in (\ref{903}) to be $N$ (cf. $(\clubsuit)$).  In order to obtain a
suitable
structure in general one considers the space ${\cal R}$ of equivalence classes
of stable holomorphic rank $m$ vector bundles V over $\Sigma_g$.  Stable
has various definitions (which we omit) and the facts used here are only that
the only global endomorphisms of $V$ should be scalars and that ${\cal R}$
is a complex manifold with $dim({\cal R})=1-m^2(1-g)$.  Then $T^*{\cal R}$
is a symplectic manifold and since $T{\cal R}$ at $[V]$ is $H^1(\Sigma,
End(V))$ Serre duality gives $T^*{\cal R}\simeq H^0(\Sigma,End(V)\otimes
K)$.  Thus locally a point in $T^*{\cal R}$ is (up to equivalence)
a VB $V$ and a holomorphic section $A$ of $End(V)\otimes K$.  Therefore
$A$ is an $m\times m$ marix with values in $K$ and its characteristic
polynomial is $p(z,w)=det(w-A(z))=w^m+a_1w^{m-1}+\cdots+a_m$ with
holomorphic coefficients $a_i\in H^0(\Sigma,K^i)$.  One can show that
$dim\,H^0(\Sigma,K)=g$ and $dim\,H^1(\Sigma,K)=1$ so by RR $deg(K)=2g-2$
($g>1$) and by Serre duality again $H^1(\Sigma,K^i)^*\simeq H^0(\Sigma,
K^{1-i})$.  But $K^{1-i}$ has negative degree and no sections for $i>1$ so
RR for $K^i$ implies $dim\,H^0(\Sigma,K^i)=(2i-1)(g-1)$.  The number of
degrees of freedom in choosing the characteristic polynomial $p(z,w)$ is then
$g+\sum_2^m(2i-1)(g-1)=1-m^2(1-g)$ and this is exactly the dimension of
${\cal R}$ or $(1/2)\times dim(T^*{\cal R})$.  This will then give rise to
a CIHS
based on $T^*{\cal R}$ and $p(z,w)=0$ determines a spectral curve $S$
in the total space of $K$ over $\Sigma$; $A$ is determined by a line bundle on
$S$ and the Hamiltonian flows are linear.  Again the construction needs
clarification (see below).
\\[3mm]\indent
We go now to \cite{oa} and
let $\Sigma_g$ be a RS
of genus $g$ and $FB(\Sigma,G)$ be the space of flat vector 
bundles (VB), $V=V_G\,\,(G=GL(N,{\bf C}))$, with smooth
connections ${\cal A}$.  Flatness means ${\bf (6A)}\,\, {\cal F}_{\cal A}
=d{\cal A}+(1/2)[{\cal A},{\cal A}]=0$.  Fixing a complex structure
on $\Sigma_g$ one has ${\cal A}\sim (A,\bar{A})$ with a system of
matrix equations
\be
(\partial +A)\psi=0;\,\,(\bar{\partial}+\bar{A})\psi=0
\label{71}
\ee
(linearization of {\bf (6A)}).
One modifies this via a parameter $\kappa\in {\bf R}$ (the level)
and uses $\kappa\partial$ instead of $\partial$ in the first equation of
(\ref{71}).  Now let $\mu$ be a Beltrami differential $\mu\in\Omega^
{(-1,1)}(\Sigma_g)$ (cf. \cite{ie}).  Then in local coordinates
${\bf (6B)}\,\,\mu=\mu(z,\bar{z})\partial_z\otimes d\bar{z}$ and
one can deform the complex structure on $\Sigma_g$ to produce
new coordinates
\be
w=z-\epsilon(z,\bar{z});\,\,\bar{w}=\bar{z};\,\,\mu=\frac{\bar{\partial}
\epsilon}{1-\partial\epsilon}
\label{72}
\ee
(note this should be called $-\mu$ to agree with $\bar{\partial}w/
\partial w$ but we retain the notation of \cite{oa}).
Then $\partial_{\bar{w}}=\bar{\partial}+\mu\partial$ annihilates
$dw$ while $\bar{\partial}$ annihilates $dz$ (note $(\bar{\partial}
+\mu\partial)(z-\epsilon)=-\bar{\partial}\epsilon+\mu-\mu\partial
\epsilon=0$).  In the new coordinates {\bf (6A)} has the form
${\bf (6C)}\,\,{\cal F}_{\cal A}=(\bar{\partial}+\partial\mu)A-
\kappa\partial\bar{A}+[\bar{A},A]=0$ and one arrives at the system
\be
(\kappa\partial+A)\psi=0;\,\,(\bar{\partial}+\mu\partial+\bar{A})\psi=0
\label{73}
\ee
Write now $\mu=\sum_{a=1}^{\ell} t_a\mu^0_a$ where $\mu^0_1,
\cdots,\mu^0_{\ell}$ is a basis in the tangent space to the moduli space
${\cal M}_g$ of complex structures on $\Sigma_g$ (here $\ell=
3g-3$ for $g>1$).  Note that $\mu\partial$ measures the deviation
of $\partial_{\bar{w}}$ from $\bar{\partial}$.  Fix now a fundamental
solution of (\ref{73}) via $\psi(z_0,\bar{z}_0)=I$.  Let $\gamma$ be a
homotopically nontrivial cycle in $\Sigma_g$ such that $(z_0,\bar{z}_0)
\in\gamma$ and set ${\bf (6D)}\,\,{\cal Y}(\gamma)=\psi(z_0,
\bar{z}_0)|_{\gamma}=P\,exp\oint_{\gamma}{\cal A}$
(monodromy - $P\sim$ path ordered product as in \cite{be} for
example where many basic ideas about connections etc. are spelled
out).  The set of matrices ${\cal Y}(\gamma)$ generates a representation
of $\Pi_1(\Sigma_g,z_0)$ in $GL(N,{\bf C})$ and
independence of the monodromy ${\cal Y}$ to deformation of complex
structure means
\be
\partial_a{\cal Y}=0\,\,\,\left(a=1,\cdots,\ell;\,\,\partial_a=\frac{\partial}
{\partial t_a}\right)
\label{74}
\ee
It follows that (\ref{74}) is consistent with (\ref{73}) if and only if
\be
\partial_aA=0;\,\,\partial\bar{A}=\frac{1}{\kappa}A\mu^0_a\,\,
(a=1,\cdots,\ell)
\label{75}
\ee
Further this system (\ref{75}) is Hamiltonian where one endows
$FB(\Sigma,G)$ with a symplectic form ${\bf (6E)}\,\,\omega^0=
\int_{\Sigma_g}<\delta A,\delta\bar{A}>$ where $<\,\,,\,\,>\sim$ 
Trace; Hamiltonians are defined via ${\bf (6F)}\,\,H_a=(1/2)\int_
{\Sigma_g}<A,A>\mu^0_a\,\,(a=1,\cdots,\ell)$.
\\[3mm]\indent
Consider now the bundle ${\cal P}$ over ${\cal M}_g$ with fiber
$FB(\Sigma,G)$; the triple $(A,\bar{A},t)$ can be used as a local
coordinate and one thinks of ${\cal P}$ as an extended phase space
with a closed 2-form ${\bf (6G)}\,\,\omega=\omega^0-(1/\kappa)\sum_a
\delta H_a \delta t$.  Although $\omega$ is degenerate on ${\cal P}$
it produces equations of motion (\ref{75}) since $\omega^0$ is 
nondegenerate along the fibers.  Gauge transformations in the deformed
complex structure have the form
\be
A\to f^{-1}\kappa\partial f+f^{-1}Af;\,\,\bar{A}\to f^{-1}(\bar
{\partial}+\mu\partial)f+f^{-1}\bar{A}f
\label{76}
\ee
The form $\omega$ is invariant under such gauge transformations
(the set of which we call ${\cal G}$) but not $\omega^0$ or 
$H_a$ independently.  If one writes now ${\bf (6H)}\,\,\bar{A}'=
\bar{A}-(1/\kappa)\mu A$ and uses $(A,\bar{A}')$ as a connection
then ${\bf (6I)}\,\,\omega=\int_{\Sigma_g}<\delta A,\delta
\bar{A}'>$.
\\[3mm]\indent
A gauge fixing via (\ref{76}) plus the flatness condition {\bf (6C)} is
in fact a symplectic reduction from the space of smooth connections to
the moduli space of flat connections $\widetilde{FB}(\Sigma,G)=
SM(\Sigma,G)//{\cal G}$ where $//$ means {\bf (6C)} plus a gauge
fixing are in force.  The flatness condition is called the moment constraint
equation.  Now fix the gauge so that the $\bar{A}$ component of ${\cal A}$
becomes antiholomorphic, i.e. ${\bf (6J)}\,\,\partial\bar{L}=0$ via
$\bar{L}=f^{-1}(\bar{\partial}+\mu\partial)f+f^{-1}\bar{A}f$.  This
can be achieved since {\bf (6J)} amounts to the classical equations of
motion for the WZW functional $S_{WZW}(f,\bar{A})$ with
gauge field $f$ in the external field $\bar{A}$ (cf. \cite{bd,kg}).
Let $L$ be the gauge transformed $A$, i.e. $L=f^{-1}\kappa\partial f+
f^{-1}Af,$ and then {\bf (6C)} takes the form ${\bf (6K)}\,\,
(\bar{\partial}+\mu\partial)L+[\bar{L},L]=0$.  Thus the moduli space
of flat connections $\widetilde{FB}(\Sigma,G)$ is characterized by the
set of solutions of {\bf (6K)} along with {\bf (6J)} and this space
has dimension ${\bf (6L)}\,\,dim(\widetilde{FB})=2(N^2-1)(g-1)$
(for $g>1$).  After gauge fixing the bundle ${\cal P}$ over ${\cal M}_g$
becomes $\tilde{{\cal P}}$ with $\widetilde{FB}$ as fibers and the
equations (\ref{73})-(\ref{74}) become
\be
(\kappa\partial+L)\psi=0;\,\,(\bar{\partial}+\mu\partial+\bar{L})\psi=0;
\,\,(\kappa\partial_a+M_a)\psi=0
\label{77}
\ee
where we have replaced $\psi$ by $f^{-1}\psi$ and $M_a=\kappa
\partial_af\,f^{-1}$.  Note that $\kappa$ is not involved in the WZW
result and thus $\kappa$ seems to be arbitrary in (\ref{77}).
The gauge transformations do not spoil the
consistency of this system and from (\ref{77}) one arrives at the Lax
form of the isomonodromy deformation equations
\be
\partial_aL-\kappa\partial M_a+[M_a,L]=0;
\label{78}
\ee
$$\kappa\partial_a\bar{L}-\mu^0_aL=(\bar{\partial}+
\mu\partial)M_a-[M_a,\bar{L}]$$
These equations play the role of (\ref{75}) and the last equation in
(\ref{78}) allows one to find $M_a$ in terms of the dynamical
variables $(L,\bar{L})$.
\\[3mm]\indent
The symplectic form $\omega$ on $\tilde{{\cal P}}$ is
\be
\omega=\int_{\Sigma_g}<\delta L,\delta\bar{L}>-\frac{1}
{\kappa}\sum_a\delta H_a\delta t_a;\,\,H_a=\frac{1}{2}
\int_{\Sigma_g}<\delta L,\delta L>\mu^0_a
\label{79}
\ee
and we introduce local coordinates $(v,u)$ in $\widetilde{FB}$ via
${\bf (6M)}\,\,(L,\bar{L})=(L,\bar{L})(v,u,t)$ with $v=(v_1,\cdots,
v_M)$ and $u=(u_1,\cdots,u_M)$ for $M=(N^2-1)(g-1)$.  Assume
for simplicity now that this leads to the canonical form on 
$\widetilde{FB}$
\be
\omega^0=\int_{\Sigma_g}<\delta L(v,u,t),\delta\bar{L}(v,u,t)>=
(\delta v,\delta u)
\label{80}
\ee
where $(\,\,,\,\,)$ is induced by the trace.  On the extended phase space
one has then
\be
\omega=(\delta v,\delta u)-\frac{1}{\kappa}\sum_a\delta K_a(v,u,t)
\delta t_a
\label{81}
\ee
and variations in the $K_a$ take the form
\be
\delta K_a=\int_{\Sigma_g}[<L,\delta L>\mu^0_a+\kappa(<\delta L,
\partial_a\bar{L}>-<\partial_aL,\delta\bar{L}>)]
\label{82}
\ee
Now because of {\bf (6K)} the Hamiltonians depend explicitly on
times and one considers the Poincar\'e-Cartan integral invariant
(cf. \cite{ad})
\be
\Theta=\delta^{-1}\omega=(v,\delta u)-\frac{1}
{\kappa}\sum_aK_a
(v,u,t)\delta t_a
\label{83}
\ee
Then there exist $3g-3=dim({\cal M}_g)$ vector fields ${\bf (6N)}\,\,
{\cal V}_a=\kappa\partial_a+\{H_a,\,\cdot\,\}\,\,\,(a=1,\cdots,\ell)$
that annihilate $\Theta$ and one can check that
($K_a\sim H_a$ here and below)
\be
\kappa\partial_sH_r-\kappa\partial_rH_s+\{H_s,H_r\}_{\omega^0}=0
\label{84}
\ee
Note from \cite{la} that ${\cal V}_a\sim\sum (A_i^a\partial_{v_i}+
B_i^a\partial_{u_i})+\partial_a$ and ${\cal V}_a\in ker(\Theta)$ means
($\partial_a\sim\partial/\partial t_a$)
\be
A_i^a+\frac{\partial H_a}{\partial u_i}=0;\,\,-B_i^a+\frac{\partial H_a}
{\partial v_i}=0;
\label{844}
\ee
$$-A_i^a\frac{\partial H_b}{\partial v_i}-B_i^a\frac{\partial H_b}
{\partial u_i}-\partial_bH_a+\partial_aH_b=0$$
so that ${\bf (6NN)}\,\,{\cal V}_a=-(\partial H_a/\partial u_i)
\partial_{v_i}+
(\partial H_a/\partial v_i)\partial_{u_i}+\partial_a$
Thereby they define the flat connection in $\tilde{{\cal P}}$ and these 
conditions are sometimes referred to as the Whitham hierarchy.
Thus for $\Sigma\in{\cal M}_g$ the Whitham equations determine
a flat connection in $\widetilde{FB}(\Sigma,G)$.
This gives an entirely new perspective for the idea of Whitham
equations; it is based on geometry and deformation theoretic
ideas (no averaging).  For a given $f(v,u,t)$ on $\tilde{{\cal P}}$
the corresponding equations take the form
\be
\frac{df(v,u,t)}{dt_s}=\kappa\frac{\partial f(v,u,t)}{\partial t_s}+\{H_s,f\}
\label{85}
\ee
called the hierarchy of isomonodromic
deformations (HID).  Both hierarchies can be derived from
variations of a prepotential $F$ on $\tilde{{\cal P}}$ where
${\bf (6O)}\,\,F(u,t)=F(u_0,t_0)+\sum\int_{u_0,t_0}^{u,t_s}{\cal L}_sdt'_s$
with ${\cal L}_s(\partial_su,u,t)=(v,\partial_su)-K_s(v,u,t)$ the
Lagrangian ($\partial_su=\delta K_s/\delta v$ - note $F\sim S$ is better
notation).  $F$ then satisfies
the Hamilton-Jacobi (HJ) equation
\be
\kappa\partial_sF+H_s\left(\frac{\delta F}{\delta u},u,t\right)=0
\label{86}
\ee
and $F\sim log(\tau)$ where $\tau$ corresponds to
a tau function of HID (the notation is clumsy however).  
Note that equations such as (\ref{86}) arise in
\cite{lg} with period integrals $a_i$ as moduli (cf. also 
Section 8 - recall that the Toda type Seiberg-Witten curves for massless
$SU(N)$ have $N-1=g$ moduli with a $1-1$ map to the $a_i$).
Singular curves are important
for low genera situations but we omit this here (cf. \cite{oa}).
\\[3mm]\indent
Consider next the moduli space ${\cal R}={\cal R}_{g,N}$ of stable
holomorphic $GL(N,{\bf C})$ vector bundles $V$ over $\Sigma=\Sigma_g$
(cf. \cite{hw,kh,pa} for stable and semistable VB and note that 
in \cite{la,lw,oa} one develops the theory for
$\Sigma\sim\Sigma_{g,n}$ with $n$ marked
points but this will be omitted here.
${\cal R}$ is a smooth
variety of dimension $\tilde{g}=N^2(g-1)+1$.  Let $T^*{\cal R}$ be 
the cotangent bundle to ${\cal R}$ with the standard symplectic form.
Then Hitchin defined a completely integrable system on $T^*{\cal R}$
(cf. \cite{hg}).  The space $T^*{\cal R}$ can be obtained by a 
symplectic reduction from $T^*{\cal R}^s_{g,N}=\{(\phi,\bar{A})\}$
where $\bar{A}$ is a smooth connection of the stable bundle 
corresponding to $\bar{\partial}+\bar{A}$ and $\phi$ is a Higgs
field $\phi\in \Omega^0(\Sigma,End(V)\otimes K)$ where $K$ is
the canonical bundle of $\Sigma$ 
($K\sim$ holomorphic cotangent bundle - cf. \cite{ce}).  
One has a symplectic form ${\bf (6P)}\,\,
\omega^0=\int_{\Sigma_g}<\delta\phi,\delta\bar{A}>$ which is 
invariant under the gauge group ${\cal G}=C^{\infty}(\Sigma,GL(N,
{\bf C}))$ where ${\bf (6Q)}\,\,\phi\to f^{-1}\phi f$ and $\bar{A}\to
f^{-1}\bar{\partial} f+f^{-1}\bar{A}f$ with ${\cal R}_{g,N}=
{\cal R}^s_{g,N}/{\cal G}$.  Let now $\rho_{sk}\partial^{k-1}_z\otimes
d\bar{z}$ be $(-k+1,1)$ differentials ($\rho_{sk}\in H^1(\Sigma,
\Gamma^{k-1}\otimes K)$ where $\Gamma^{k-1}\sim k-1$ times
differentiable sections, and $s$ enumerates the basis in $H^1(\Sigma,
\Gamma^{k-1}\otimes K)$ so $\rho_{s,2}\sim\mu_s$).  By
Riemann-Roch $dim(H^1(\Sigma,\Gamma^{k-1}\otimes K))=
(2k-1)(g-1)$ and one can define gauge invariant Hamiltonians
\be
H_{s,k}=\frac{1}{k}\int_{\Sigma}<\phi^k>\rho_{s,k}\,\,\,
(k=1,\cdots,N;\,s=1,\cdots,(2k-1)(g-1))
\label{87}
\ee
where the Hamiltonian equations are
\be
\partial_a\phi=0\,\,\,\left(\partial_a= \frac{\partial}{\partial t_a};\,\,
a=(s,k)\right);\,\,\partial_a\bar{A}=\phi^{k-1}\rho_{s,k}
\label{88}
\ee
The gauge action produces a moment map $\mu:\,\,T^*{\cal R}^s\to
\widetilde{gl}^*(N,{\bf C})$ and from {\bf (6P)}-{\bf (6Q)} one has
${\bf (6R)}\,\,\mu=\bar{\partial}\phi+[\bar{A},\phi]$.  The reduced
phase space is the cotangent bundle $T^*{\cal R}_{g,N}\sim
T^*{\cal R}^s//{\cal G}=\mu^{-1}(0)/{\cal G}$ 
(gauge fixing and flatness) and finally the
Hitchin hierarchy (HH) is the set of Hamiltonians (\ref{87}) on
$T^*{\cal R}_{g,N}$.  Note also that the number of $H_{s,k}$ is
$\sum_1^N(2k-1)(g-1)=N^2(g-1)+1=
\tilde{g}=dim(T^*{\cal R}_{g,N})$.
Since they are independent and Poisson commuting (HH) is a set of
completely integrable Hamiltonian systems on $T^*{\cal R}_{g,N}$. 
To obtain equations
of motion for (HH) fix the gauge of $\bar{A}$ via $\bar{A}=f\bar
{\partial}f^{-1}+f\bar{L}f^{-1}$ so that $L=f^{-1}\phi f$ is a
solution of the moment constraint equation ${\bf (6S)}\,\,
\bar{\partial}L+[\bar{L},L]=0$ (flatness condition).  The space of
solutions of {\bf (6S)} is isomorphic to $H^0(\Sigma,End(V)
\otimes K)\sim$ cotangent space to the moduli space ${\cal R}_{g,N}$.
The gauge term $f$ defines the element $M_a=\partial_af\,f^{-1}\in
\tilde{gl}(N,{\bf C})$ while the equations
\be
(\lambda+L)Y=0;\,\,(\partial_a+M_a)Y=0;
\label{89}
\ee
$$(\bar{\partial}+\sum_{s,k}\lambda^{k-1}t_{s,k}\rho_{s,k}+
\bar{L})Y=0$$
are consistent and give the equations of motion for (HH).  To prove this
one checks that consistency follows from {\bf (6S)} and in terms of
$L$ the equations (\ref{78}) take the form
\be
\partial_aL+[M_a,L]=0\,\,(Lax);\,\,\partial_a\bar{L}-\bar{\partial}
M_a+[M_a,\bar{L}]=L^{k-1}\rho_{s,k}
\label{90}
\ee
where $a=(s,k)$.  The Lax equation provides consistency of the first
two equations in (\ref{89}) while the second equation in (\ref{90})
plays the same role for the second and third equations in (\ref{89}); this
equation also allows one to determine $M_a$ from $L$ and $\bar{L}$.
\\[3mm]\indent
Due to the Liouville theorem the phase flows of (HH) are restricted
to the Abelian varieties corresponding to a level set of the Hamiltonians
$H_{s,k}=c_{s,k}$.  This becomes simple in terms of action-angle
coordinates defined so that the angle type coordinates are angular 
coordinates on the Abelian variety and the Hamiltonians depend only
on the actions.  To describe this consider
\be
P(\lambda,z)=det(\lambda-L)=\lambda^N+b_1\lambda^{N-1}+\cdots
+b_j\lambda^{N-j}+\cdots + b_N
\label{91}
\ee
where $b_j=\sum Min_j$ where $Min_j\sim$ principal minors of
order $j$, and $b_1=Tr(L)$ with $b_N=det(L)$.
The spectral curve $C\subset T^*\Sigma$ is defined via ${\bf (6T)}\,\,
C=\{P(\lambda,z)=0\}$ and this is well defined since the $b_j$ are
gauge invariant.  Since $L\in H^0(\Sigma,End(V)\otimes K)$ the 
coefficients $b_j\in H^0(\Sigma_g,K^j)$ and one has a map
${\bf (6U)}\,\,p:\,T^*{\cal R}_{g,N}\to B=\oplus_1^NH^0(\Sigma,
K^j)$.  The space $B$ can be considered as the moduli space of the
family of spectral curves parametrized by the Hamiltonians 
$H_{s,k}$; the fibers of $p$ are Lagrangian subvarieties of 
$T^*{\cal R}_{g,N}$ and the spectral curve $C$ is the $N$-fold
covering of the base curve $\Sigma$:  $\pi:\,\,C\to\Sigma$.
One can say that the genus
of $C$ is the dimension $\tilde{g}$ of ${\cal R}_{g,N}$
(recall $\tilde{g}=N^2(g-1)+1$).
There is a line bundle ${\cal L}$ with an eigenspace of $L(z)$
corresponding to the eigenvalue $\lambda$ as a fiber over a generic
point $(\lambda,z)$; thus ${\bf (6X)}\,\,{\cal L}\subset ker(\lambda+L)
\subset \pi^*(V)$.  It defines a point of the Jacobian $Jac(C)$, the
Liouville variety of dimension $\tilde{g}=g(C)$.
Conversely if
$z\in \Sigma_g$ is not a branch point one can reconstruct $V$ for
a given line bundle on $C$ as ${\bf (6Y)}\,\,V_z=\oplus_
{v\in\pi^{-1}(z)}{\cal L}_v$.
Let now $\omega_j$ where $j=1,\cdots,\tilde{g}$,
be the canonical holomorphic differentials on $C$ such that for the
cycles $\alpha_1,\cdots,\alpha_{\tilde{g}},\beta_1,\cdots,\beta_{\tilde{g}}$
with $\alpha_i\cdot\alpha_j=\beta_i\cdot\beta_j=0$ one has
$\oint_{\alpha_i}\omega_j=\delta_{ij}$.  Then the symplectic form
{\bf (6P)} can be written in the form
\be
\omega^0=\int_{\Sigma}<\delta L,\delta\bar{L}>=\sum_1^N
\delta\lambda_j\delta\xi_j
\label{96}
\ee
Here $\xi_j$ are diagonal elements of ${\cal S}\bar{L}{\cal S}$ 
where ${\cal S}L{\cal S}^{-1}=diag(\lambda_1,\cdots,\lambda_N)$.
There results ${\bf (6Z)}\,\,\omega^0=\int_C\delta\lambda\delta\xi$.
Since $\lambda$ is a holomorphic 1-form on $C$ it can be decomposed
as $\lambda=\sum_1^{\tilde{g}}a_j\omega_j$ and consequently
${\bf (I)}\,\,
\omega^0=\sum_1^{\tilde{g}}\delta a_j\int\omega_j\delta\xi$.  The
action variables can be identified with ${\bf (II)}\,\,a_j=\oint_{\alpha_j}
\lambda$.  To define the angle variables one puts locally
$\xi=\bar{\partial}log(\psi)$; then if $(p_m)$ is a divisor of $\psi$
\be
\int_C\omega_j\delta\xi=\sum_m\int_{p_0}^{p_m}\omega_j
log(\psi)=\delta\phi_j
\label{97}
\ee
The $\phi_j$ are linear coordinates on $Jac(C)$ and ${\bf (III)}
\,\,\omega^0=\sum_1^{\tilde{g}}\delta a_j\delta\phi_j$.
\\[3mm]\indent
{\bf REMARK 6.2.}$\,\,$  For completeness let us give here some 
comments on the above following \cite{de,dy,dx}.  As before
take ${\cal R}$ for the moduli space of stable holomorphic
$GL(N,{\bf C})$ vector bundles $V$ over $\Sigma=\Sigma_g$ with
$dim({\cal R})=\tilde{g}=N^2(g-1)+1$ (note $rank(V)=N$ and
write $d=deg(V)$.  The
cotangent space to ${\cal R}$ at a point $V$ is $T^*_V{\cal R}=H^0
(End(V)\otimes K)$ (note $H^0(End(V)\otimes K)\sim H^0(\Sigma,
End(V)\otimes K)$) and Hitchin's theory says $T^*{\cal R}$ is an
algebraically completely integrable system (ACIS).  This means there
is a map $h:\,\,T^*{\cal R}\to B$, for a $\tilde{g}$ dimensional vector
space $B$, that is Lagrangian with respect to the natural symplectic
structure on $T^*{\cal R}$ (i.e. the tangent space to a general fiber
$h^{-1}(a)$ for $a\in B$ is a maximal isotropic subspace relative to
the symplectic form).  Then by contraction with the symplectic form
one obtains a trivialization of the tangent bundle $T_{h^{-1}(a)}
\simeq {\cal O}_{h^{-1}(a)}\otimes T^*_aB$.  This gives a family
of Hamiltonian vector fields on $h^{-1}(a)$, parametrized by $T^*_aB$,
and the flows generated by these fields on $h^{-1}(a)$ all commute.
Algebraic complete integrability means in addition that the fibers
$h^{-1}(a)$ are Zariski open subsets of Abelian varieties on which
the flows are linear (i.e. the vector fields are constant).  In a slightly 
more general framework let $K$ be the canonical bundle as before
with total space ${\bf K}=T^*\Sigma_g$ and think of a $K$ valued 
Higgs pair $(V,\phi:\,V\to V\otimes K)$ where $V$ is a VB on 
$\Sigma_g$ and $\phi$ is a $K$ valued endomorphism.  Imposing
a stability condition this leads to moduli spaces ${\cal R}_K$
(resp. ${\cal R}_K^s$) parametrizing equivalence classes of 
semistable bundles (resp. isomorphism classes of stable bundles).
Let $B=B_K$ be the vector space parametrizing polynomial maps
$p_a:\,\,{\bf K}\to {\bf K}^N$ where $p_a=x^N+a_1x^{N-1}+
\cdots + a_N$ with $a_i\in H^0(K^{\otimes i})$ (i.e. $B=B_K=
\oplus_1^NH^0(K^{\otimes i})$).  The assignment $(V,\phi)\to
det(xI-\phi)$ gives a morphism $h_K:\,\,{\cal R}_K\to B_K$ 
(to the coefficients of $det(xI-\phi)$).  Then
the map $h$ is the restriction of $h_K$ to $T^*{\cal R}$ which is 
an open subset of ${\cal R}_K^s$ and $dim(B)=\tilde{g}$ 
(miraculously - see Remark 6.1).  
The spectral curve $\tilde{\Sigma}=\tilde{\Sigma}_a$
defined by $a\in B_K$ is the inverse image in ${\bf K}=T^*\Sigma$
of the zero section of ${\bf K}^{\otimes N}$ under $p_a:\,\,{\bf K}
\to {\bf K}^N$.  It is finite over $\Sigma$ of degree $N$ and for 
$\tilde{\Sigma}_a$ nonsingular the general fiber of $h_K$ is the
Abelian variety $Jac(\tilde{\Sigma})$.
\\[3mm]\indent
{\bf REMARK 6.3.}
$\,\,$  We omit discussion from \cite{hg,hz,hx,hw} since the notation becomes
complicated and refer also to \cite{ae,de,dz,dy,dx,gl,hh,kh,mv,pa,sf,vb}
for more on all of this (\cite{mx} is an excellent reference for symplectic
matters).
\\[3mm]\indent
{\bf REMARK 6.4.}$\,\,$  There are two other versions of the Hitchin
approach related to physics, namely those of \cite{de,da} and 
\cite{fh,gz}.  Both serve as a short cut to some kind of understanding
but both are flawed as to details (cf. \cite{dx,mv} for more detail
but with a lack of concern regarding connection ideas).  Let us look
at \cite{fh,gz} as the most revealing, especially \cite{gz}.  Thus
let $\Sigma$ be a compace RS of genus $g$ and let $G$ be a complex
Lie group which can be assumed simple, connected, and simply
connected.  Let ${\cal A}$ be the space of $\tilde{g}$ valued
$(0,1)$ gauge fields $\bar{A}=A_{\bar{z}}d\bar{z}$ on $\Sigma$
and take for $T^*{\cal A}$ the symplectic manifold of pairs $(\bar{A},
\phi)$ where $\phi=\Phi_zdz
=\phi dz$ is a $\tilde{g}$ valued $(1,0)$ Higgs
field.  The holomorphic symplectic form on $T^*{\cal A}$ is 
${\bf (XXIV)}\,\,\int_{\Sigma}Tr\,\delta\phi\,\delta\bar{A}$
where $Tr$ stands for the Killing form on $\tilde{g}$ 
suitably normalized.  The local gauge transformations $h\in
{\cal G}\equiv\,Map(\Sigma,G)$ act on $T^*{\cal A}$ via
\be
\bar{A}\to\bar{A}^h=h\bar{A}h^{-1}+h\bar{\partial}h^{-1};\,\,\phi\to
\phi^h\equiv h\phi h^{-1}
\label{959}
\ee
and preserve the symplectic form.  The corresponding moment map
$\mu:\,T^*{\cal A}\to\tilde{g}^*\simeq \wedge^2(\Sigma)\otimes
\tilde{g}$ takes the form ${\bf (XXV)}\,\,\mu(\bar{A},\phi)=
\bar{\partial}\phi+\bar{A}\phi+\phi\bar{a}$ (note here that
$\bar{A}d\bar{z}\wedge\phi dz+\phi dz\wedge\bar{A}d\bar{z}=
[\bar{A},\phi]d\bar{z}\wedge dz$ so one has adjoint action).
The symplectic reduction gives the reduced phase space 
${\bf (XXVI)}\,\,{\cal P}=\mu^{-1}(\{0\})/{\cal G}$ with the
symplectic structure induced from that of $T^*{\cal A}$.
As before ${\cal P}$ can be identified with the complex cotangent
bundle $T^*{\cal N}$ to the orbit space ${\cal N}={\cal A}/{\cal G}$
where ${\cal N}$ is the moduli space of holomorphic $G$ bundles on
$\Sigma$ (of course the identification here really should be
restricted to gauge fields $\bar{A}$ leading to stable or perhaps
semi-stable $G$ bundles but \cite{gz} is rather cavalier about
such matters and we are delighted to follow suit).  Now the Hitchin
system will have ${\cal P}$ as its phase space and the Hamiltonians
are obtained via ${\bf (XXVII)}\,\,h_p(\bar{A},\phi)=p(\phi)=
p(\Phi_z)(dz)^{d_p}$ where $p$ is a homogeneous $Ad$ invariant
polynomial on $\tilde{g}$ of degree $d_p$.  Since $h_p$ is
constant on the orbits of ${\cal G}$ it descends to the reduced phase
space ${\bf (XXVIII)}\,\, h_p:\,{\cal P}\to H^0(K^{d_p})$.  Again
$K$ is the canonical bundle of covectors proportional to $dz$ and
$H^0(K^{d_p})$ is the finite dimensional vector space of holomorphic
$d_p$ differentials on $\Sigma$.
The components of $h_p$ Poisson-commute (they Poisson-commute
already as functions on $T^*{\cal A}$ since they depend only on
the ``momenta" $\phi$).  The point of Hitchin's construction is that
by taking a complete system of polynomials $p$ one obtains on
${\cal P}$ a complete system of Hamiltonians in involution.  For the
matrix groups the values of $h_p$ at a point of ${\cal P}$ can be
encoded in the spectral curve ${\cal C}$ defined by ${\bf (XXIX)}\,\,
det(\phi-\xi)=0$ where $\xi\in K$.  This spectral curve of eigenvalues
$\xi$ is a ramified cover of $\Sigma$; the corresponding eigenspaces of
$\phi$ form a holomorphic line bundle over ${\cal C}$ belonging
to a subspace of $Jac({\cal C})$ on which the $h_p$ induce linear
flows.  For example for the quadratic polynomial $p_2=(1/2)Tr$ 
the map $h_{p_2}$ takes values in the space of holomorphic quadratic
differentials $H^0(K^2)$.  This is the space cotangent to the moduli
space ${\cal M}$ of complex curves $\Sigma$.  Variations of the complex
structure of $\Sigma$ are described by Beltrami differentials $\delta 
\mu=\delta\mu_{\bar{z}}^z\partial_zd\bar{z}$ such that $z'=z+\delta z$
with $\partial_{\bar{z}}\delta z=\delta\mu_{\bar{z}}^z$ gives new
complex coordinates (see below for connections to the notation in
(\ref{72})).  The Beltrami differentials $\delta\mu$ may be paired with
holomorphic quadratic differentials 
$\beta\sim \beta\,dz^2$ via ${\bf (XXX)}\,\,(\beta,\delta\mu)=
\int_{\Sigma}\beta\,\delta\mu$.  The differentials $\delta\mu=\bar
{\partial}(\delta\xi)$ for $\delta\xi$ a vector field on $\Sigma$ describe
variations of the complex structure due to diffeomorphisms of $\Sigma$
and they pair to zero with $\beta$.  The quotient space $H^1(K^{-1})$
of differentials $\delta\mu$ modulo $\bar{\partial}(\delta\xi)$ is the
tangent space to the moduli space ${\cal M}$ and $H^0(K^2)$ is its
dual.  The pairing {\bf (XXX)} defines then for each $[\delta\mu]
\in H^1(K^{-1})$ a Hamiltonian ${\bf (XXXI)}\,\,h_{\delta\mu}\equiv
h_{p_2}\delta\mu$ and these commute for different $\delta\mu$.
\\[3mm]\indent
{\bf REMARK 6.5.}
In (\ref{72}) one has $\mu=\mu(z,\bar{z})\partial_z\otimes d\bar{z}$
with $w=z-\epsilon(z,\bar{z}),\,\,\bar{w}=\bar{z}$, and $\mu=\bar
{\partial}\epsilon/(1-\partial\epsilon)$ (which should be $\mu=
-\bar{\partial}\epsilon/(1-\partial\epsilon)=\bar{\partial}w/\partial w$)
while from \cite{gz} the notation is $z'=z+\delta z$ and $\delta\mu
=\delta\mu^z_{\bar{z}}\partial_zd\bar{z}$ with $\bar{\partial}
(\delta z)=\delta\mu^z_{\bar{z}}$.  Thus $\delta z\sim-\epsilon$ and one
surely must think of $\delta\mu^z_{\bar{z}}\sim\mu(z,\bar{z})$ in
which case $-\bar{\partial}\epsilon\sim\mu$ (adequate for small
$\partial\epsilon$).  Thus for small $\partial\epsilon$ the notations
of (\ref{72})$\sim$ \cite{oa} and \cite{gz} can be compared at least.
To compare ideas with the Kodaira-Spencer theory of deformations
(cf. \cite{kz,mu}) one can extract from \cite{ie} (cf. also \cite{ez,
hu,hw}).  Thus the space of infinitesimal deformations of $\Sigma$
is determined by $H^1(\Sigma,\Theta)$ where $\Theta$ is the sheaf
of germs of holomorphic vector fields on $\Sigma$.  This in turn can
be identified with the tangent space $T_0({\cal T}(\Sigma))$ of the
Teichm\"uller space ${\cal T}(\Sigma)$ (or ${\cal T}_g$) at the
base point $(\Sigma,id.)$.  We further recall that the moduli space 
${\cal M}_g$ corresponds to ${\cal T}(\Sigma)/Mod(\Sigma)$ where 
$Mod(\Sigma)$ is the set of homotopy classes of orientation preserving
diffeomorphisms $\Sigma\to\Sigma$ (modular group).  
Recall ${\cal M}_g$ is the set of biholomorphic equivalence classes
of compact RS of genus $g$ with dimension $3g-3$.  Now,
more precisely, going to \cite{ie} for notation etc., for a given RS
$\Sigma$ consider pairs $(R,f)$ with orientation preserving
diffeomorphism $f:\,\,\Sigma\to R$.  Set $(R,f)\equiv (S,g)$ if
$g\circ f^{-1}:\,\,R\to S$ is homotopic to a biholomorphic map 
$h:\,\,R\to S$.  Then write $[(R,f)]$ for the equivalence class and
the set of such $[(R,f)]$ is ${\cal T}(\Sigma)$.  For $[(R,f)]\in
{\cal T}(\Sigma)$ work locally:  $(U,z)\to (V,w)$ with $f(U)\subset
V$, and write $F=w\circ f\circ z^{-1}$ with $\mu_f=F_{\bar{z}}/
F_z=\bar{\partial}F/\partial F$.  This is the Beltrami coefficient and 
one can write $\mu=\bar{\partial f}/\partial f$ in a standard notation.
There are transition functions for coordinate changes leading to an
expression $\mu_f=\mu(d\bar{z}/dz)$ where $\mu_f$ is a $(-1,1)$ form.
In terms of the space $M(\Sigma)$ of Riemannian metrics on
$\Sigma$ one has ${\cal T}(\Sigma)\simeq M(\Sigma)/Diff_0(\Sigma)$
and ${\cal M}_g\simeq M(\Sigma)/Diff_{+}(\Sigma)$ where $Diff_{+}\sim$
orientation preserving diffeomorphisms of $\Sigma$ and $Diff_0\sim$
elements in $Diff_{+}$ homotopic to the identity.  Next one writes
${\cal A}_2(\Sigma)$ for the space of holomorphic quadratic
differentials $\phi=\phi(z)dz^2$ and $\phi\in {\cal A}_2(\Sigma)_1$
if $\|\phi\|_1=2\int_{\Sigma}|\phi|^2dxdy<1$ (cf. below for the factor
of 2).  Then ${\cal T}(\Sigma)$ is homeomorphic to ${\cal A}_2
(\Sigma)$ (and hence to ${\bf R}^{6n-6}$).  One notes again the
natural pairing {\bf (XXX)} of ${\cal A}_2$ with Beltrami forms via
$(\phi,\mu_f)=\int_{\Sigma}\mu\phi\,dzd\bar{z}$ formally (recall
for $z=x+iy$ and $\bar{z}=x-iy$ one has $dx\wedge dy=(i/2)dz\wedge
d\bar{z}$ so strictly $2\int g\,dxdy=i\int g\,dz\wedge d\bar{z}$).
Now it is easy to show that there is an isomorphism
\be
\delta^*:\,\,\frac{H^0(\Sigma,{\cal E}^{0,1}(K^{-1}))}{\bar
{\partial} H^0(\Sigma,{\cal E}^{0,0}(K^{-1}))}\to H^1(\Sigma,
\Theta)
\label{960}
\ee
and $\Theta={\cal O}(K^{-1})$ with ${\cal A}_2(\Sigma)=
H^0(\Sigma,{\cal O}^{1,0}(K))$ (recall $K\sim$ holomorphic
cotangent space of $\Sigma$).  We note that $H^0(\Sigma,{\cal E}^
{0,1}(K^{-1}))\sim$ Beltrami differentials $\{\mu_j(d\bar{z}_j/dz_j\}$
while $v\in H^0(\Sigma,{\cal E}^{0,0}(K^{-1}))$ corresponds
to a $C^{\infty}$ vector field $v\sim\{v_j(\partial/\partial z_j\}$ so
$\bar{\partial}v=\{\mu_j(d\bar{z}_j/dz_j\}$ where $\mu_j=\partial
v_j/\partial\bar{z}_j$.  Consequently there is a canonical isomorphism
${\bf (XXXII)}\,\,\Lambda\circ (\delta^*)^{-1}:\,\,H^1(\Sigma,\Theta)
\to {\cal A}_2(\Sigma)$ where $\Lambda[\mu](\phi)=\int_{\Sigma}
\mu\phi\,dxdy$ for $\mu\in H^0(\Sigma,{\cal E}^{0,1}(K^{-1}))$ and
$\phi\in {\cal A}_2(\Sigma)$.  This gives the isomorphism $H^1
(\Sigma,\Theta)\simeq T_0({\cal T}(\Sigma))$ mentioned above.
These facts will help clarify some remarks already made above 
(note $H^1(K^{-1})\sim H^1(\Sigma,\Theta)$ and
its dual $H^0(K^2)\sim {\cal A}_2$).
\\[3mm]\indent
We return now to \cite{oa} again and
consider HID in the (scaling) limit $\kappa\to 0$ (critical
value).  One can prove that on the critical level HID coincides with the
part of HH relating to the quadratic Hamiltonians in (\ref{87}).
Note first that in this limit the $A$ connection is transformed into
the Higgs field ($A\to \phi$ as $\kappa\to 0$ - cf. (\ref{76}) and recall
$L=f^{-1}\phi f$)
and therefore
$\widetilde{FB}(\Sigma,G)\to T^*{\cal R}_{g,N}$
(perhaps modulo questions of stability). 
But the form
$\omega$ on the extended phase space ${\cal P}$ appears to be
singular (cf. {\bf (6G)} and (\ref{79}))
and to get around this one rescales the times ${\bf (IV)}\,\,
t=T+\kappa t^H$ where $t^H$ are the fast (Hitchin) times and 
$T$ the slow times.  Assume that only the fast times are dynamical,
which means ${\bf (V)}\,\,\delta\mu(t)=\kappa\sum_s\mu_s^0\delta
t_s^H$ where one writes $\mu_s^0=\bar{\partial}n_s$.  After this
rescaling the forms {\bf (6G)} and (\ref{79}) become regular.
The rescaling procedure means that we blow up a vicinity of the
fixed point $\mu^0_s$ in ${\cal M}_{g,n}$ and the whole dynamics
is developed in this vicinity.  The fixed point is defined by the complex
coordinates ${\bf (VI)}\,\,w_0=z-\sum_sT_s\epsilon_s(z,\bar{z})$
with $\bar{w}_0=\bar{z}$.  Now compare the BA function
$\psi$ of HID in (\ref{77}) with the BA function $Y$ of HH in (\ref{89}).
Using a WKB approximation we assume 
${\bf (VII)}\,\,\psi=\Phi exp[(S^0/\kappa)+S^1]$ where $\Phi$ is a
group valued function and $S^0,\,S^1$ are diagonal matrices.
Put this in the linear system (\ref{77}) and if ${\bf (VIII)}\,\,\partial
S^0/\partial\bar{w}_0=0=\partial S^0/\partial t^H_s$ then there are
no terms of order $\kappa^{-1}$.  It follows from the definition of
the fixed point in the moduli of complex structures {\bf (VI)} that
a slow time dependent $S_0$ emerges in the form
\be
S^0=S^0\left(T_1,\cdots,T_{\ell}|
z-\sum_sT_s\epsilon_s(z,\bar{z})\right)
\label{98}
\ee
In the quasiclassical limit put ${\bf (IX)}\,\,\partial S^0=\lambda$.
so in the zero order approximation we arrive at the linear system of HH
(namely (\ref{77})) defined by the Hamiltonians $H_{s,k}\,\,(k=1,2)$
and the BA function $Y$ takes the form ${\bf (X)}\,\,Y=\Phi exp
[\sum_st^H_s(\partial S^0/\partial T_s)]$.  The goal now is the inverse
problem, i.e. to construct the dependence on the slow times $T$
starting from solutions of HH.  Since $T$ is a vector in the tangent
space to the moduli space
of curves ${\cal M}_g$ it defines a deformation
of the spectral curve in the space $B$ (cf. {\bf (6U)}).  Solutions $Y$
of the linear system (\ref{89}) take the form $Y=\Phi 
exp[\sum_st^H_s\Omega_s]$
where the $\Omega_s$ are diagonal matrices.  Their entries are 
primitive functions of meromorphic differentials with singularities
matching the corresponding poles of $L$.  Then in accord with 
{\bf (X)} we can assume that ${\bf (XI)}\,\,\partial dS/\partial 
T_s=d\Omega_s$ so the $T_s$ correspond to Whitham times
(along with the $t_s$ from (\ref{84}), (\ref{85}), and (\ref{86}) - note
$t_s=T_s+\kappa t^H_s$ so $t_s$ and $T_s$ are comparable in this spirit).  
These equations define the approximation to the phase of 
$\psi$ in the linear problems (\ref{77}) of HID along with
${\bf (XII)}\,\,\partial dS/\partial a_j=d\omega_j$.  The differential
$dS$ plays the role of the SW differential and an important point
here is that only a portion of the spectral moduli, namely those
connected with $H_{s,k}$ for $k=1,2$, are deformed.  As a result
there is no matching between the action parameters of the spectral
curve $a_j\,\,(j=1,\cdots,\tilde{g})$ in {\bf (II)} and deformed
Hamiltonians (cf. \cite{ti} and Section 3).
\\[3mm]\indent
Next the KZB equations (Knizhnik, Zamolodchikov, and 
Bernard) are the system of differential equations having the form
of non-stationary Schr\"odinger equations with the times coming
from ${\cal M}_{g,n}$ (cf. \cite{id}).  They arise in the geometric
quantization of the moduli of flat bundles $\widetilde{FB}(\Sigma,
G)$.  Thus let $V=V_1\otimes\cdots\otimes V_n$ be associated 
with the marked points and the Hilbert space of the quantum system
is a space of sections of the bundle ${\cal E}_{V,\kappa^{quant}}
(\Sigma_{g,n})$ with fibers $\widetilde{FB}(\Sigma,G)$ depending on a
number $\kappa^{quant}$; it is the space of conformal
blocks of the WZW theory on $\Sigma_{g,n}$.  The Hitchin systems
are the classical limit of the KZB equations on the critical level
where classical limit means that one replaces operators by their symbols
and generators of finite dimensional representations in the vertex 
operators acting in the spaces $V_j$ by the corresponding elements
of coadjoint orbits.  To pass to the classical limit in the KZB
equations ${\bf (XIII)}\,\,(\kappa^{quant}\partial_s+\hat{H}_s)F=0$
(note these are kind of flat connection equations)
one replaces the conformal block by its quasi-classical expression
${\bf (XIV)}\,\,F=exp({\cal F}/\hbar)$ where 
$\hbar=(\kappa^{quant})^{-1}$ (where $\kappa=\kappa^{quant}/\hbar$ 
as in \cite{la}) and consider the
classical limit $\kappa^{quant}\to 0$ which leads to HID as in
(\ref{86}) so {\bf (XIII)} is a quantum counterpart of the Whitham
equations).  One
assumes that the limiting values involving 
Casimirs $C_a^i\,\,(i=1,\cdots,rank(G))$
and $a=1,\cdots,n$), corresponding to the irreducible representations
defining the vertex operators, remain finite and this allows one to
fix the coadjoint orbits at the marked points.  In this classical limit
{\bf (XIII)} then is transformed into the HJ equation for the action
${\cal F}=log(\tau)$ of HID, namely (\ref{86})
(note $\kappa^{quant}\partial_sF\to\kappa\partial_s{\cal F}$ etc.).
The integral
representations of conformal blocks are known for WZW theories
over rational and elliptic curves (cf. \cite{ib,fh,fi})
so {\bf (XIV)} then determines
the prepotential ${\cal F}$ of HID.  The KZB operators {\bf (XIII)}
play the role of flat connections in the bundle ${\cal P}^{quant}$
over ${\cal M}_{g,n}$ with fibers ${\cal E}_{V,\kappa^{quant}}
(\Sigma_{g,n})$ (cf. \cite{fd,hz}) with
\be
[\kappa^{quant}\partial_s+\hat{H}_s,\kappa^{quant}\partial_r
+\hat{H}_r]=0
\label{99}
\ee
In fact these equations are the quantum counterpart of the Whitham
hierarchy (\ref{84}).

\section{WHITHAM AND SEIBERG-WITTEN}
\renewcommand{\theequation}{7.\arabic{equation}}
\setcounter{equation}{0}

We mention first that a general abstract theory of Whitham equations
has been developed in \cite{dc,hb,ha,ka,kc,ko} and this was
summarized in part and reviewed in \cite{cc,cb,ca}.  One deals with
algebraic curves $\Sigma_{g,N}$ having punctures at points
$P_{\alpha}\,\,(\alpha=1,\cdots,N)$ and a collection of Whitham times
$T_A$ and corresponding differentials $d\Omega_A$ is constructed.
The Whitham equations ${\bf (XXXVIII)}\,\,\partial_Ad\Omega_B=
\partial_Bd\Omega_A$ persist as in (\ref{BM}).  The construction
of differentials becomes complicated however and we will not deal 
with this here.  Rather we will follow the approach of \cite{en,ec,ea,
gc,gf,ia,na,tc} (summarized as in \cite{cc,cb,cu,ca}).  The most
revealing and accurate development follows \cite{gf,tc} (as recorded
in \cite{cb,cu}) and we will extract here from \cite{cb}.

\subsection{Background from \cite{gf}}

We take a SW situation following \cite{bz,cc,cb,cu,de,da,ea,gc,gf,hb,
ha,ia,ko,mn,na,sd} 
and recall the SW curves $\Sigma_g\,\,
(g=N-1)$ for a pure $SU(N)$ susy YM theory
\be
det_{N\times N}[L(w)-\lambda]=0;\,\,
P(\lambda)=\Lambda^N\left(w+
\frac{1}{w}\right);
\label{558}
\ee
$$
P(\lambda)=\lambda^N-\sum_2^Nu_k\lambda^{N-k}=\prod_1^N(\lambda-\lambda_j);
\,\,u_k=(-1)^k\sum_{i_1<\cdots<i_k}\lambda_{i_1}\cdots\lambda_{i_k}$$
Here the $u_k$ are Schur polynomials of $h_k=(1/k)\sum_1^N\lambda_i^k$ via
the formula
${\bf (A)}\,\,log(\lambda^{-N}P(\lambda))=-\sum_k(h_k/\lambda^k)$;
there are $g=N-1$ moduli $u_k$ and we refer to \cite{ia} for the Lax
operator $L$.
Thus $u_0=1,\,\,u_1=0,\,\,u_2=h_2,\,\,u_3=h_3,\,\,u_4=
h_4-(1/2)h_2^2,$ etc. ($h_1=0$ for $SU(N)$).
One also has the representation
\be
y^2=P^2(\lambda)-4\Lambda^{2N};\,\,y=\Lambda^N\left(w-\frac{1}{w}\right)
\label{660}
\ee
giving a two fold covering of the punctured Riemann sphere with parameter
$\lambda$.  Such Toda chain curves are characterized by a function
${\bf (B)}\,\,2\Lambda^Nw=(P+y)$.
Note also from (\ref{558}) - (\ref{660}) one obtains 
\be
\delta P+P'\delta\lambda=NP\delta log(\Lambda)+y\frac{\delta w}{w};\,\,
\delta P=-\sum\lambda^{N-k}\delta u_k;\,\,P'=\frac{\partial P}
{\partial\lambda}
\label{661}
\ee 
On a given curve (fixed $u_k$ and $\lambda$)
\be
\frac{dw}{w}=\frac{P'd\lambda}{y};\,\,dS_{SW}=\lambda\frac{dw}{w}=
\frac{\lambda dP}{y}
\label{662}
\ee
\be
\left.\frac{\partial dS_{SW}}{\partial u_k}\right|_{w=c}=\frac
{\lambda^{N-k}}{P'}\frac{dw}{w}=\frac{\lambda^{N-k}d\lambda}{y}=dv^k;
\,\,k=2,\cdots, N
\label{663}
\ee
where the $dv^k$ are $g=N-1$ holomorphic one forms with 
\be
a_i=\oint_{A_i}dS_{SW};\,\,
\sigma^{ik}=\oint_{A_i}dv^k=\frac{\partial a_i}{\partial u_k}
\label{664}
\ee
and $d\omega_i=(\sigma_{ik})^{-1}dv^k$ are the canonical holomorphic
differentials with ${\bf (C)}\,\,
\oint_{A_i}d\omega_j=\delta_{ij};\,\,\oint_{B_i}d\omega_j=B_{ij}$.
Note in (\ref{663}) it is necessary to assume $w$ is constant when the
moduli $u_k$ are varied in order to have $\partial dS_{SW}/\partial u_k=dv^k$ 
holomorphic.  
Now the periods $a_i=\oint_{A_i}dS_{SW}$ define the $a_i$ as 
functions of $u_k$ (i.e. $h_k$) and $\Lambda$, or inversely, the $u_k$
as functions of $a_i$ and $\Lambda$.
One proves for example (see (\ref{669}) below)
\be
\frac{\partial u_k}{\partial\,log(\Lambda)}=
ku_k-a_i\frac{\partial u_k}{\partial 
a_i}
\label{770}
\ee
and generally $\Lambda$ and $T_1$ can be identified after suitable
scaling (cf. \cite{bw,bz,cb,ea,gf}).  This is in keeping with the
idea in \cite{ea} that Whitham times are used to restore the homogeneity
of the prepotential when it is disturbed by renormalization (cf. also
\cite{cb}).
\\[3mm]\indent
For the prepotential
one goes to \cite{ia,na} for example and
defines differentials ${\bf (D)}\,\,
d\Omega_n\sim (\xi^{-n-1}+O(1))d\xi$ for $n\geq 1$ with
$\oint_{A_i}d\Omega_n=0$ (pick one puncture momentarily).  
This leads to the generating function of differentials
\be
W(\xi,\zeta)=\sum_1^{\infty}n\zeta^{n-1}d\zeta d\Omega_n(\xi)=\partial_{\xi}
\partial_{\zeta}E(\xi,\zeta)
\label{771}
\ee
where $E$ is the prime form (cf. \cite{fz}) and one has
\be
W(\xi,\zeta)\sim\frac{d\xi d\zeta}{(\xi-\zeta)^2}+O(1)=\sum_1^{\infty}
n\frac{d\xi}{\xi^{n+1}}\zeta^{n-1}d\zeta +O(1)
\label{772}
\ee
(cf. here \cite{cb,cf} for comparison to the kernel $K(\mu,\lambda)=1/(P(\mu)-
P(\lambda))$ of Section 2.4).
One can also
impose a condition of the form
${\bf (E)}\,\,\partial d\hat{\Omega}/\partial\,(moduli)=
holomorphic$ on differentials $d\hat{\Omega}_n$
and use a generating functional (note we distinguish $dS$ and
$dS_{SW}$)
\be
dS=\sum_1^{\infty}T_nd\hat{\Omega}_n=\sum_1^g\alpha_id\omega_i+\sum_0^
{\infty}T_nd\Omega_n
\label{773}
\ee
(we have added a $T_0d\Omega_0$ term here even though $T_0=0$ in the
pure $SU(N)$ theory - cf. \cite{ea,oe}).  Note that there
is a possible confusion in notation with $d\Omega_n$ since we will
choose the $d\hat{\Omega}_n$ below to have poles at $\infty_{\pm}$
and the poles must balance in (\ref{773}).  The matter is clarified by noting
that {\bf (D)} determines singularities at $\xi=0$ and for our curve
$\xi=0\sim \infty_{\pm}$ via $\xi=w^{\mp 1/N}$ which in turn corresponds
to $P(\lambda)^{1/N}$ for $\Lambda =1$
(see (\ref{883}), (\ref{995}), and remarks before
(\ref{880}) - cf. also (\ref{996}) for $d\Omega_n^{\pm}$).  Thus in a certain
sense $d\Omega_n$ here must correspond to $d\Omega_n^{+}+d\Omega_n^{-}$ 
in a hyperelliptic parametrization (cf. (\ref{996})) and
$T_n\sim T_n^{+}=T_n^{-}$ (after adjustment for the 
singular coefficient at
$\infty_{\pm}$).
The periods $\alpha_i=\oint_{A_i}dS$ can be considered as coordinates
on the moduli space (note these are not the $\alpha_i$ of \cite{cc,na}).  
They are not just the same as the $a_i$ but are
defined as functions of $h_k$ and $T_n$ (or alternatively $h_k$ can
be defined as functions of $\alpha_i$ and $T_n$ so that derivatives
$\partial h_k/\partial T_n$ for example are nontrivial.  One will consider
the variables $\alpha_i$ and $T_n=
-(1/n)Res_{\xi=0}\xi^ndS(\xi)$ as independent
so that 
\be
\frac{\partial dS}{\partial \alpha_i}=d\omega_i;\,\,\frac{\partial dS}
{\partial T_n}=d\Omega_n
\label{774}
\ee
Next one can introduce the prepotential $F(\alpha_i,T_n)$ via an analogue
of $a_i^D=\partial{\cal F}/\partial a_i$, namely
\be
\frac{\partial F}{\partial \alpha_i}=\oint_{B_i}dS;\,\,\frac{\partial F}
{\partial T_n}=\frac{1}{2\pi in}Res_0\xi^{-n}dS
\label{775}
\ee
Then one notes the formulas
\be
\frac{\partial^2 F}{\partial T_m\partial T_n}=\frac{1}{2\pi in}
Res_0\xi^{-n}\frac{\partial dS}{\partial T_m}=\frac{1}{2\pi in}
Res_0\xi^{-n}d\Omega_m=\frac{1}{2\pi im}Res_0\xi^{-m}d\Omega_n
\label{776}
\ee
(the choice of $\xi$ is restricted to $w^{\pm 1/N}$ in the situation of
(\ref{660}))
and factors
like $n^{-1}$ arise since $\xi^{-n-1}d\xi=-d(\xi^{-n}/n)$.  Below
we use also a slightly different normalization $d\Omega_n\sim
\pm w^{\pm n/N}(dw/w)=(N/n)dw^{\pm n/N}$ near $\infty_{\pm}$
so that residues in (\ref{775}) and (\ref{776})
will be multiplied by $N/n$ instead of $1/n$.  
Note also that in accord with the remarks above
$Res_{\xi=0}$ will correspond to the sum of residues at 
$\infty_{\pm}$ involving $\xi=w^{\mp 1/N}$ which in turn corresponds to
$P(\lambda)^{1/N}$. 
By definition
$F$ is a homogeneous function of $\alpha_i$ and $T_n$ of degree two, so that
\be
2F=\alpha_i\frac{\partial F}{\partial \alpha_i}+T_n\frac{\partial F}
{\partial T_n}=\alpha_i\alpha_j\frac{\partial^2F}{\partial\alpha_i\partial
\alpha_j}+2\alpha_iT_n\frac{\partial^2F}{\partial\alpha_i\partial T_n}
+T_nT_m\frac{\partial^2F}{\partial T_n\partial T_m}
\label{778}
\ee
Note however 
that $F$ is not just a quadratic function of $\alpha_i$ and $T_n$; a
nontrivial dependence on these variables arises through the dependence
of $d\omega_i$ and $d\Omega_n$ on the moduli (such as $u_k$ or $h_k$)
which in turn depend on $\alpha_i$ and $T_n$.  The dependence is described
by a version of Whitham equations, obtained for example by substituting
(\ref{773}) into (\ref{774}).  Thus
\be
\frac{\partial dS}{\partial T_n}=
d\hat{\Omega}_n+T_m\frac{\partial d\hat{\Omega}_m}{\partial u_{\ell}}
\frac{\partial u_{\ell}}{\partial T_n}=d\Omega_n\Rightarrow
\left(\sum_{m,\ell}T_m\frac{\partial u_{\ell}}{\partial T_n}
\oint_{A_i}\frac{\partial d\hat{\Omega}_m}{\partial u_{\ell}}
\right)=
-\oint_{A_i}d\hat{\Omega}_n
\label{779}
\ee
(since $\oint_{A_I}d\Omega_n=0$). 
Now (cf. \cite{cc,ia}) to achieve {\bf (E)} one can specify
$\partial d\hat{\Omega}_m/\partial u_{\ell}=\sum\beta^m_{\ell j}
d\omega_j$ so $\oint_{A_i}(\partial d\hat{\Omega}_m/\partial u_{\ell})=
\oint_{A_i}\sum\beta^m_{\ell j}d\omega_j=\beta^m_{\ell i}$ and since
$d\hat{\Omega}_n=d\Omega_n+\sum c_j^nd\omega_j$ of necessity
there results ${\bf (F)}\,\,\sum (\partial u_{\ell}/\partial T_n)\sum T_m
\beta^m_{\ell i}=\sum (\partial u_{\ell}/\partial T_n)\sigma_{\ell i}
=-c^n_i$ 
which furnishes
the Whitham dynamics for $u_p$ in the form $\partial u_p/\partial T_n=
-\sum c^n_i\sigma^{ip}$ (cf. \cite{cc,ia} - the formulas in \cite{gf}
and copied in an earlier version of \cite{cb} were too rushed).
\\[3mm]\indent
Now the SW spectral curves (\ref{660}) are related to Toda hierarchies
with two punctures.  We recall (\ref{558}) - (\ref{660}) and note
therefrom that ${\bf (G)}\,\,w^{\pm 1}=(1/2\Lambda^N)(P\pm y)
\sim (1/\Lambda^N)
P(\lambda)(1+O(\lambda^{-2N}))$ near $\lambda=\infty_{\pm}$ since
from $y^2=P^2-4\Lambda^{2N}$ we have $(y/P)=[1-(4\Lambda^{2N}/P^2)]^{1/2}
=(1+O(\lambda^{-2N}))$ so $w^{\pm 1}=(P/2\Lambda^N)[1\pm (y/P)]=
(P/2\Lambda^N)[2+O(\lambda^{-2N})]$.
One writes
$w(\lambda=\infty_{+})=\infty$ and $w(\lambda=\infty_{-})=0$ with
$\xi\sim w^{\mp1/N}$ (i.e. $\xi=w^{-1/N}\sim\lambda^{-1}$ at $\infty_{+}$ and
$\xi=w^{1/N}\sim \lambda^{-1}$ at $\infty_{-}$).  
Near $\lambda=\pm\infty$ by {\bf (G)} one can write then $w^{\pm 1/N}\sim
P(\lambda)^{1/N}$ (for $\Lambda=1$)
in calculations involving $w^{\pm n/N}$ with $n<2N$.
The $w$ parametrization
is of course not hyperelliptic and we note that {\bf (D)} applies for
$\xi=w^{\pm 1/N}$ with all $d\Omega_n$ and later only for $\xi=\lambda^{-1}$ 
in certain differentials $d\tilde{\Omega}_n$.
It would now be possible to envision
differentials
$d\Omega_n^{\pm}$ and $d\hat{\Omega}_n^{\pm}$ but
$d\hat{\Omega}_n=d\hat{\Omega}_n^{+}+d\hat{\Omega}_n^{-}$ is then clearly
the only admissible object (i.e. $d\hat{\Omega}_n$ must have poles at
both punctures); this is suggested
by the form $dw/w$ in (\ref{662}) and  
the
coefficients of $w^{n/N}$ at $\infty_{+}$ and of $w^{-n/N}$ at $\infty_{-}$
must be equal (see also remarks above about $d\Omega_n$ in (\ref{773})).
This corresponds to the Toda chain situation with the same
dependence on plus and minus times.  Moreover 
one takes differentials $d\hat{\Omega}_n$ for (\ref{660})
($\Lambda=1$ here for awhile to simplify formulas)
\be
d\hat{\Omega}_n=R_n(\lambda)\frac{dw}{w}=P_{+}^{n/N}(\lambda)\frac{dw}{w}
\label{880}
\ee
These differentials satisfy {\bf (E)} provided the moduli derivatives
are taken at constant $w$ (not $\lambda$) and we can use the formalism
developed above for $\xi=w^{\mp 1/N}$.  
Note the poles of $d\hat{\Omega}_n$ balance those of $d\Omega_n
=d\hat{\Omega}_n-(\sum_1^g(\oint_{A_j}d\hat{\Omega}_n)
d\omega_j$ and we have
\be
dS=\sum T_nd\hat{\Omega}_n=\sum T_nd\Omega_n +\sum_1^g
\left(\sum_nT_n\oint_{A_j}d\hat{\Omega}_n\right)d\omega_j
\label{6633}
\ee
where we keep $n<2N$ for technical reasons (cf. \cite{tc}).
The SW differential $dS_{SW}$ is
then simply $dS_{SW}=d\hat{\Omega}_1$, i.e.
\be
\left. dS\right|_{T_n=\delta_{n,1}}=dS_{SW};\,\,\left.\alpha_i\right|_{T_n=
\delta_{n,1}}=a_i;\,\,\left.\alpha_i^D\right|_{T_n=\delta_{n,1}}=a_i^D
\label{882}
\ee
With this preparation we can now write
for $n<2N$, where $w^{\pm 1/N}\sim P(\lambda)^{1/N}$,
with $d\hat{\Omega}_n$ as indicated
in (\ref{880}),
\be
\frac{\partial F}{\partial T_n}=\frac{N}{2\pi in}\left(Res_{\infty_{+}}
w^{n/N}dS+Res_{\infty_{-}}w^{-n/N}dS\right)=
\label{883}
\ee
$$=\frac{N}{2\pi in}\left(Res_{\infty_{+}}w^{n/N}+
Res_{\infty_{-}}w^{-n/N}\right)\left(\sum_m
T_mP_{+}^{m/N}(\lambda)\right)\frac{dw}{w}=$$
$$=\frac{N^2}{i\pi n^2}\sum_mT_mRes_{\infty}\left(P_{+}^{m/N}(\lambda)dP^
{n/N}(\lambda)\right)=-\frac{N^2}{i\pi n^2}\sum_mT_mRes_{\infty}
\left(P^{n/N}(\lambda)dP_{+}^{m/N}(\lambda)\right)$$
One can introduce Hamiltonians here of great importance in the general
theory (cf. \cite{cb,dv,gf,ia,lg}) but we only indicate a few relations
since our present concerns lie elsewhere.  Then
evaluating at $T_n=\delta_{n,1}$ (\ref{883})
becomes
\be
\frac{\partial F}{\partial T_n}=-\frac{N^2}{i\pi n^2}Res_{\infty}
P^{n/N}(\lambda)d\lambda=\frac{N}{i\pi n}{\cal H}_{n+1}
\label{885}
\ee
where 
$${\cal H}_{n+1}=-\frac{N}{n}Res_{\infty}P^{n/N}d\lambda=\sum_{k\geq 1}
\frac{(-1)^{k-1}}{k!}\left(\frac{n}{N}\right)^{k-1}\sum_{i_1+\cdots +i_k=n+1}
h_{i_1}\cdots h_{i_k}=$$
\be
=h_{n+1}-\frac{n}{2N}\sum_{i+j=n+1}h_ih_j +O(h^3)
\label{886}
\ee
This can be rephrased as
\be
\frac{\partial F}{\partial T_n}=\frac{\beta}{2\pi in}
\sum_mmT_m{\cal H}_{m+1,n+1}=
\frac{\beta}{2\pi in}T_1{\cal H}_{n+1}+O(T_2,T_3,\cdots)
\label{887}
\ee
where
\be
{\cal H}_{m+1,n+1}=-\frac{N}{mn}Res_{\infty}\left(P^{n/N}dP_{+}^{m/N}\right)=
-{\cal H}_{n+1,m+1};
\label{991}
\ee
$${\cal H}_{n+1}\equiv {\cal H}_{n+1,2}=-\frac{N}{n}Res_{\infty}
P^{n/N}d\lambda=h_{n+1}+O(h^2)$$
\\[3mm]\indent
Now for the mixed derivatives one writes
\be
\frac{\partial^2F}{\partial \alpha_i\partial T_n}=\oint_{B_i}d\Omega_n=
\frac{1}{2\pi in}Res_0\xi^{-n}d\omega_i=
\label{888}
\ee
$$=\frac{N}{2\pi in}\left(Res_{\infty_{+}}w^{n/N}d\omega_i+Res_{\infty_{-}}
w^{-n/N}d\omega_i\right)=\frac{N}{i\pi n}Res_{\infty}P^{n/N}d\omega_i$$
Next set ${\bf (H)}\,\,P^{n/N}=\sum_{-\infty}^{\infty}p^N_{nk}
\lambda^k$ so that ${\bf (I)}\,\,Res_{\infty}P^{n/N}d\omega_i=\sum_{-\infty}^n
p^N_{nk}Res_{\infty}\lambda^kd\omega_i$.  Then e.g.
$$
d\omega_j(\lambda)=\sigma_{jk}^{-1}dv^k(\lambda)=
\sigma_{jk}^{-1}\frac{\lambda^{N-k}d\lambda}{y(\lambda)}=
\sigma_{jk}^{-1}\frac{\lambda^{N-k}d\lambda}{P(\lambda)}\left(1+O(\lambda^
{-2N})\right)=$$
\be
=-\sigma_{jk}^{-1}\frac{\partial\,log\,P(\lambda)}{\partial u_k}
d\lambda\left(1+O(\lambda^{-2N})\right)
\label{889}
\ee
From {\bf (A)} and $\sigma_{jk}^{-1}=\partial u_k/\partial a_j$ one
obtains then
\be
d\omega_j(\lambda)\left(1+O(\lambda^{-2N})\right)=\sum_{n\geq 2}\sigma_{jk}^
{-1}\frac{\partial h_n}{\partial u_k}\frac{d\lambda}{\lambda^n}=
\sum_{n\geq 1}\frac{\partial h_{n+1}}{\partial a_i}\frac{d\lambda}
{\lambda^{n+1}}
\label{990}
\ee
so for $k <2N,\,\,{\bf (J)}\,\,Res_{\infty}\lambda^kd\omega_i=
\partial h_{k+1}/\partial a_i$.  
Further analysis yields 
${\bf (K)}\,\,Res_{\infty}w^{n/N}d\omega_i
\newline
=Res_{\infty}P^{n/N}
d\omega_i=\partial {\cal H}_{n+1}/\partial a_i$ leading to
\be
\frac{\partial^2F}{\partial \alpha_i\partial T_n}=\frac{N}{i\pi n}Res_
{\infty}P(\lambda)^{n/N}d\omega_i=\frac{N}{i\pi n}\frac{\partial
{\cal H}_{n+1}}{\partial \alpha_i}
\label{994}
\ee
For the second $T$ derivatives one uses the general 
formula (\ref{776}) written as
\be
\frac{\partial^2F}{\partial T_n\partial T_m}=\frac{1}{2\pi in}Res_0
\xi^{-n}d\Omega_m=\frac{N}{2\pi in}\left(Res_{\infty_{+}}w^{n/N}d\Omega_m
+Res_{\infty_{-}}w^{-n/N}d\Omega_m\right)
\label{995}
\ee
while for the second $\alpha$ derivatives one has evidently
\be
\frac{\partial^2F}{\partial\alpha_i\partial\alpha_j}=
\oint_{B_i}d\omega_j=B_{ij}
\label{300}
\ee
Note that one can also use differentials $d\tilde{\Omega}$ defined
by {\bf (D)} with $\xi=\lambda^{-1}$ 
(not $\xi=w^{\mp 1/N}$); recall $\infty_{\pm}\sim (\pm,\lambda\to\infty)$
in the hyperelliptic parametrization. 
This
leads to
\be
d\Omega_n^{\pm}\sim\pm\left(w^{\pm n/N}+O(1)\right)\frac{dw}{w}=
\frac{N}{n}dw^{\pm n/N}+\cdots =
\label{996}
\ee
$$=\frac{N}{n}dP^{n/N}+\cdots =\frac{N}{n}\sum_1^nkp^N_{nk}\lambda^{k-1}
d\lambda+\cdots =\frac{N}{n}\sum_1^nkp^N_{nk}d\tilde{\Omega}_k^{\pm}$$
Putting 
{\bf (H)} and (\ref{996}) into (\ref{995}) gives then
\be
\frac{\partial^2F}{\partial T_m\partial T_n}=-\frac{N^2}{i\pi mn}
\sum_1^m\ell p^N_{m\ell}Res_{\infty}w^{n/N}d\tilde{\Omega}_{\ell}
\label{997}
\ee
where $d\tilde{\Omega}_{\ell}=d\tilde{\Omega}_{\ell}^{+}+
d\tilde{\Omega}_{\ell}^{-}$.
\\[3mm]\indent
Further analysis in \cite{gf} involves theta functions and the Szeg\"o
kernel (cf. \cite{fz,gf}).  Thus let $E$ be the 
even theta characteristic associated
with the distinguished separation of ramification points into two equal
sets $P(\lambda)\pm 2\Lambda^N=\prod_1^N(\lambda-r_{\alpha}^{\pm})$.  This
allows one to write the square of the corresponding Szeg\"o kernel as
\be
\Psi_E^2(\lambda,\mu)=\frac{P(\lambda)P(\mu)-4\Lambda^{2N}+y(\lambda)
y(\mu)}{2y(\lambda)y(\mu)}\frac{d\lambda d\mu}{(\lambda-\mu)^2}
\label{999}
\ee
We can write (cf. \cite{gf})
\be
\Psi_E^2(\lambda,\mu)=\sum_{n\geq 1}\hat{\Psi}_E^2(\lambda)\frac{n\lambda^
{n-1}d\mu}{\mu^{n+1}}\left(1+O(P^{-1}(\mu)\right);
\label{1102}
\ee
$$\hat{\Psi}_E^{\pm}(\lambda)\equiv\frac{P\pm y}{2y}d\lambda=\left\{
\begin{array}{cc}
(1+O(\lambda^{-2N}))d\lambda & near\,\,\infty_{\pm}\\
O(\lambda^{-2N}d\lambda & near\,\,\infty_{\mp}
\end{array}\right.$$
and
utilize the formula
\be
\Psi_E(\xi,\zeta)\Psi_{-E}(\xi,\zeta)=W(\xi,\zeta)+d\omega_i(\xi)d\omega_j
(\zeta)\frac{\partial^2}{\partial z_i\partial z_j}log\,\theta_E(\vec{0}|B)
\label{1103}
\ee
(cf. \cite{fz,gf}).  Here one uses (\ref{990}) and (\ref{771}) to get
($1\leq n <2N,\,\,\zeta\sim 1/\mu$)
$$
d\omega_j(\mu)=\sum_{n\geq 1}\frac{nd\mu}{\mu^{n+1}}\left(\frac{1}{n}
\frac{\partial h_{n+1}}{\partial a_j}\right);\,\,d\tilde{\Omega}_n^{\pm}
(\lambda)=\lambda^{n-1}\hat{\Psi}_E^{2\pm}(\lambda)-\rho_n^id\omega_i(\lambda);
$$
\be
d\tilde{\Omega}_n(\lambda)=\lambda^{n-1}(\hat{\Psi}_E^{2+}
(\lambda)+\hat{\Psi}_
E^{2-}(\lambda))-2\rho_n^id\omega_i(\lambda)
\label{1104}
\ee
where ${\bf (L)}\,\,\rho_n^i=(1/n)(\partial h_{n+1}/\partial a_j)
\partial^2_{ij}log\,\theta_E(\vec{0}|B)$.  
From this one can deduce with some calculation
\be
\frac{\partial^2{\cal F}}{\partial T^m\partial T^n}=-\frac{N}{\pi in}
\left({\cal H}_{m+1,n+1}+\frac{2N}{mn}\frac{\partial{\cal H}_{m+1}}
{\partial a_i}\frac{\partial{\cal H}_{n+1}}{\partial a_j}\partial^2_{ij}
log\,\theta_E(\vec{0}|B)\right)
\label{777}
\ee
Next one notes
\be
\sum_k\left(\left.\frac{\partial u_k}{\partial\,log(\Lambda)}\right|_
{a_i=c}\right)\oint_{A_i}\frac{\partial dS_{SW}}{\partial u_k}+\oint_{A_i}
\frac{\partial dS_{SW}}{\partial\,log(\Lambda)}=0
\label{667}
\ee
Then there results
\be
\sum_k\frac{\partial u_k}{\partial\,log(\Lambda)}\frac{\partial a_i}
{\partial u_k}=-\oint_{A_i}\frac{\partial dS_{SW}}{\partial\,log(\Lambda)}=
-N\oint_{A_i}\frac{P}{P'}\frac{dw}{w}=-N\oint_{A_i}\frac{Pd\lambda}{y}=
\label{668}
\ee
$$=-N\oint_{A_i}\frac{P+y}{y}d\lambda=-2N\Lambda^N\oint_{A_i}\frac
{wd\lambda}{y}=-2N\Lambda^N\oint_{A_i}wdv^N$$
Here one is taking $dS_{SW}=\lambda dw/w$ and using (\ref{661}) in the form
($\delta P=\delta w=0$)
$\delta dS_{SW}/\delta\,log(\Lambda)=
\newline
\delta\lambda(dw/w)/\delta\,log)\Lambda)=
(NP/P')(dw/w)$.  Then (\ref{662})
gives $NPd\lambda/y$ and the next step involves $\oint_{A_i}d\lambda=0$.  
Next {\bf (B)} is used along with (\ref{663}).  Note also from
${\bf (M)}\,\,\lambda dP=\lambda[N\lambda^{N-1}-\sum (N-k)u_k\lambda^{N-k-1}]
d\lambda=NPd\lambda+\sum ku_k\lambda^{N-k}d\lambda$ one obtains via
(\ref{662}), (\ref{663}), and (\ref{668}) (middle term)
\be
-\sum\frac{\partial u_k}{\partial\,log(\Lambda)}\frac{\partial a_i}{\partial
u_k}=N\oint_{A_i}\frac{Pd\lambda}{y}=
\label{669}
\ee
$$= \oint_{A_i}\left(\frac{\lambda dP}{y}-\sum ku_k\lambda^{N-k}\frac{d\lambda}
{y}\right)=a_i-\sum ku_k\frac{\partial a_i}{\partial u_k}$$
which evidently implies (\ref{770}).  Now from (\ref{668}) there results
\be
-\frac{\partial u_k}{\partial \,log(\Lambda)}\frac{\partial a_i}{\partial u_k}
=2N\Lambda^N\oint_{A_i}\frac{wd\lambda}{y}=N\oint_{A_i}\frac{P+y}{y}
d\lambda=
\label{1105}
\ee
$$= 2N\oint_{A_i}\hat{\Psi}^2_E(\lambda)=2N\rho_1^i=2N\frac{\partial h_2}
{\partial a_j}\partial^2_{ij}log\,\theta_E(\vec{0}|B)$$
from which
\be
\frac{\partial u_k}{\partial\,log(\Lambda)}=-2N\frac{\partial u_k}{\partial
a_i}\frac{\partial u_2}{\partial a_j}\partial^2_{ij}log\,\theta_E
(\vec{0}|B)
\label{1106}
\ee
Here one can replace $u_k$ by any function of $u_k$ alone such as
$h_k$ or ${\cal H}_{n+1}$ (note $u_2=h_2$).  
Note also (cf. \cite{cb,ea,gf,ha,mn}) that identifying
$\Lambda$ and $T_1$
(after
appropriate rescaling $h_k\to T_i^kh_k$ and ${\cal H}_k\to T_1^k{\cal H}_k$)
one has ($\beta=2N$)
\be
\frac{\partial F_{SW}}{\partial\,log(\Lambda)}=\frac{\beta}{2\pi i}(T_1^2h_2)
\label{1100}
\ee
(this equation for $\partial F/\partial\,log(\Lambda)$ also follows
directly from (\ref{885}) - (\ref{887}) when $T_n=0$ for $n\geq 2$ 
since ${\cal H}_2=h_2$).
\\[3mm]\indent
Finally consider (\ref{661}) in the form ${\bf (N)}\,\,
P'\delta\lambda -\sum_k\lambda^{N-k}\delta u_k=NP\delta\,log(\Lambda)$.
There results for $\delta a_i=0$ (cf. (\ref{668}))
\be
\delta a_i=\oint_{A_i}\delta\lambda\frac{dw}{w}=\sum_k\delta u_k
\oint_{A_i}\frac{\lambda^{N-k}}{P'}\frac{dw}{w}+N\delta\,log(\Lambda)
\oint_{A_i}\frac{P}{P'}\frac{dw}{w};
\label{2200}
\ee
$$\sum_k\oint_{A_i}dv^k\left(\left.\frac{\partial u_k}{\partial\,
log(\Lambda)}\right|_{a=\hat{c}}\right)=-N\oint_{A_i}\frac{P}{P'}\frac{dw}{w}
=-N\oint_{A_i}\frac{Pd\lambda}{y}$$
On the other hand for $\alpha_i=T_1a_i+O(T_2,T_3,\cdots)$
\be
\delta\alpha_i=\alpha_i\delta\,log(T_1)+T_1\oint_{A_i}\delta\lambda
\frac{dw}{w}+O(T_2,T_3,\cdots)
\label{2201}
\ee
so for constant $\Lambda$ with $T_n=0$ for $n\geq 2$
(while $\alpha_i$ and $T_n$ are independent) $\delta\alpha_i=0$ implies
\be
\sum_k\oint_{A_i}dv^k\left(\left.\frac{\partial u_k}{\partial\,log(T_1)}
\right|_{\alpha=c}\right)=-\frac{\alpha_i}{T_1}=
-\oint_{A_i}\frac{\lambda dP}{y}
\label{2202}
\ee
Since $\lambda dP=NPd\lambda+\sum_kku_k\lambda^{N-k}d\lambda$ it follows that
(cf. (\ref{770}))
\be
\left.\frac{\partial u_k}{\partial\,log(T_1)}\right|_{\alpha=c}=
\left.\frac{\partial u_k}{\partial\,log(\Lambda)}\right|_{a=\hat{c}}-ku_k=
-a_i\frac{\partial u_k}{\partial a_i}
\label{2203}
\ee
(cf. (\ref{770}) - note the evaluation points are different and
$\alpha_i=T_1a_i+O(T_2,T_3,\cdots)$).  
This relation is true for any homogeneous algebraic
combination of the $u_k$ (e.g. for $h_k$ and ${\cal H}_k$).
We will return to the $log(\Lambda)$ derivatives later.  
For further relations involving Whitham theory
and $\Lambda$ derivatives see \cite{bw,bz,cb,cu,ea,gf,hb,
ha,ia} and references there.
\\[3mm]\indent
{\bf THEOREM  7.1.}$\,\,$  Given the RS (\ref{558})- (\ref{660})
one can determine $\partial F/\partial T_n$ from (\ref{885}),
$\partial^2F/\partial\alpha_i\partial T_n$ from (\ref{994}),
$\partial^2F/\partial T_n\partial T_m$ from (\ref{995})
or (\ref{777}), and 
$\partial^2F/\partial\alpha_i\partial\alpha_j$ from (\ref{300}) (also
for $T_n=\delta_{n,1},\,\,\alpha_i=a_i$ as in (\ref{882})
with $\Lambda =1$).  Finally
the equation (\ref{1100}) for $\partial F/\partial\,log(\Lambda)$
corresponds to an identification $T_1\sim\Lambda$ and follows from
(\ref{885}) - (\ref{887}) when $T_n=0$ for $n\geq 2$.  Therefore,
since $F\sim F_{SW}$ for $T_1=1$ we see that all derivatives of 
$F_{SW}$ are determined by the RS alone so up to a normalization the 
prepotential is completely determined by the RS.  We emphasize
that $F_{SW}$ involves basically $T_n=\delta_{n,1}$ 
only, with no higher $T_n$, and
$\alpha_i=a_i$, whereas $F$ involving $\alpha_i$ and $T_n$ is defined
for all $T_n$ (cf. here \cite{cu,gf}).
In fact it is really essential to distinguish between $F_{SW}=F^{SW}$ and
general $F=F^W=F_W=F_{Whit}$ and this distinction is developed further
in \cite{cu}.  The identification of $\Lambda$ and $T_1$ is rather cavalier
in \cite{gf} and it would be better not to set $\Lambda=1$ in various
calculations (the rescaling idea is decpetive although correct).  Thus since
$F^{SW}$ arises from $F^W$ by setting $T_n=\delta_{1,n}$ it is impossible
to compare $\Lambda\partial_{\Lambda}F^{SW}$ with
$T_1\partial_1F^W$ directly.  Indeed a statement like equation
(\ref{1100}) is impossible
as such since $T_1$ doesn't appear in $F^{SW}$.
In \cite{cu} we took a simple example (elliptic curve) and computed everything 
explicitly; the correct statement was then shown to be 
$T_1\partial_1F^W=\Lambda
\partial_{\Lambda}F^W$.  In addition it turns out in this simple example that
the Whitham dynamics lead directly to homogeneity equations for various
moduli and the prepotential.

\subsection{Connections to \cite{na}}

The formulation in Section 7.1, based on \cite{gf}, differs from
\cite{ea,ia,na} in certain respects and we want to clarify
the connections here (for \cite{hb,ha,ka,kc,ko} we refer to the original
papers and to \cite{cb,ca}). 
Thus first we sketch very briefly some of the
development in \cite{na} (cf. also \cite{bd,cc}).  
Toda wave functions with a discrete parameter $n$ lead via $T_k=\epsilon
t_k,\,\,\bar{T}_k=\epsilon \bar{t}_k,\,\,T_0=-\epsilon n$, and $a_j
=i\epsilon\theta_j\,\,(=\oint_{A_j}dS)$ to a quasiclassical 
(or averaged) situation where (note $\bar{T}$ does not mean complex
conjugate)
\be
dS=\sum_1^ga_id\omega_i+\sum_{n\geq 0}T_nd\Omega_n+\sum_{n\geq 1}\bar{T}_n
d\bar{\Omega}_n
\label{A1}
\ee
\be
F=\frac{1}{2}\left(\sum_1^ga_j\frac{\partial F}{\partial a_j}+
\sum_{n\geq 0}T_n\frac{\partial F}{\partial T_n}+\sum_{n\geq 1}
\bar{T}_n\frac{\partial F}{\partial \bar{T}_n}\right)
\label{A2}
\ee
where $d\Omega_n\sim d\Omega_n^{+},\,\,d\bar{\Omega}_n\sim d\Omega_n^{-},
\,\,\bar{T}_n\sim T_{-n},$ and near $P_{+}$
\be
d\Omega_n^{+}=\left[-nz^{-n-1}-\sum_1^{\infty}q_{mn}z^{m-1}\right]dz\,\,
(n\geq 1);
\label{A130}
\ee
$$d\Omega^{-}_n=\left[\delta_{n0}z^{-1}-\sum_1^{\infty}r_{mn}z^{m-1}\right]
dz\,\,(n\geq 0)$$
while near $P_{-}$
\be
d\Omega^{+}_n=\left[-\delta_{n0}z^{-1}-\sum_1^{\infty}\bar{r}_{mn}z^{m-1}
\right]dz\,\,(n\geq 0);
\label{A131}
\ee
$$d\Omega^{-}_n=\left[-nz^{-n-1}-\sum_1^{\infty}\bar{q}_{mn}z^{m-1}\right]
dz\,\,(n\geq 1)$$
Here $d\Omega_0$ has simple poles at $P_{\pm}$ with residues $\pm 1$ and
is holomorphic elswhere; further $d\Omega_0^{+}=d\Omega_0^{-}=d\Omega_0$
is stipulated.  In addition the Abelian differentials 
$d\Omega_n^{\pm}$ for $n\geq 0$
are normalized to have zero $A_j$ periods and for the
holomorphic differentials $d\omega_j$ we write at $P_{\pm}$ respectively
\be
d\omega_j=-\sum_{m\geq 1}\sigma_{jm}z^{m-1}dz;\,\,d\omega_j=-\sum_{m\geq 1}
\bar{\sigma}_{jm}z^{m-1}dz
\label{A3}
\ee
where $z$ is a local coordinate at $P_{\pm}$.
Further for the SW
situation where ($g=N-1$)
\be
dS=\frac{\lambda dP}{y}=\frac{\lambda P'd\lambda}{y};\,\,y^2=P^2-\Lambda^{2N};
\,\,P(\lambda)=\lambda^N+\sum_0^{N-2}u_{N-k}\lambda^k
\label{A4}
\ee
(cf. (\ref{558}) where the notation is slightly different)
one can write near $P_{\pm}$ respectively
\be
dS=\left(-\sum_{n\geq 1}nT_nz^{-n-1}+T_0z^{-1}-\sum_{n\geq 1}\frac{\partial F}
{\partial T_n}z^{n-1}\right)dz;
\label{A5}
\ee
$$dS=\left(-\sum_{n\geq 1}n\bar{T}_nz^{-n-1}-T_0z^{-1}-\sum_{n\geq 1}
\frac{\partial F}{\partial\bar{T}_n}z^{n-1}\right)dz$$
leading to
\be
F=\frac{1}{2}\left(\sum_1^{N-1}\frac{a_j}{2\pi i}\oint_{B_j}dS-\sum_{n\geq 1}
T_nRes_{+}z^{-n}dS-\right.
\label{A6}
\ee
$$-\left.\sum_{n\geq 1}\bar{T}_nRes_{-}z^{-n}dS-T_0
[Res_{+}log(z)dS-Res_{-}log(z)dS]\right)
$$
(the $2\pi i$ is awkward but let's keep it - note one defines 
$a_j^D=\oint_{B_j}dS$).
In the notation of \cite{na} one can write now 
$(\bullet)\,\,h=y+P,\,\,\tilde{h}
=-y+P,$ and $h\tilde{h}=\Lambda^{2N}$ with $h^{-1}\sim z^N$ at $P_{+}$
and $z^N\sim \tilde{h}^{-1}$ at $P_{-}$  
(evidently $h\sim w$ of Section 2).  Note also 
$(\bullet\bullet)\,\,h+(\Lambda^{2N}/h)
=2P$ and $2y=h-(\Lambda^{2N}/h)$ yielding $y^2=P^2-\Lambda^{2N}$ and
calculations in \cite{na} give $(\bullet\bullet\bullet)\,\,dS=
\lambda dP/y=\lambda dy/P=\lambda dh/h$ (so $h\sim w$ in Section 7.1). 
Further the holomorphic
$d\omega_i$ can be written as linear combinations of holomorphic
differentials ($g=N-1$)
\be
dv_k=\frac{\lambda^{k-1}d\lambda}{y}\,\,(k=1,\cdots,g);\,\,y^2=P^2-1=\prod_1^
{2g+2}(\lambda-\lambda_{\alpha})
\label{A121}
\ee
(cf. (\ref{663}) where the notation
differs slightly).  Note also from (\ref{A121}) that
$
2ydy=\sum_1^{2N}\prod_{\alpha\not=\beta}(\lambda-\lambda_{\alpha})d\lambda$
so $d\lambda=0$ corresponds to $y=0$.
From the theory of \cite{na} (cf. also \cite{cc}) one has then Whitham
equations
\be
\frac{\partial d\omega_j}{\partial a_i}=\frac{\partial d\omega_i}{\partial a_j};
\,\,\frac{\partial d\omega_i}{\partial T_A}=
\frac{\partial d\Omega_A}{\partial a_i};\,\,\frac{\partial d\Omega_B}
{\partial T_A}=\frac{\partial d\Omega_A}{\partial T_B}
\label{A8}
\ee
along with structural equations
\be
\frac{\partial dS}{\partial a_i}=d\omega_i;\,\,\frac{\partial dS}{\partial
T_n}=d\Omega_n^{+};\,\,\frac{\partial dS}{\partial \bar{T}_n}=d\Omega_n^{-};
\,\,\frac{\partial dS}{\partial T_0}=d\Omega_0
\label{A9}
\ee
Finally we note that in \cite{na} one presents a case (cf. also \cite{cb})
for identifying $N=2$ susy Yang-Mills (SYM) with a coupled system of
two topological string models based on the $A_{N-1}$ string.  This seems
to be related to the idea of $t\bar{t}$ fusion (cf. \cite{ds}).

\section{SOFT SUSY BREAKING AND WHITHAM}
\renewcommand{\theequation}
{8.\arabic{equation}}\setcounter{equation}{0}

The idea here is to describe briefly some work of Edelstein, G\'omez-Reino,
Mari\~no, and Mas about the promotion of Whitham times to spurion
superfields and subsequent soft susy breaking ${\cal N}=2\to {\cal N}=0$.

\subsection{Remarks on susy}

We begin by sketching some ideas from \cite{bi,bj} where a nice
discussion of susy and gauge field theory can be found.  In particular
these books are a good source where all of the relevant notation
is exhibited in a coherent manner.  One says that a quantum field
theory (QFT) is renormalizable if it is rendered finite by the 
renormalization of only the parameters and fields appearing in the
bare Lagrangian (for renormalization we refer also e.g. to \cite{bi,bj}
and the bibliography of \cite{cb}).  We denote by $\Lambda$ the 
renormalization scale parameter.  One notes that renormalization of
bare parameters occurs as a quantum effect of interaction and the shifts
thus generated are infinite, which means that the bare parameters were
also infinite, in order to produce a finite measured value.  Further
in order to implement gauge invariance in weak interactions for
example one must find a method of generating gauge vector boson
masses without destroying renormalizability.  Any such mass term 
breaks the gauge symmetry and the only known way of doing this
in a renormalizable manner is called spontaneous symmetry breaking.
This arises when e.g. when there are nonzero ground states or vacua
which are not invariant under the same symmetries as the Lagrangian
or Hamiltonian.  Once such a vacuum is chosen (perhaps 
spontaneously by the system ``settling down") the symmetry is
broken.  One can then define new fields centered around the vacuum
which have zero vev (vacuum expectation value) and the Lagrangian
expressed in the new fields will no longer have the same symmetry
as before.  Such new fields (with nonzero vev) have to be scalar
(not vector or spinor) and are called Higgs fields; this kind of
spontaneous symmetry breaking  is nonperturbative (the vevs are zero
in all orders of perturbation theory).  In the case of continuous global
symmetry in the Lagrangian there can be a subspace of degenerate
ground states and massless modes called Goldstone bosons arise.
In any event the idea now is to break a local gauge invariance
spontaneously in the hope that the break will induce gauge boson
masses while the (now hidden) symmetry will protect renormalizability.
This is referred to as the Higgs mechanism and what happens is e.g.
that the Goldstone bosons are ``eaten" by the gauge transformed
massive boson field and a scalar Higgs field remains (we recall that
massive terms are quadratic in the Lagrangian).  In the case of
nonabelian gauge theories one includes Yukawa couplings
of fermions to the scalar fields in order to have fermion masses emerge
under spontaneous symmetry breaking.  Further magnetic monopoles
may arise in spontaneously broken nonabelian gauge theories and 
instantons arise in general (which are classical gauge configurations
not necessarily related to spontaneously broken symmetry); we omit
any discussion of these here.
\\[3mm]\indent
Now, turning to susy (following \cite{bj}) one introduces a spinor
geometry to supplement the bosonic generators of the Poincar\'e
group.  This leads to a natural description of fermions and in a susy
theory the vanishing of the vacuum energy is a necessary and
sufficient condition for the existence of a unique vacuum.  Further,
every representation has an equal number of equal mass bosonic and
fermionic states.  Generally the nonzero masses of observed particles
are generated by susy breaking effects so one looks first at representations
(and their TCP conjugate representations) of the $N=1$ susy algebra
that can be realized by massless ``one" particle states.  This leads to
supermultiplets for $N=1$ involving ($\lambda\sim$ helicity)
{\bf (A)} chiral:  quarks, leptons, Higgsinos ($\lambda=1/2$) with
squarks, sleptons, Higgs  particles for $\lambda=0$ (scalar
particles) {\bf (B)} vector:  gauge bosons ($\lambda=1$) with
gauginos ($\lambda=1/2$) {\bf (C)} gravity:  graviton ($\lambda=2$)
with gravitino ($\lambda=3/2$) along with TCP conjugate
representations ${\bf (A)'}\,\,\lambda=-1/2$ and $\lambda=0\,\,\,
{\bf (B')}\,\, \lambda=-1$ and 
$\lambda=-1/2\,\,\,{\bf (C')}\,\,\lambda=-2$
and $\lambda=-3/2$.  For $N=2$ susy one has supermultiplets 
involving 4-D real representations of $U(2)$ (in the absence of
central charges) {\bf (D)} vector:  $\lambda=1$, double $\lambda=1/2$,
and $\lambda=0$ {(\bf E)} hypermultiplet:  $\lambda=1/2$,
double $\lambda=0$, and $\lambda=-1/2$ (TCP self-conjugate)
{\bf (F)} gravity:  $\lambda=2,$ double $\lambda=3/2$, and $\lambda
=1$ along with TCP conjugations for {\bf (D)} and {\bf (F)}.  
\\[3mm]\indent
One goes then to superfields $S(x,\theta,\bar{\theta})$ with Grassman
variables $\theta,\,\bar{\theta}$ (for which a nice discussion is given
in \cite{bj}).  There will be expansions
\be
\Phi(x^{\mu},\theta,\bar{\theta})=\phi+\sqrt{2}\theta\psi+\theta
\theta F+i\partial_{\mu}\phi\theta\sigma^{\mu}\bar{\theta}-\frac
{i}{\sqrt{2}}\theta\theta\partial_{\mu}\psi\sigma^{\mu}\bar{\theta}-
\frac{1}{4}\partial_{\mu}\partial^{\mu}\phi\theta\theta\bar{\theta}
\bar{\theta};
\label{B1}
\ee
$$\Phi^{\dagger}=\phi^{\dagger}+\sqrt{2}\bar{\theta}\bar{\psi}+
\bar{\theta}\bar{\theta}F^{\dagger}-i\partial_{\mu}\phi^{\dagger}
\theta\sigma^{\mu}\bar{\theta}+\frac{i}{\sqrt{2}}\bar{\theta}
\bar{\theta}\theta\sigma^{\mu}\partial_{\mu}\bar{\psi}
-\frac{1}{4}\partial_{\mu}\partial^{\mu}\phi^{\dagger}\theta\theta
\bar{\theta}\bar{\theta}$$
for chiral superfields where $\psi\sim$ left handed Weyl spinor,
$\phi,\,F\sim$ complex scalar fields, and $\sigma^0=I_2$ with
$\sigma^i\sim$ Pauli matrices ($i=1,2,3$).  It is useful to note also
that $\delta F$ is a total divergence under susy transformations.
Similarly vector superfields can be written
\be
V(x,\theta,\bar{\theta})=C(x)+i\theta\chi(x)-i\bar{\theta}\bar{\chi}(x)
+\frac{i\theta\theta}{2}\left[M(x)+iN(x)\right]-\frac{i\bar{\theta}
\bar{\theta}}{2}\left[M(x)-iN(x)\right]+
\label{B2}
\ee
$$+\theta\sigma^{\mu}\bar{\theta}V_{\mu}(x)+i\theta\theta\bar
{\theta}\left[\bar{\lambda}(x)+\frac{i}{2}\bar{\sigma}^{\mu}
\partial_{\mu}\chi(x)\right]-i\bar{\theta}\bar{\theta}\theta\left[
\lambda(x)+\frac{i}{2}\sigma^{\mu}\partial_{\mu}\bar{\chi}(x)\right]
+\frac{1}{2}\theta\theta\bar{\theta}\bar{\theta}\left[D-\frac{1}{2}
\partial_{\mu}\partial^{\mu}C\right]$$
and $\delta D$ will be a total divergence (along with $\delta(\partial_
{\mu}\partial^{\mu}C)$).  Here $\chi,\,\lambda$ are Weyl spinor
fields, $V_{\mu}$ is a real vector field, and $C,\,M,\,N,\,D$ are real
scalar fields.  Generally one refers to the coefficient of $\theta\theta$
in product expansions $\Phi_i\Phi_j$ or $\Phi_i\Phi_j\Phi_k$ for 
example as an $F$ term and the coefficient of $\bar{\theta}\bar
{\theta}\theta\theta$ as a $D$ term.  Then susy Lagrangians involving
chiral superfields will have the form
\be
{\cal L}=\sum[\Phi^{\dagger}_i\Phi_i]_D+([W(\Phi)]_F+HC)\sim
\int d^4\theta\sum\Phi_i^{\dagger}\Phi_i+
\left(\int d^2\theta W(\Phi)+HC\right)
\label{B3}
\ee
($HC\sim$ Hermitian conjugate)
where $W$ is called a superpotential and 
involves powers of $\Phi_i$ only up to order three for renormalizability.
The latter equations arises since the superspace integration projects
out $D$ and $F$ terms (recall $\int d\theta\theta=1,\,\,\int d\theta=0,
\,\,\int d^2\theta\theta\theta=1,\,\,(d/d\theta)f(\theta)=\int d\theta
f(\theta)$, 
etc.).  Note that, apart from a possible tadpole
term linear in the $\Phi_i$, one will have
\be
W(\Phi)=\frac{1}{2}m_{ij}\Phi_i\Phi_j+\frac{1}{3}\lambda_{ijk}
\Phi_i\Phi_j\Phi_k
\label{B4}
\ee
There is then a theorem which states that the superpotential for
$N=1$ susy is not renormalizable, except by finite amounts,
in any order of perturbation theory, other than by wave function
renormalization.  Regarding susy breaking one must evidently have
this since we do not see scalar particles accompanied by their
susy associated fermions.  To recognize when susy is spontaneously
broken we need a vacuum $|0>$ which is not invariant under susy
(or alternatively $|0>$ should not be annihilated by all the susy
generators).  As a consequence whenever a susy vacuum exists
as a local minimum of the effective potential it is also a global
minimum.  For the global minimum of the effective potential 
(physical vacuum) to be non susy it is therefore necessary for the 
effective potential to possess no susy minimum.  In theories
of chiral superfields one needs $<0|F_i|0>\not=0$ for spontaneous
susy breaking where the tree level effective potential is $V=F_i^
{\dagger}F_i=|F_i|^2$ and $F_i^{\dagger}=-\partial W(\phi)/
\partial\phi_i$
(here $F_i$ is an $F$ term arising in (\ref{3})).
Once spontaneous susy breaking occurs a massless Goldstone fermion
appears which for $F_i$ will be the spinor $\psi_i$ in the 
supermultiplet to which $F_i$ belongs.  When global susy becomes
local susy in supergravity (sugra) theories the Goldstone fermion
is eaten by the gravitino to give the gravitino a mass.  For theories
involving vector superfields there is also another possibility when there
is a $D$ term with $<0|D(x)|0>=\phi\not=0$.  Finally one notes
that the renormalized coupling constants necessarily depend on the
mass scale $\Lambda$ but the physics described by the bare
Lagrangian is independent of $\Lambda$ so the coupling constants
must ``run" with $\Lambda$.  The RG equations specify how these
coupling constants vary.
\\[3mm]\indent
Now to couple with sugra one recalls first the Noether procedure
for deriving an action with a local symmetry from an action with
a global symmetry.  For example given $S_0=i\int d^4x\bar{\psi}
\gamma^{\mu}\partial_{\mu}\psi$ invariant under the global symmetry
$\psi\to exp(-i\epsilon)\psi$ one lets $\epsilon=\epsilon(x)$ and 
considers $(\clubsuit)\,\,\psi\to exp(-i\epsilon(x))\psi$.  Then
$\delta S_0=\int d^4x\bar{\psi}\gamma^{\mu}\psi\partial_{\mu}
\epsilon=\int d^4xj^{\mu}\partial_{\mu}\epsilon$ where $j_{\mu}=
\bar{\psi}\gamma^{\mu}\psi$ is the Noether current.  To restore
invariance a gauge field $A^{\mu}$ is introduced transforming
under $(\clubsuit)$ as $(\spadesuit)\,\,A_{\mu}\to A_{\mu}
+\partial_{\mu}\epsilon$ and a coupling term is added to $S_0$ to
obtain $S=S_0-\int d^4xj^{\mu}A_{\mu}=\int d^4x i\bar{\psi}\gamma^
{\mu}(\partial_{\mu}+iA_{\mu})\psi$.  This $S$ is invariant under
$(\clubsuit)$ and $(\spadesuit)$.  One can use this technique to 
construct a locally susy action from the global susy action for the sugra
multiplet.  Next one extends the pure sugra Lagrangian of the graviton
and gravitino to include couplings with matter fields.  Recall that
whereas in global susy theories susy breaking manifests itself in the
appearance of a massless Goldstone fermion, in locally susy theories
the corresponding effect is the appearance of a mass for the gravitino
which is the gauge particle of local susy.  The most general global
susy Lagrangian for chiral superfields is 
\be
{\cal L}_{Glob}=\int d^4\theta K(\Phi^{\dagger},\Phi)+\int d^2
\theta (W(\Phi)+HC)
\label{B5}
\ee
where $K$ is a general function (since nonrenormalizable kinetic
terms cannot be excluded in the presence of gravity).  Similarly
the superpotential may contain arbitrary powers of the $\Phi_i$.  The
sugra Lagrangian turns out to depend only on a single function of
$\phi_i^*$ and $\phi_i$, namely $(\bullet)\,\,G(\phi^*,\phi)=J(\phi^*,
\phi)+log\,|W|^2;\,\,J=-3log(-K/3)$, where $G$ (or $J$) is called
the K\"ahler potential.  Note $G$ is invariant under $(\bullet\bullet)
\,\,J\to J+h(\phi)+h^*(\phi^*);\,\,W\to exp(-h)W$.  The sugra
Lagrangian ${\cal L}$ may be written as ${\cal L}={\cal L}_B
+{\cal L}_{FK}+{\cal L}_F$ where ${\cal L}_B$ contains only
bosonic fields, ${\cal L}_{FK}$ contains fermionic fields and 
covariant derivatives (supplying the fermionic kinetic energy terms),
and ${\cal L}_F$ has fermionic fields but no covariant derivatives
(see \cite{bj} for details).  Now regarding spontaneous susy breaking,
for theories with local susy breaking the vacuum energy is no longer 
positive semidefinite.  There are in particular the following possibilities.
First at least one of the fields in the theory must have a vev not
invariant under susy.  Under certain assumptions one has 
then an $F$ term 
generalization
involving $\partial W/\partial\phi+\phi^*W\not=0$ or a $D$ term
generalization
involving $G^i(T_a)_{ij}\phi_j\not=0$ where 
$G^i=(\phi^i)^*+(1/W)\partial W/\partial \phi_i$ (here
$(T_a)_{ij}\sim$ generators of the gauge group in the appropriate
representation).
There are also
other possibilities (cf. \cite{bj}).
\\[3mm]\indent
For string theory both IIA and IIB are unsuitable to describe the real
world but the heterotic string is perhaps tenable, with the extra 16
left mover dimensions providing the 
gauge group for the resulting 10-D theory 
(upon compactifying on a 16-D torus or variations of this).
Further compactification of 6 dimensions is then still necessary leading
at first to an $N=4$ susy theory if toroidal compactification is used
(which is unsuitable since $N\geq 2$ susy models are nonchiral).
However orbifold compactifications will yield a 4-D theory having 
$N=1$ susy and then some symmetry breaking must be induced.
One can also compactify on Calabi-Yau (CY) manifolds and 
modular invariance can be achieved.  We leave \cite{bi,bj} in what
follows in order to concentrate on material more directly related to
the projects at hand.
\\[3mm]\indent
In \cite{lh} one takes now an $N=1$ Yang-Mills (YM) action
\be
\frac{1}{8\pi}\Im\left[\tau\int d^4x\int d^2\theta TrW^{\alpha}W_{\alpha}
\right]=-\frac{\theta_{YM}}{32\pi^2}\int d^4x Tr\,F_{mn}\tilde{F}^
{mn}+
\label{B12}
\ee
$$+\frac{1}{g^2}\int d^4x Tr\left[-\frac{1}{4}F_{mn}F^
{mn}-i\lambda\sigma^m\nabla_m\bar{\lambda}+\frac{1}{2}
D^2\right]$$
where $(\spadesuit\spadesuit)\,\,\tau=(\theta_{YM}/2\pi)+(4\pi i/g^2)$. 
Here the $W_{\alpha}$ are chiral spinor superfields of the form
\be
W_{\alpha}=-i\lambda_{\alpha}(y)+\theta_{\alpha}D(y)-\frac{i}{2}
(\sigma^m\bar{\sigma}^n\theta)_{\alpha}(\partial_mv_n-\partial_n
v_m)(y)+(\theta\theta)\sigma^m_{\alpha\dot{\beta}}\partial_m
\bar{\lambda}^{\dot{\beta}}(y)
\label{B13}
\ee
$$W^{\alpha}=-i\lambda^{\alpha}(y)+\theta^{\alpha}D(y)+\theta^
{\beta}\sigma_{\beta}^{mn\alpha}F_{mn}(y)-(\theta\theta)\bar
{\sigma}^{m\dot{\beta}\alpha}\nabla_m\bar{\lambda}_
{\dot{\beta}}(y)$$
where $\tilde{F}^{mn}=(1/2)\epsilon^{mnpq}F_{pq}$ 
(dual field strength) and 
\be
F_{mn}=\partial_mv_n-\partial_nv_m+i[v_m,v_n];\,\,\nabla_m
\bar{\lambda}^{\dot{\beta}}=\partial_m\bar{\lambda}^
{\dot{\beta}}+i[v_m,\bar{\lambda}^{\dot{\beta}}]
\label{B14}
\ee
We omit the background considerations involving the Wess-Zumino
(WZ) gauge etc. (cf. also \cite{bj}).  For $N=2$ susy one thinks
of $\theta$ and $\tilde{\theta}$ with $D_{\alpha}\to D_{\alpha},
\tilde{D}_{\alpha}$ and $\int d^2\theta\to \int d^2\theta d^2
\tilde{\theta}$, etc.  Then an $N=2$ chiral superfield is an
$N=2$ scalar superfield which is a singlet under global $SU(2)$
and satisfies
\be
\bar{D}_{\dot{\alpha}}\Psi(x,\theta,\bar{\theta},\tilde{\theta},
\bar{\tilde{\theta}})=0;\,\,\bar{\tilde{D}}_{\dot{\alpha}}\Psi=0
\label{B16}
\ee
where e.g.
\be
D_{\alpha}=\partial_{\alpha}+2i\sigma^m_{\alpha\dot{\beta}}\bar
{\theta}^{\dot{\beta}}\partial_m;\,\,\bar{D}_{\dot{\alpha}}=-\bar
{\partial}_{\dot{\alpha}}
\label{B8}
\ee
Set also $(\clubsuit\clubsuit\clubsuit)\,\,\tilde{y}^m=
x^m+i\theta\sigma^m\bar{\theta}+i\tilde{\theta}\sigma^m\bar{\tilde
{\theta}}$.  Then expanding an $N=2$ chiral superfield in powers
of $\tilde{\theta}$ the components are $N=1$ chiral superfields.
Thus
\be
\Psi=\Phi(\tilde{y},\theta)+i\sqrt{2}\tilde{\theta}^{\alpha}
W_{\alpha}(\tilde{y},\theta)+\tilde{\theta}\tilde{\theta}G(\tilde{y},
\theta)
\label{B17}
\ee
($W_{\alpha}$ is an $N=1$ chiral spinor superfield).  For $N=2$
YM, if one forgets about renormalizability, there arises
\be
{\cal L}=
\frac{1}{4\pi}\Im\left[\int d^4x\int d^2\theta d^2\tilde{\theta} Tr
{\cal F}(\Psi)\right]
\label{B18}
\ee
with ${\cal F}=(1/2)\tau\Psi^2$ and constraints on $\Psi$ of the form
$(\spadesuit\spadesuit\spadesuit)\,\,(D^{a\alpha}D^b_{\alpha})\Psi
=(\bar{D}^a_{\dot{\alpha}}\bar{D}^{b\dot{\alpha}})\Psi^{\dagger}$
where $a,b$ are $SU(2)$ indices.
Writing ${\cal F}_a(\Phi)=\partial{\cal F}/\partial\Phi_a$
and ${\cal F}_{ab}=
\partial^2{\cal F}/\partial\Phi_a\partial\Phi_b$ the Lagrangian
(\ref{B18}) can be written in terms of $N=1$ superfields as
\be
{\cal L}=\frac{1}{4\pi}\Im\left[\frac{1}{2}\int d^2\theta
{\cal F}_{ab}(\Phi)W^{\alpha a}W^b_{\alpha}+\int d^4\theta
(\Phi^{\dagger}e^{2V})^a{\cal F}_a(\Phi)\right]
\label{B19}
\ee
where $V$ will be clarified below.

\subsection{Soft susy breaking and spurion fields}

We go now to \cite{en,ec,ed,mo} and one works upon the foundation of
\cite{gf} sketched in Section 7.1.  The idea of soft susy breaking goes back
to \cite{gb} for example and was developed in a form relevant here in
\cite{ag,ah,ai,mh}.  From \cite{en,ec,mo} one extracts the following
philosophical
comments:  Softly broken susy models offer the best phenomenological
candidates to solve the hierarchy problem in grand unified theories.  The 
spurion formalism of \cite{gb} provides a tool to generate soft susy
breaking in a neat and controlled manner (i.e. no uncontrolled divergences
arise).  To illustrate the method, start from a susy Lagrangian
$L(\Phi_0,\Phi_1,
\cdots)$ with some set of chiral superfields, and single out a particular one,
say $\Phi_0$.  If you let this superfield acquire a constant vev along a given
direction in superspace, such as e.g. $<\Phi_0>=c_0+\theta^2F_0$, it will
induce soft susy breaking terms and a vacuum energy of order $|F_0|^2$.
Turning the argument around, you could promote $\underline{any}$ parameter
in your Lagrangian to a chiral superfield, and then freeze it along a susy
breaking direction in superspace giving a vev to its highest component
(the $F_n$ terms below).
In the embedding of the SW solution within the Toda-Whitham framework
we have obtained an analytic dependence of the prepotential on some new
parameters $T_n$ (the Whitham or slow times).  Then these slow times 
can be interpreted as parameters of a non-supersymmetric family of theories
by promoting them to be spurion superfields.  In \cite{ag,ah,ai,mh} this 
program was initiated with the scale parameter $\Lambda$ and the masses
of additional hypermultiplets $m_i$ as the the only sources for spurions.
To deal with the times $T_n$ now one writes
\be
\hat{T}_n=T_nT_1^{-n};\,\,\hat{u}_k=T_1^ku_k;\,\,\alpha_i
(u_k,T_n)=T_1a_i(u_k,\Lambda=1)+O(T_{n>1});
\label{B32}
\ee
$$\hat{a}_i=\alpha_i(u_k,T_1,\hat{T}_{n>1}=0)=T_1a_i(u_k,
\Lambda=1)=a_i(\hat{u}_k,\Lambda=T_1)$$
Then the Whitham times $T_n$ (or $\hat{T}_n$) are promoted to
spurion superfields via
\be
s_1=-ilog(\Lambda);\,\,s_n=-i\hat{T}_n;\,\,{\cal S}_1=s_1+\theta^2F_1;
\label{B33}
\ee
$$V_1=\frac{1}{2}D_1\theta^2\bar{\theta}^2;\,\,{\cal S}_n=s_n+
\theta^2F_n;\,\,V_n=\frac{1}{2}D_n\theta^2\bar{\theta}^2$$
\be
\Lambda=exp(is_1);\,\,s_1=\frac{\pi\tau}{N};\,\,\tau=\frac{\theta}
{2\pi}+\frac{4\pi i}{g^2};\,\,\Lambda^{2N}\sim exp(2\pi i\tau)
\label{B34}
\ee
(note the $\theta$ in $\tau$ has a different meaning from the
Grassman $\theta$ in the superfields).  
One also writes $\hat{{\cal H}}_{m+1,n+1}=T_1^{m+n}{\cal H}_
{m+1,n+1}$ and in the manifold $T_{n>1}=0$ the $\hat{T}_n$
are dual to the $\hat{{\cal H}}_{n+1}$ via 
$$ {\bf (DUAL)}\,\,\,\frac{\partial {\cal F}}
{\partial log(\Lambda)}=\frac{N}{\pi i}\hat{{\cal H}}_2;
\,\,\left.\frac{\partial{\cal F}}{\partial\hat{T}_n}\right|_{T_m=\delta_{m,1}}
=\frac{N}{\pi in}\hat{{\cal H}}_
{n+1}$$
(these are special cases of more general formulas below and in Section 7.1 - cf.
(\ref{885})).
Further the $\hat{{\cal H}}_{n+1}$ are homogeneous combinations of the 
Casimir operators of the group; this means that one can parametrize soft
susy breaking terms induced by all the Casimirs of the group and not just
the quadratic one (associated to $\Lambda$).  In this way one extends to
${\cal N}=0$ the family of ${\cal N}=1$ susy breaking terms first considered
in \cite{aj}.  Note the $u_k$ in the SW curve (\ref{558}) can be written
as $u_k=<{\cal O}_k>$ for ${\cal O}_k=(1/k)Tr\,\phi^k+$ lower order
terms (basic observables for a complex scalar field $\phi$ as in \cite{mo}
for example - see Section 10); the ${\cal H}_{m,n}$ are certain
homogeneous polynomials in the $u_k$ and the (Casimir) moduli
$h_k$ of (\ref{558}) refer basically to a background Toda dynamics (cf.
\cite{ia}).
Thus there are at least two points of view regarding the role of the
$T_n$:  (1)  The philosophy of (\ref{B33}) implies that $T_n$ or
$\hat{T}_n$ correspond to coupling constants while (2) The 
philosophy of looking at moduli 
dynamics of $u_k$ or $h_n$ depending on $T_n$ 
puts them in the role of deformation parameters.  We should
probably always treat $T_n$ and $\alpha_j$ in parallel, either
as coupling constants or deformation parameters (this does not
preclude treating $a_j=a_j(T_n)$ however).  
\\[3mm]\indent
We indicate now
some formulas arising from \cite{en,gf} which are discussed
further in \cite{tc} in connection with \cite{lg,me,mg}.  Thus, in
the notation of \cite{en,ec}, one defines Hamiltonians 
$\hat{{\cal H}}_{m+1,n+1}=T_1^{m+n}{\cal H}_{m+1,n+1}$ 
with ${\cal H}_{m+1}={\cal H}_{m+1,2}$ (homogeneous
polynomials in the $\hat{u}_k$ (or $h_k$) via
\be
{\cal H}_{m+1,n+1}=\frac{N}{mn}Res_{\infty}\left(P^{m/N}d
P_{+}^{n/N}\right)={\cal H}_{n+1,m+1};\,\,{\cal H}_{n+1}=\frac
{N}{n}Res_{\infty}\left(P^{m/N}d\lambda\right)
\label{B44}
\ee
and setting $s_n^D=\partial F/\partial s_n$ write
\be
\tau_{ij}=\frac{\partial^2F}{\partial\alpha_i\alpha_j};\,\,\tau^n_i
=\frac{\partial^2F}{\partial\alpha_i\partial s_n};\,\,
\tau^{mn}=\frac{\partial^2F}{\partial s_m\partial s_n}
\label{B45}
\ee
There results
\be
s^D_1=\frac{\beta}{2\pi}\left[\hat{{\cal H}}_2+
i\sum_{m\geq 2}ms_m\hat{{\cal H}}_{m+1}
-\sum_{m,n\geq 2}ms_ms_n\hat{{\cal H}}_{m+1,n+1}\right];
\label{B46}
\ee
$$s^D_n=\frac{\beta}{2\pi n}\left[\hat{{\cal H}}_{n+1}+
i\sum_{m\geq 2}ms_m
\hat{{\cal H}}_{m+1,n+1}\right];\,\,\tau^1_i=\frac{\beta}{2\pi}\left[\frac
{\partial\hat{{\cal H}}_2}{\partial a_i}+i\sum_{n\geq 2}s_n\frac{\partial
\hat{{\cal H}}_{n+1}}{\partial a_i}\right];$$
$$\tau^n_i=\frac{\beta}{2\pi n}\frac{\partial\hat{{\cal H}}_{n+1} }
{\partial a_i};\,\,
\tau^{11}=-2\tau^1_i\tau^1_j\partial_{\tau_{ij}}log\Theta_E(0|E);$$
$$\tau^{1n}=-2\tau^1_i\tau^n_j\partial_{\tau_{ij}}
log\Theta_E(0|\tau);\,\,
\tau^{mn}=\frac{\beta}{2\pi i}\hat{{\cal H}}_{m+1,n+1}-2\tau^n_i\tau^m_j
\partial_{\tau_{ij}}log\Theta_E(0|\tau)$$
Here $\Theta_E$ designates 
\be
\Theta_E(0|\tau)=\Theta[\vec{\alpha},\vec{\beta}](t\vec{V}|\tau)=
\sum exp[i\pi\tau_{ij}n_in_j+itV_in_i-i\pi\sum n^i]
\label{B47}
\ee
where $V_i=\partial u_2/\partial a^i$ and $\vec{\alpha}=(0,\cdots,0)$
with $\vec{\beta}=(1/2,\cdots,1/2)$.  
\\[3mm]\indent
{\bf ALERT!}$\,\,$ Henceforth we assume all ${\cal H}_p,\,\,
 a_k,\,\,u_i$, etc.
have hats but we remove them for notational convenience.
\\[3mm]\indent
Now the spurion superfield
${\cal S}_1$ appears in the classical prepotential as ${\bf (D)}\,\,
{\cal F}=(N/\pi){\cal S}_1{\cal H}_2$ and one obtains the
microscopic Lagrangian by turning on the scalar and auxillary
components of ${\cal S}_1$.  Next the remaining ${\cal S}_n$
are included and one expands the prepotential 
around $s_2=\cdots=s_{N-1}=0$; the $D_n$ and $F_n$ will be the soft
susy breaking parameters (more on this below).  The microscopic
Lagrangian is then determined by 
\be
{\cal F}=\frac{N}{\pi}\sum_1^{N-1}\frac{1}{n}{\cal S}_n
{\cal H}_{n+1}+\frac{N}{2\pi i}\sum_{m,n\geq 2}{\cal S}_n
{\cal S}_m{\cal H}_{n+1,m+1}
\label{B48}
\ee
(with $s_n=0$) and one is primarily interested in
\be
{\cal F}^{red}=\frac{N}{n}\sum_1^{N-1}\frac{1}{n}{\cal S}_n
{\cal H}_{n+1}
\label{B49}
\ee
which in fact is the relevant prepotential for Donaldson-Witten
(DW) theory.  Note
\be
\left.\frac{\partial^2{\cal F}^{red}}{\partial {\cal S}_m\partial
{\cal S}_m}\right|_{s_2=\cdots=s_{N-1}=0}=\frac{2N^2}{\pi imn}
\frac{\partial{\cal H}_{m+1}}{\partial a_i}\frac{\partial{\cal H}_{n+1}}
{\partial a_j}\frac{1}{i\pi}\partial_{\tau_{ij}}log\Theta_E
(0|\tau)
\label{B50}
\ee
are essentially the contact terms of \cite{lg} (cf. Section 9.1).  
Expanding (\ref{B49})
in superspace one has a microscopic Lagrangian and this gives an
exact effective potential at leading order for the ${\cal N}=0$ theory,
allowing one to determine the vacuum structure.  Detailed calculations
for $SU(3)$ theory are given in \cite{en,ec}.  
\\[3mm]\indent
So are we dealing with coupling constants $T_n$ or deformation
parameters?  One answer is ``both" and we refer to Section 9 for more
on this.
The spurion variables parametrize deformations of the
SW differential and the $\tau^{mn}$ and $\tau^n_i$ of (\ref{B46})
have nice transformation properties under $Sp(2(N-1),{\bf Z})$;
the $s_n$ behave like the $\alpha_j$ in many ways.  Promotion of
$T_n\to {\cal S}_n$ for susy breaking should however correspond
to a coupling constant role; the $T_n$ are parameters of a non-susy
family of theories; 
thus role of $s_n$ is parallel to $\alpha_j$.
The prepotential determines the Lagrangian via ${\cal F}_A,\,\,
{\cal F}_{AB},\,\,{\cal F}_{ABC}$, the $D_n$ and $F_n$,
$\lambda,\,\,\psi$ gluinos, $\phi=$ scalar component of ${\cal N}=2$ 
superfield, etc.  Thus ${\cal L}=L_{kin}+L_{int}$ with
\be
L_{kin}=\frac{1}{4\pi}\Im\left[(\nabla_{\mu}\phi)^{\dagger}_a
(\nabla^{\mu}{\cal F})^a+i(\nabla_{\mu}\psi)_a^{\dagger}
\bar{\sigma}^{\mu}\psi^b{\cal F}^a_b-\right. 
\label{B51}
\ee
$$-\left.i{\cal F}_{ab}\lambda^a
\sigma^{\mu}(\nabla_{\mu}\bar{\lambda})^b-\frac{1}{4}
{\cal F}_{ab}(F^a_{\mu\nu}F^{b\mu\nu}+iF^a_{\mu\nu}
\tilde{F}^{b\mu\nu})\right];$$
$$L_{int}=\frac{1}{4\pi}\Im\left[{\cal F}_{AB}F^A(F^*)^B-\frac
{1}{2}{\cal F}_{abC}\left((\psi^a\psi^b)(F^*)^C+(\lambda^a
\lambda^b)F^C+i\sqrt{2}(\psi^a\lambda^b)D^C\right)\right.+$$
$$+\left.\frac{1}{2}{\cal F}_{AB}
D^AD^B+ig\left(\phi^*_af^a_{bc}D_b
{\cal F}^c+\sqrt{2}\left\{(\phi^*\lambda)_a{\cal F}^a_b\psi^b-
(\bar{\psi}\bar{\lambda})_a{\cal F}^a\right\}\right)\right]$$
Here $\lambda$ and $\psi$ are the gluinos and $\phi$ is the scalar
component of the ${\cal N}=2$ vector superfield.  The $f^a_{bc}$
are structure constants of the Lie algebra.  The indices $a,b,c,\cdots$ 
belong to the adjoint representation of $SU(N)$ and are raised and
lowered with the invariant metric.  Indices $A,B,\cdots$ run over 
both indices in the adjoint and over the slow times ($m,n,\cdots$).  Since
all spurions corresponding to higher Casimirs are purely auxillary
superfields the Lagrangian can be simplified as follows.
Set $D^a=-(b^{-1}_{class})^{ac}((b^{class})^m_cD_m+\Re(g
\phi^*_bf^b_{ca}{\cal F}^a))$ with $F^a=-(b^{-1}_{class})^{ac}
(b^{class})_c^mF_m$ where the classical matrix of couplings
$b^{class}_{AB}$ is defined via $b^{class}=(1/4\pi)\tau^{class}$
where
\be
\tau^{class}_{ab}=\tau\delta_{ab};\,\,\tau^{mn}_{class}=0;\,\,
(\tau^{class})^m_a=\frac{N}{\pi im}\frac
{\partial{\cal H}_{m+1}^{class}}{\partial\phi^a}=\frac{N}{\pi im}
Tr\left(\phi^m\hat{T}_a\right)+\cdots
\label{B52}
\ee
where the dots denote the derivative with respect to $\phi^a$ of lower
order Casimir operators.  This leads to
\be
{\cal L}={\cal L}_{{\cal N}=2}-B^{mn}_{class}\left(F_mF^*_n+\frac
{1}{2}D_mD_n\right)+f^e_{bc}(b^{class})^m_a(b^{class})^{-1}_{ae}
D_m\phi^b\bar{\phi}^c+
\label{B53}
\ee
$$+\frac{1}{8\pi}\Im\frac{\partial(\tau^{class})_b^m}{\partial \phi^a}\left[
(\psi^a\psi^b)F^*_m+(\lambda^a\lambda^b)F_m+i\sqrt{2}
(\lambda^a\psi^b)D_m\right]$$

\section{RENORMALIZATION}
\renewcommand{\theequation}{9.\arabic{equation}}
\setcounter{equation}{0}

We extract here from \cite{cb} where a number of additional topics concerning
renormalization also appear.  In particular we omit here the work in
\cite{hb,ha}
(sketched in \cite{cb} and partially subsumed in the formulation of Section
7.1) and other work of various authors on the Zamolodchikov C theorem.
The formulas in Section 7.1 involving derivatives of the prepotential all have
some connection to renormalization of course and to indicate this briefly we
refer to the original elliptic curve situation (cf.
\cite{cu,ea,gc,ia,mn,na,sd} for 
example).  Thus consider ${\bf (CC)}\,\,y^2=(\lambda-\Lambda^2)(\lambda
+\Lambda^2)(\lambda-u)$ for example with $a=(\sqrt{2}/\pi)\int^{\Lambda^2}_
{\Lambda^2}[(\lambda-u)/(\lambda^2-\Lambda^4)]^{1/2}d\lambda$.  One
will have then for $F\sim F^{SW}$
\be
2F=aF_a-\frac{2iu}{\pi};\,\,\Lambda F_{\Lambda}=-\frac{2iu}{\pi}=-8\pi i
b_1u
\label{C1}
\ee
where $b_1=1/4\pi^2$ is the coefficient of the 1-loop beta function.  This is
the only renormalization term here and (in the more enlightened notation of
Section 7.1) $\Lambda F^W_{\Lambda}=T_1F_1^W$ shows that an important
role of the Whitham times is to restore the homogeneity of $F^W$ which can be
disturbed by renormalization (cf. \cite{cb,cu,ea} for more on this - and see
below for more details about renormalization).  We remark also that beta
functions
for this situation are often defined via
\be
\beta(\tau)=\left.\Lambda\partial_{\Lambda}\tau\right|_{u=c};\,\,\beta^a
(\tau)=\left.\Lambda\partial_{\Lambda}\tau\right|_{a=c}
\label{C2}
\ee
where $\tau$ is the curve modulus (e.g. $\tau= F^{SW}_{aa}$ - cf. 
\cite{bw,bz,cu}).
\\[3mm]\indent
In any event renormalization
is a venerable subject and we make no attempt to survey it here
(for renormalization in susy gauge
theories see e.g. \cite{bw,bz,dt,dh,di,hb,ha,li,mn,rb,sc}.
In particular there are various geometrical ideas which can be introduced
in the space of theories $\equiv$ the space of coupling constants
(cf. here \cite{dj,dh,di,dk,dl,dm,lj,od,sa,sb}).  We extract here
now mainly from \cite{dt} where
it is argued that RG (= renormalization group) flow can be 
interpreted as a Hamiltonian vector flow on a phase space which consists
of the couplings of the theory and their conjugate ``momenta", which are
the vacuum expectation values of the corresponding composite operators.
For theories with massive couplings the identity operator plays a central
role and its associated coupling gives rise to a potential in the flow
equations.  The evolution of any quantity under RG flow can be obtained
from its Poisson bracket with the Hamiltonian.  Ward identities can be
represented as constants of motion which act as symmetry generators on the
phase space via the Poisson bracket structure.  
For moduli $g$ regarded as coupling
constants one could obtain beta functions via $\kappa_n(\partial g/
\partial\kappa_n)=\partial_ng$ for $T_n=log(\kappa_n)$ and 
$\partial/\partial T_n=\kappa_n(\partial/\partial\kappa_n)$.  Whitham
dynamics on moduli spaces corresponds then to RG flows and one
obtains e.g. $\partial h_k/\partial T_n$ or $\partial u_k/
\partial T_n$ which could represent beta functions $\beta^k_n$.
Thus following \cite{cb,dt} (revised for \cite{cm})
we may consider the moduli space
of $u_k$ or $h_n$ as a coupling constant space ${\cal M}$ of 
elements $g^a$ (space of theories) and interpret RG flows as
a Hamilonian vector flow on a phase space $T^*{\cal M}$
(in this context $\alpha_j$ should also be regarded as a deformation
parameter but we will ignore it here).  Take $T=log(\kappa)$
and set $\beta^a(g)=\kappa\partial_{\kappa}g^a$ (note, corresponding
to $\Lambda=exp(is_1)$ in (\ref{B33}) with $is_1=log(\Lambda)$ one
would obtain $\beta^a\sim\Lambda\partial_{\Lambda}g^a$ and recall
that $\beta=\Lambda\partial_{\Lambda}\tau$ is a standard beta
function).  We have now a tangent bundle $T{\cal M}\sim (g^a,
\beta^a)$ with $T^*{\cal M}\sim (g^a,\phi_a)$ where $\phi_a=
\partial w(g,t)/\partial g^a$ for some free energy $W=-log(Z)$
with $W\sim \int wd^Dx$.  Here $Z$ could correspond
to $\int{\cal D}\phi exp(-S(\phi))$ and $1=\int{\cal D}\phi
exp[-S(\phi)+W]$ implies $dW=<dS>=\int{\cal D}\phi dS(\phi)
exp[-S+W]$ (and $S\sim \int{\cal L}d^Dx$ for some Lagrangian
so an underlying D-dimensional space is envisioned).  The
approach here of \cite{dt} is field theoretic and 
modifications are perhaps indicated for the ${\cal N}=2$ susy YM theory; thus
the constructions are heuristic.  Now one can display a Hamiltonian
${\bf (HH)}\,\,H(g,\phi)=
\sum\beta^a(g)\phi_a+\beta^{\Gamma}
(g,\Gamma)\phi_{\Gamma}$ which governs the RG 
evolution of $(g^a,\phi_a)$ via
\be
\frac{dg^a}{dT}=\left.\frac{\partial H}{\partial\phi_a}\right|_g;\,\,
\frac{d\phi_a}{dT}=-\left.\frac{\partial H}{\partial g^a}\right|_
{\phi}
\label{C35}
\ee
Here $\Gamma$ (cosmological constant) is a coupling associated
with the identity $I$ whose conjugate momentum is the expectation
value of the identity - in fact one can take heuristically
\be
g^{\Gamma}=\Gamma;\,\,\beta^{\Gamma}(g,\Gamma)=
\frac{d\Gamma}{dT}=-D\Gamma+U^{\Gamma}(g);\,\,\phi_{\Gamma}=
\kappa^D
\label{C36}
\ee
where $T=log(\kappa)$ could refer to any $T_n\,\,(\sim\kappa_n)$. 
One has then a symplectic structure and a
Hamilton-Jacobi (HJ) equation
\be
\frac{\partial w}{\partial T}+H\left(g,\frac{\partial w}{\partial g}
\right)=0=\frac{\partial w}{\partial T}+\sum\beta^a(g)\phi_a+
\beta^{\Gamma}\phi_{\Gamma}
\label{C37}
\ee
For $\beta^a=\beta^a(g,T)$ one could take $T$  
as an additional coupling and work
on $\hat{{\cal M}}=(g^a,\Gamma,T)$ with $\beta^T=1$ and
$\phi_T=\partial_Tw=-H(g,\phi,T)$ ($T\sim log(\kappa)$ implies
$\kappa\partial_{\kappa}T=\beta^T=1$). 
\\[3mm]\indent
Now identify $w\sim F+\Gamma\kappa^D$ 
so $\phi_T=w_T=F_T+\Gamma D\kappa^D$ so that
(\ref{C37}) says e.g. (with $g^a\sim h_a$)
\be
\frac{\partial F}{\partial T}+\sum\frac{\partial h_k}{\partial T}\frac
{\partial F}{\partial h_k}+U^{\Gamma}\kappa^D=0
\label{C38}
\ee
(note $\beta^{\Gamma}\phi_{\Gamma}\sim (-D\Gamma+U^{\Gamma})\kappa^D
\sim -D\Gamma\kappa^D+U^{\Gamma}\kappa^D$ and the $D\Gamma
\kappa^D$ term cancels).
We know by Whitham dynamics that $h_k=h_k(\alpha_j,T_n)$
satisfies some homogeneity equations ${\bf (A)}\,\,\sum\alpha_j
(\partial h_k/\partial\alpha_j)+\sum T_n\partial_nh_k=0$ so a
typical equation like ${\bf (EE)}\,\,
2F=\sum\alpha_j(\partial F/\partial\alpha_j)+\sum T_n\partial_nF$
(cf. (\ref{778})) plus (\ref{C38}) for $T\sim T_n$ implies
(we think now of $w_n=F+\Gamma\kappa_n^D$ for fixed $F$ while
$U^{\Gamma}$ is also held fixed for different $T_n$)
$$2F=\sum T_n\partial_nF+\sum\alpha_j\left(\sum\frac{\partial F}
{\partial h_k}\frac{\partial h_k}{\partial \alpha_j}\right)=\sum
T_n\partial_nF-\sum\frac{\partial F}{\partial h_k}\left(\sum T_n
\partial_nh_k\right)=$$
\be
=\sum T_n\partial_nF-\sum T_n\left(-\partial_nF-U^{\Gamma}
\kappa_n^D\right)
=2\sum T_n\partial_nF+\left(\sum T_n\kappa_n^D\right)U^{\Gamma}
\label{C39}
\ee
This seems to say (for $\sum_1^MT_nexp(DT_n)={\cal T}$)
\be
F-\sum T_n\partial_nF=\frac{1}{2}{\cal T}U^{\Gamma}
\label{C40}
\ee
where $F=F(T_n,\alpha_j)$ and $h_k=h_k(T_n,\alpha_j)$ 
with $T_n$ and $\alpha_j$ independent allow us to set $\alpha_j
=\alpha_j(h_k)$ for $T_n$ fixed.  In particular from (\ref{778}) and
(\ref{C38}) with say $T_n=\delta_{n,1}$ and $T_1\sim \Lambda$
(there is an implicit switch here to the notation of (\ref{B32})) we
see that (\ref{778}) says $2F=\Lambda\partial_{\Lambda}F+
\sum a_j(\partial F/\partial a_j)$ (and $F=F^{SW}$ now) while
(\ref{C38}) says (for $T\sim log(\Lambda)$
as in (\ref{B33})) and no other $T_n$
${\bf (C)}
\,\,\Lambda\partial_{\Lambda}F+\sum \Lambda
(\partial h_k/\partial\Lambda)(\partial F/\partial h_k)+U^{\Gamma}
\Lambda^D=0$.
This means (cf. (\ref{C1}) with more variables inserted)
\be
2F-\sum a_j\frac{\partial F}{\partial a_j}=\Lambda\partial_{\Lambda}F=
-\frac{2iu_2}{\pi}=-U^{\Gamma}
\Lambda^D-
\label{C41}
\ee
$$-\Lambda\sum\frac{\partial F}{\partial h_k}\frac{\partial h_k}
{\partial\Lambda}=-U^{\Gamma}\Lambda^D-\sum\beta_{\Lambda}^
k\frac{\partial F}{\partial h_k}$$
Note $u_2\sim h_2\sim {\cal H}_2$ (cf. below for more such notation)
is a Hamiltonian and we see that it has the form {\bf (HH)}
(after shifting $W\sim F+\Gamma\Lambda^D$).
In addition $U^{\Gamma}$ is determined as
\be
\Lambda^DU^{\Gamma}=\frac{2i}{\pi}u_2-\sum\beta_{\Lambda}^k
\frac{\partial F}{\partial h_k}
\label{C42}
\ee
One notes that HJ equations 
reminiscent of (\ref{C37}) appear in \cite{lg} in the form
\be
\frac{\partial{\cal F}}{\partial t}=
-H\left(a,\frac{\partial{\cal F}}{\partial a},t\right)
\label{C43}
\ee
in connection with Lagrangian submanifolds (with $H$ somewhat unclear)
and this is surely connected to formulas such as (\ref{86}) referring to
Whitham flows.
\\[3mm]\indent
Thus our heuristic picture of renormalization based on \cite{cb,dt} seems
suggestive at least, modulo various concerns over terms like $\Gamma,\,\,
W=\int wd^Dx$ with $w\sim F+\Gamma\kappa^D$ for $t\sim log(\kappa)$, etc.
We remark that in \cite{dt} one has a dictionary of correspondence between
QFT or statistical mechanics with RG and classical mechanics.  This has the
form
\be
\begin{array}{cc}
QFT\sim Statistical\,\, Mechanics & Classical\,\,Mechanics\\
Couplings\sim g^a(t) & Coordinates\sim q^a(t)\\
Beta\,\,functions\sim\beta^a(t) & Velocities\sim \dot{q}^a(t)\\
vev's\sim \phi_a(t) & Momenta\sim p_a(t)\\
Bare\,\,couplings\sim(g_0^a,\phi^0_a) & Initial\,\,values\sim (q^a_0,p^0_a)\\
Generating\,\,functional\sim w(g(t),g_0,t) & Action\sim S(q(t),q_0,t)\\
H(g,\phi,t)=\beta^a(g,t)\phi_a & H=\frac{1}{2m}g^{ab}(q)p_ap_b+U(q,t)\\
\beta^a=\frac{\partial H}{\partial\phi_a};\,\,\dot{\phi}_a=-\frac{\partial H}
{\partial g^a} & \dot{q}^a=\frac{\partial H}{\partial p_a};\,\,\dot{p}_a=-
\frac{\partial H}{\partial q^a}\\
U(g,t)=\beta^{\Lambda}+D\Lambda & U(q,t)\\
\frac{d\phi}{dt}=-dU & \frac{dp}{dt}=-dU\\
w_t+H\left(g(t),\frac{\partial w}{\partial g},t\right)=0 & S_t+
H\left(q(t),\frac{\partial S}{\partial q},t\right)=0\\
No\,\,explicit\,\,\kappa\,\,dependence\,\,in\,\,\beta & Conservative\,\,system\\
Anomalous\,\,dimensions & Pseudo-forces\,\,(Coriolis)\\
RG\,\,invariant\,\,\{\theta,H\}=0 & Constant\,\,of\,\,motion\,\,\{\theta,H\}=0
\end{array}
\label{C3}
\ee
In \cite{pc} one also finds a dictionary of comparisons between QFT and 
magnetic systems based on the effective action and we expand on this 
idea of effective action as
follows.
\\[3mm]\indent
{\bf REMARK 9.1.}$\,\,$  In fact the effective action is isolated by 
A. Morozov in \cite{md} as the source of integrability (cf. also
\cite{gc,gf,gg,ia,mj}).  Indeed effective action of the form
\be
exp[S_{eff}(t|\phi)]=Z(t|\phi)=\int {\cal D}\phi\,exp[S(t|\phi)]
\label{C4}
\ee
in matrix models for example will correspond to a tau function of integrable
systems such as KP and the ``time" variables $t$ can be thought of as
coupling constants.  Further investigation of related generalized Kontsevich
models (GKM) leads to effective action involving both KP times $t_n$ along
with ``Whitham" times $T_k$ which have a 
rather different origin.  Thus for $D=n^2$,
modulo a few details ($C$ is defined below)
\be
Z_{GKM}(L|V_{p+1})=C\int d^DX\,exp[Tr(-V_{p+1}(X)+XL)]
\label{C5}
\ee
where $L$ is an $n\times n$ Hermitian matrix, $V_{p+1}$ is a polynomial
of degree $p+1$, and $W_p=V'_{p+1}$ is a polynomial of degree $p$.  Further
$C$ in (\ref{C5}) is a prefactor used to cancel the quasiclassical contribution
to the integral around the saddle point $X=\Lambda$, namely, modulo
inessential details,
\be
C=exp[Tr(V_{p+1}(\Lambda))-Tr(\Lambda V'_{p+1}(\Lambda))]det^{1/2}
[\partial^2V_{p+1}(\Lambda)]
\label{C6}
\ee
The time variables are introduced in order to parametrize the $L$ dependence
and the shape of $V_{p+1}$ via
\be
T_k=\frac{1}{k}Tr(\Lambda^{-k});\,\,\tilde{T}_k=\frac{1}{k}
Tr(\tilde{\Lambda}^{-k});
\label{C7}
\ee
$$L=W_p(\Lambda)=\tilde{\Lambda}^p;\,\,t_k=\frac{p}{k(p-k)}
Res_{\mu}W_p^{1-(k/p)}(\mu)d\mu$$
Then one can write
\be
Z_{GKM}=exp[-{\cal F}_p(\tilde{T}_k|t_n)]\tau_p(\tilde{T}_k+t_k)
\label{C8}
\ee
where $\tau_p$ is a $p$-reduced KP tau function and 
\be
{\cal F}_p=\frac{1}{2}\sum A_{ij}(t)(\tilde{T}_i+t_i)(\tilde{T}_j+t_j);\,\,
A_{ij}=Res_{\infty}W^{i/p}(\lambda)dW_{+}^{j/p}(\lambda)
\label{C9}
\ee
is a quasiclassical (Whitham) tau function (or the logarithm thereof). 
In addition such a GKM 
prepotential is a natural genus zero component in the SW prepotential
(cf. \cite{gf} and compare (\ref{C9}) with (\ref{991}) for example).
\\[3mm]\indent
{\bf REMARK 9.2.}$\,\,$  It is possible now to apprehend how the Whitham
times can play two apparently different roles, namely {\bf (1)}$\,\,$ coupling
constants, and {\bf (2)}$\,\,$ deformation parameters.  We follow the Russian
school of Gorsky, Marshakov, Mironov, and Morozov (\cite {gc,gf,gg,
gx,gy,ky,md,mj,mk,mm})
and especially the latter as in \cite{md,mj} (cf. also \cite{ia,lg}).  The comments
to follow are mainly physical and heuristic and the matter seems to be 
summarized in a statement from \cite{md}, namely:$\,\,$ The time variables
(for SW theory), associated with the low energy correlators (i.e. the 
renormalized coupling constants) are Whitham times (the deformations of 
symplectic structure).  Thus both roles {\bf (1)} and {\bf (2)}
appear but further clarification seems required and is possible.
One relevant theme here is described in \cite{ia,md} 
roughly as follows. 
Given a classical dynamical system one can think
of two ways to proceed after exact action-angle variables are somehow found.
One can quantize the system or alternatively one can average over fast 
fluctuations of angle variables and get some effective slow dynamics on the
space of integrals of motion (Whitham dynamics). 
Although seemingly different, these are exactly
the same problems, at least in the first approximation (nonlinear WKB).
Basically the reason is that quantum wave functions appear from averaging 
along the classical trajectories - very much in the spirit of ergodicity
theorems.
In string theory, the classical system in question arises after some first
quantized problem is exactly solved with its effective action (generating
function of all the correlators in the given background field) being a tau
function of some underlying loop-group symmetry for example.  The two 
above mentioned problems concern deformation of classical into quantum 
symmetry and renormalization group flow to the low energy (topological)
field theory.  The effective action arising after averaging over fast 
fluctuations (at the end point of RG flow) is somewhat different from the
original one (which is a generating functional of all the matrix elements of
some group); the ``quasiclassical" tau function at the present moment does
not have any nice group theoretical interpretation.  The general principle
is in any case that the Whitham method is essentially the same as 
quantization, but with a considerable change in the nature of the variables;
the quantized model lives on the moduli space (the one of zero-modes or
collective coordinates), and not on the original configuration space.
\\[3mm]\indent
Note that in higher dimensional field theories the functional integrals depend
on the normalization point $\mu$ (IR cutoff) and effective actions describe
the effective dynamics of excitations with wavelengths exceeding $\mu^{-1}$.
The low energy effective action arises when $\mu\to 0$ and only a finite
number of excitations (zero modes of massless fields) remain relevant.
Such low energy effective actions are pertinent for universality classes and 
one could say that ${\cal N}=2$ susy YM models belong to the universality
class of which the simplest examples are $(0+1)$-dimensional integrable
systems.
We could supplement the above comments with further remarks based on
the fundamental paper \cite{gc} (some such extractions appear in 
\cite{cb}).  Let us rather try to summarize matters in the following manner,
based on \cite{gc,md}.  First think of a field theory with partition function
$Z(t|\phi)=\int_{\phi_0}{\cal D}\phi\,exp[S(t|\phi)]$ as in (\ref{C4}).
The dynamics in space time is replaced by the effective dynamics in the
space of coupling constants $g^{-2},\,\theta,$ and $t_i$ via Ward identities
which generate Virasoro conditons and imply that $Z$ is the tau function
of some KP-Toda type theory.  The parameter space is therefore a spectral
curve for such a theory and the family of vacua ($\sim\phi_0$) is associated
with the family of spectral curves (i.e. with the moduli space).  Now the 
averaging process or passage to Whitham level corresponds to {\bf (A)}
$\,\,$ Quantization of the effective dynamics corresponding to some sort
of renormalization process creating renormalized coupling constants
$T_i$ for example, and {\bf (B)}$\,\,$ Creating  RG type slow dynamics
on the Casimirs $h_k$ of the KP-Toda theory; then via a $1-1$ map
$(h_k)\to (u_m)$ to moduli one deals with moduli as coupling constants
while the $T_n$ are deformation parameters.  In this
spirit it seems that in \cite{en,ec,ed} both features are used.  Promoting 
$\hat{T}_n$ to spurion superfields corresponds to a coupling constant role
for the $T_n$ (as does treating the prepotential as a generating
function of correlators)
while dealing with the derivative of the prepotential with
respect to the $\hat{T}_n$ in the spirit of renormalization
involves a deformation parameter aspect.
\\[3mm]\indent
{\bf REMARK 9.3.}$\,\,$  We remark that there is also a brane picture involving
RG flows and Whitham times (cf. \cite{gk,gd,ge,gx,gj,gy,ky} for example)
in which Whitham dynamics arises from the motion of certain $D-4$ branes.
This dynamics generates conditions for the approximate invariance of the
spectral
curve under RG flow and provides the validity of the classical equations of
motion in the unperturbed theory if suitable first order perturbation is
allowed.
It is also conjectured that both Hitchin spin chains and Whitham theories
can be regarded as RG equations (Hitchin times are to be identified with the
space RG scale - $t\sim log(r)$ - and the fast Toda-Calogero system - motion
of certain $D-0$ branes - corresponds to RG flows on a hidden Higgs branch,
providing susy invariant renormalization for the nonperturbative effects,
while the Whitham system is operating on the Coulomb branch).  In this
spirit the spectral curve is the RG invariant and the very meaning of
integrability is to provide the regularization of nonperturbative effects
consistent
with the RG flows.

\subsection{Contact terms}

Another arena where Whitham times play an important role involves
the structure of
contact terms in the topological twisted ${\cal N}=2$ susy gauge
theory on a 4-manifold $X$ (cf. \cite{en,ec,lg,me,mo,mg,tc}).  
We will not try to cover the background here.  An earlier version of this
section (written in 1998) made some attempt at this but it would have to
be considerably enlarged and revised to be instructive now.  Since enlarged
versions already exist in the references above there seems to be no point
in an inadequate sketch.  Thus we follow mainly \cite{me,mo,tc} here for the
$SU(N)$ theory (as in Sections 7 and 8) and go directly to the u-plane
integral (for $b_{2+}=1$)
\be
Z_u=\int_{{\cal M}_{Coulomb}}[da\,d\bar{a}]A(u_k)^{\chi}B(u_k)^{\sigma}
exp\left(\sum p_ku_k+S^2\sum f_k
f_mT_{k,m}\right)\Psi
\label{10A}
\ee
Here $S\in H_2(X,{\bf Z})$ and the $T_{k,m}$ are so called contact terms.
We omit discussion of the other terms.  The contact terms can be derived via
blowup procedures and this introduces the tau function of a periodic Toda
lattice.  In fact one can show that (notation as in Sections 7 and 8)
\be
T_{k+1,m+1}=\frac{\pi ikm}{4N^2}\left.\left(\frac{\partial^2{\cal F}^{red}}
{\partial T_k\partial T_m}\right)\right|_{T_{n\geq 2}=0}
\label{10B}
\ee
(cf. (\ref{B50})) where ${\cal F}^{red}$ is defined in (\ref{B49}).  The key
idea is that the slow times $\hat{T}_k$ are dual to the $\hat{{\cal H}}_{k+1}$
in the subspace $T_{n\geq 2}=0$ (cf. {\bf (DUAL)} after (\ref{B34}));
note also ${\cal H}_{k+1}=u_{k+1}+{\cal O}(u_k)$.  Further 
the $f_k$ in (\ref{10A}) 
are proportional to the fast Toda times $t_k$.  The development in
\cite{me,mo,tc}
shows how this is all a very natural development.

\section{WHITHAM, WDVV, and PICARD-FUCHS}
\renewcommand{\theequation}{10.\arabic{equation}}\setcounter{equation}{0}

We add a few comments now relating Whitham times to the flat times
of Frobenius manifold theory and the WDVV equations.  This involves
connections to TFT, Landau-Ginzburg (LG) models and Hurwitz spaces,
and Picard-Fuchs (PF) equations.

\subsection{ADE and LG approach}

Connections of TFT, ADE, and LG models abound 
(cf. \cite{aa,ci,dn,dc,kc,ko,ta,ya}) and for $N=2$ susy YM we go to 
\cite{ib} 
(cf. also \cite{bw,ey}).  First we extract from \cite{ib} as in \cite{cb}.  
Thus one evaluates integrals $a_i=\oint_{A_i}
\lambda_{SW}$ and $a_i^D=\oint_{B_i}\lambda_{SW}$ using Picard-Fuchs
(PF) equations.  One considers $P_R(u,x_i)=det(x-\Phi_R)$ where 
$R\sim$ an irreducible representation of $G$ and $\Phi_R$ is a representation
matrix.  Let $u_i\,\,(1\leq i\leq r)$ be Casimirs built from $\Phi_R$ of
degree $e_i+1$ where $e_i$ is the $i^{th}$ exponent of $G$ (see below).
In particular $u_1\sim$ quadratic Casimir and $u_r\sim$ top Casimir of 
degree $h$ where $h$ is the dual Coxeter number of $G$ ($h=r+1$ for 
$A_r$).  The quantum SW curve is then
\be
\tilde{P}_R(x,z,u_i)\equiv P_R\left(x,u_i+\delta_{i,r}\left[z+
\frac{\mu^2}{z}\right]\right)=0
\label{11D}
\ee
where $\mu^2=\Lambda^{2h}$ with $\Lambda\sim$ the dynamical scale and the
$u_i$ are considered as gauge invariant moduli parameters in the Coulomb
branch.  This curve is viewed as a multisheeted foliation $x(z)$ over
${\bf CP^1}$ and the SW differential is $\lambda_{SW}=x(dz/z)$.  The physics
of $N=2$ YM is described generally
by a complex $rank(G)$ dimensional subvariety of
the Jacobian which is a special Prym variety (cf. \cite{da,mm}).  Now
one writes (\ref{11D}) in the form
\be
z+\frac{\mu^2}{z}+u_r=\tilde{W}_G^R(x,u_1,\cdots,u_{r-1})
\label{11E}
\ee
For the fundamental representation of $A_r$ 
for example one has (we have organized the indexing to conform
more closely to \cite{de,gf,mm,na})
\be
\tilde{W}_{A_r}^{r+1}=x^{r+1}-u_1x^{r-1}-\cdots-u_{r-1}x;\,\,
\label{11F}
\ee
and setting ${\bf (SP)}\,\,W_G^R(x,u_1,\cdots,u_r)=\tilde{W}^R_G(x,
u_1,\cdots,u_{r-1})-u_r$ it follows that $W^{r+1}_{A_r}$ 
is the fundamental LG superpotentials for $A_r$ type topological
minimal models (cf. also \cite{dn,dc,ey}).
The $u_i$ can be thought of as coordinates on the space of TFT.
We will concentrate on $A_r$ but $D_r$ and other groups are discussed in
\cite{ib}.
For comparison to \cite{gf} we recall (cf. \cite{cb})
that for a pure $SU(N)$ susy YM theory
\be
det_{N\times N}[L(w)-\lambda]=0;\,\,
P(\lambda)=\Lambda^N\left(w+
\frac{1}{w}\right);
\label{E58}
\ee
$$
P(\lambda)=\lambda^N-\sum_2^Nv_k\lambda^{N-k}=\prod_1^N(\lambda-\lambda_j);
\,\,v_k=(-1)^k\sum_{i_1<\cdots<i_k}\lambda_{i_1}\cdots\lambda_{i_k}$$
Here the $v_k$ are Schur polynomials of $h_k=(1/k)\sum_1^N\lambda_i^k$ via
the formula
${\bf (A)}\,\,log(\lambda^{-N}P(\lambda))=-\sum_k(h_k/\lambda^k)$;
there are $g=N-1$ moduli $u_k$ and we refer to \cite{ia} for the Lax
operator $L$.
Thus $v_0=1,\,\,v_1=0,\,\,v_2=h_2,\,\,v_3=h_3,\,\,v_4=
h_4-(1/2)h_2^2,$ etc. ($h_1=0$ for $SU(N)$).
One also has the representation
\be
y^2=P^2(\lambda)-4\Lambda^{2N};\,\,y=\Lambda^N\left(w-\frac{1}{w}\right)
\label{E60}
\ee
giving a two fold covering of the punctured Riemann sphere with parameter
$\lambda$. 
Thus $r+1\sim N\sim h,\,\,g=N-1,\,\,\lambda\sim x$, and $u_k\sim v_{k+1}$
with $u_r\sim v_N$ while
$P\sim W^{r+1}_{A_r}=W^N_{A_{N-1}}$ where $A_{N-1}\sim SU(N)$ or
$SL(N)$.  To clarify the curve correspondence write
\be
\hat{w}=z+\mu^2z^{-1};\,\,\hat{w}=P(x);\,\,y=z-\mu^2z^{-1};y^2=P^2-4\mu^2
\label{11G}
\ee
with $\mu^2=\Lambda^{2N}$ leading to the equivalence of curves
$y^2=P^2-4\Lambda^{2N}$.  Recall also from Section 7 (\ref{662}), etc.,
that $dS_{SW}=\lambda(dw/w)=\lambda(dP/y)=\lambda(dy/P)$. 
For the moment we ignore relations between
$w$ and $\hat{w}$ and refer to \cite{de,mm} for discussion of the 
algebraic geometry picture for the $\hat{w}$ parametrization.
\\[3mm]\indent
Now in 2-D TFT of LG type $A_r$ with superpotential {\bf (SP)} the flat
time coordinates for the moduli space are given via
\be
T_i=c_i\oint dx\,W^R_G(x,u)^{e_i/h}\,\,\,(i=1,\cdots,r)
\label{11H}
\ee
(the $e_i$ will be discussed later and recall $h=r+1$).
Note that no Riemann surface is involved here since the integral corresponds
to a residue calculation for $x\sim p$ at $\infty$ (cf. Remark 10.2).  Thus
the TFT here depends only on $W^R_G(x,u)$ with $x$ a formal parameter.
These are residue calculations for times $T_i$ 
(with no reference to Whitham theory) which
will be polynomials in the $u_j$ (the normalization
constants $c_i$ are specified below.  One defines primary fields
\be
\phi_i^R(x)=\frac{\partial W^R_G(x,u)}{\partial T_i}\,\,\,(i=1,\cdots,r)
\label{11I}
\ee
where $\phi_r^R=1$ is the identity $\sim$ puncture operator.  The one
point functions of the gravitational descendents $\sigma_n(\phi_i^R)$
(cf. \cite{aa,dn,dc}) are evaluated via
\be
<\sigma_n(\phi_i^R)>=b_{n,i}\sum_1^r\eta_{ij}\oint W_G^R(x,u)^
{(e_j/h)+n+1}\,\,\,(n=0,1,\cdots)
\label{11J}
\ee
for certain constants $b_{n,i}$ (cf. \cite{ib} for details).  The
topological metric $\eta_{ij}$ is given by
\be
\eta_{ij}=<\phi_i^R\phi_j^RP>=b_{0,r}\frac{\partial^2}{\partial T_i
\partial T_j}\oint W^R_G(x,u)^{1+(1/h)}
\label{11K}
\ee
and $\eta_{ij}=\delta_{e_i+e_j,h}$ can be obtained by adjustment of
$c_i$ and $b_{n,i}$.  The primary fields generate the closed operator
algebra
\be
\phi_i^R(x)\phi_j^R(x)=\sum_1^rC^k_{ij}(T)\phi^R_k(x)+Q^R_{ij}(x)
\partial_xW^R_G(x)
\label{11L}
\ee
where
\be
\frac{\partial^2W^R_G(x)}{\partial T_i\partial T_j}=\partial_xQ^R_{ij}(x)
\label{11M}
\ee
(for details on (\ref{11L}) - (\ref{11M}) we refer to \cite{aa,dn,do,ex,ew,
ho,wc}).
Note again $x\sim p$ is a formal parameter and $X\sim T_r$ would be a natural
identification.
At this point it is not evident (nor is it true) 
that the times $T_i$ of (\ref{11H}) correspond to the
Whitham times of Section 7 based on the Toda curve $y^2=
P^2-4\Lambda^{2N}$.  Indeed we note from (\ref{773}) and e.g. (\ref{A5}),
that $T_k=-(1/k)Res_{\xi=0}\xi^kdS$ is more or less tautological whereas
a prescription (\ref{11H}) determines the $T_k$ as functions of $u_j$.
Now the structure constants $c^k_{ij}$ are 
independent of $R$ since $c_{ijk}=c^{\ell}_{ij}\eta_{\ell k}$ is given
via $c_{ijk}(T)=<\phi_i^R\phi_j^R\phi_k^R>$ which are topologically
invariant physical observables.  
In 2-D TFT one then has
a free energy $F$ such that $c_{ijk}=\partial^3F/\partial T_i\partial T_j
\partial T_k$ (cf. \cite{aa,cy,dn,dc}).
\\[3mm]\indent
Now for the Picard-Fuchs (PF) equations, the SW differential
$\lambda_{SW}\sim\lambda=(xdz/z)$ can be written as (cf. 
\cite{cb,gf})
$\lambda_{SW}=[x\partial_xW/\sqrt{W^2-4\mu^2}]dx$ (for $W\sim W_G^R$) and 
one has then, for $\mu$ fixed
\be
\frac{\partial\lambda_{SW}}{\partial T_i}=-\frac{1}{\sqrt{W^2-4\mu^2}}\frac
{\partial W}{\partial T_i}dx+d\left(\frac{x}{\sqrt{W^2-4\mu^2}}
\frac{\partial W}{\partial T_i}\right)
\label{11N}
\ee
(total derivative terms will then be suppressed).
Suppose $W$ is quasihomogeneous leading to
\be
x\partial_xW+\sum_1^rq_iT_i\frac{\partial W}{\partial T_i}=hW\,\,
(i.e.\,\,W(tx,t^{q_i}T_i)=t^hW(x,T_i))
\label{11O}
\ee
($q_i=e_i+1$ is the degree of $T_i$).  Then
\be
\lambda_{SW}-\sum_1^rq_iT_i\frac{\partial\lambda_{SW}}{\partial T_i}
=\frac{hWdx}{\sqrt{W^2-4\mu^2}}
\label{11P}
\ee
and applying the Euler derivative
$\sum q_jT_j(\partial/\partial T_j)$ to both sides yields
(after some calculation)
\be
\left(\sum_1^rq_iT_i\frac{\partial}{\partial T_i}-1\right)^2\lambda_{SW}-
4\mu^2h^2\frac{\partial^2\lambda_{SW}}{\partial T_r^2}=0
\label{11Q}
\ee
The calculation goes as follows; first
\be
\sum q_jT_j\partial_j\lambda-(\sum q_jT_j\partial_j)\sum q_iT_i\partial_i
\lambda=hdx\sum q_jT_j\partial_j\frac{W}{\sqrt{W^2-4\mu^2}}\equiv
\label{11R}
\ee
$$\equiv \sum q_iq_jT_iT_j\lambda_{ij}+\sum q_j(q_j-1)T_j\lambda_j-
\sum q_jT_j\lambda_j+\lambda-\frac{hWdx}{\sqrt{W^2-4\mu^2}}=$$
$$=-hdx\left[
\sum\frac{q_jT_jW_j}{\sqrt{W^2-4\mu^2}}+\frac{W^2\sum q_jT_jW_j}
{(W^2-4\mu^2)^{3/2}}\right]\equiv$$
$$\equiv \sum q_iq_jT_iT_j\lambda_{ij}+\sum q_j(q_j-2)T_j\lambda_j
+\lambda=hdx\left[\frac{W}{\sqrt{W^2-4\mu^2}}+\frac{4\mu^2(hW-xW_x)}
{(W^2-4\mu^2)^{3/3}}\right]$$
Next note that 
\be
hd\left(\frac{xW}{\sqrt{W^2-4\mu^2}}\right)=hdx\left[\frac{W}{\sqrt
{W^2-4\mu^2}}-\frac{4\mu^2 xW_x}{(W^2-4\mu^2)^{3/2}}\right]
\label{11S}
\ee
so the right side of (\ref{11R}) can be written as $4W\mu^2h^2dx/
(W^2-4\mu^2)^{3/2} +hd\left(xW/\sqrt{W^2-4\mu^2}\right)$.  We note 
also that by degree counting (see below for more on this)
\be
\frac{\partial}{\partial T_r}=\sum\frac{\partial u_i}{\partial T_r}\frac
{\partial}{\partial u_i}=-\frac{\partial}{\partial u_r}\Rightarrow 
\frac{\partial W}{\partial T_r}=1
\label{11T}
\ee
since $\partial W/\partial u_r=-1$ (note here $\partial u_k/\partial T_r=-
\delta_{kr}\Rightarrow$ (\ref{11T}))
and thus from (\ref{11N})
($W_r=1,\,\,W_{xr}=0$)
\be
\frac{\partial\lambda}{\partial T_r}=-\frac{dx}{\sqrt{W^2-4\mu^2}}+d\left(\frac
{x}{\sqrt{W^2-4\mu^2}}\right);
\label{11U}
\ee
$$\frac{\partial^2\lambda}{\partial T_r^2}=\frac{Wdx}{(W^2-4\mu^2)^{3/2}}
+d\left(\partial_r\frac{x}{\sqrt{W^2-4\mu^2}}\right)\sim\frac{Wdx}
{(W^2-4\mu^2)^{3/2}}$$
Consequently the right side of (\ref{11R}) is equivalent to
$4\mu^2h^2\partial^2\lambda/\partial T_r^2$ and (\ref{11R}) becomes
\be
-4\mu^2h^2\frac{\partial^2\lambda}{\partial T_r^2}+\sum q_iq_jT_iT_j
\frac{\partial^2\lambda}{\partial T_i\partial T_j}+\sum q_j(q_j-2)T_j
\frac{\partial\lambda}{\partial T_j}+\lambda=0
\label{11V}
\ee
which is equivalent to (\ref{11Q}).
We note that the
second term in (\ref{11Q})
represents the scaling violation due to $\mu^2=\Lambda^{2h}$
since (\ref{11Q}) reduces to the scaling relation for $\lambda_{SW}$ in the
classical limit $\mu^2\to 0$.  Note that $\lambda_{SW}(T_i,\mu)$ is of
degree one (equal to the mass dimension) which implies 
$(\spadesuit\spadesuit\spadesuit)\,\,\left(
\sum_1^rq_iT_i(\partial/\partial T_i)+h\mu(\partial/\partial\mu)-1\right)
\lambda_{SW}=0$ 
(see Remark 10.1 below) from which (\ref{11Q}) can also be obtained. 
In this respect we note that
\be
\partial_{\mu}\lambda=\frac{4x\mu W_xdx}{(W^2-4\mu^2)^{3/2}}\Rightarrow
(\sum q_iT_i\partial_i-1)\lambda=-h\mu\partial_{\mu}\lambda=
-\frac{4h\mu^2xW_xdx}{(W^2-4\mu^2)^{3/2}}
\label{11W}
\ee
Then, using (\ref{11S})
$$
\left(\sum q_iT_i\partial_i -1\right)\lambda=-h\mu\partial_{\mu}\lambda=
-\frac{4h\mu^2xW_xdx}{(W^2-4\mu^2)^{3/2}}=$$
\be
=hd\left(\frac{xW}{\sqrt
{W^2-4\mu^2}}\right)-\left(\frac{Whdx}
{\sqrt{W^2-4\mu^2}}\right)
\label{11X}
\ee
and this implies, via (\ref{11U})
\be
\left(\sum q_iT_i\partial_i - 1\right)^2\lambda=-h\mu d\left(\partial_{\mu}
\frac{xW}{\sqrt{W^2-4\mu^2}}\right)+\frac{4h^2\mu^2Wdx}{(W^2-4\mu^2)^{3/2}}
\sim 4h^2\mu^2\frac{\partial^2\lambda}{\partial T_r^2}
\label{11Y}
\ee
\indent
{\bf REMARK 10.1.}$\,\,$ In connection with scaing we note from 
\be
W(tx,t^{n+1}u_n)=t^Nx^N-u_1t^2(xt)^{N-2}-\cdots- t^Nu_{N-1}=t^NW(x,u_n)
\label{11c}
\ee
that (corresponding to (\ref{11O}))
\be
x\partial_xW+\sum (n+1)u_n\left(\frac{\partial W}{\partial u_n}\right)=NW
\label{11d}
\ee
Alternatively we can write $W(x,u_n)=x^N-u_2x^{N-2}-\cdots -u_N$ with
$W(tx,t^nu_n)=t^NW(x,u_n)$ leading to $[x\partial_x+\sum nu_n(\partial/
\partial u_n)]W=NW$.
Similarly we see that
$W_x=Nx^{N-1}-(N-2)u_1x^{N-3}-\cdots -u_{N-2}$ implies 
$W_x(tx,t^{n+1}u_n)=t^{N-1}W_x(x,u_n)$ and 
consequently for $\lambda=xW_xdx/
\sqrt{W^2-4\mu^2}$ one gets
$$
\lambda(tx,t^{n+1}u_n,t^N\mu)=\frac{txt^{N-1}W_xtdx}{(t^{2N}W^2-
4t^{2N}\mu^2)^{1/2}}=t\lambda(x,u_n,\mu)\Rightarrow$$
\be
\Rightarrow \left(x\partial_x+\sum (n+1)u_n\frac{\partial}{\partial u_n}
+N\mu\partial_{\mu}\right)\lambda=\lambda
\label{11e}
\ee
Next from (\ref{11H}) we have ($x\to t\xi$)
\be
T_i(t^{n+1}u_n)=c_i\oint dxW(t^{n+1}u_n,x)^{e_i/N}=
\label{11f}
\ee
$$= c_it^{e_i+1}\oint d\xi W(u_n,\xi)=t^{q_i}T_i(u_n)\Rightarrow
\sum (n+1)u_n\frac{\partial T_i}{\partial u_n}=q_iT_i$$
Now to confirm (\ref{11O}) write $W(tx,t^{q_i}T_i\mu)\sim
W(tx,t^{n+1}u_n,\mu)=t^NW(x,T_i,\mu)$ (via (\ref{11c}) 
and (\ref{11f})).  Further
note (via (\ref{11e})
$$
\lambda(tx,t^{q_i}T_i,t^N\mu)\sim \lambda(tx,t^{n+1}u_n,t^N\mu)=t\lambda
(x,T_i,\mu)\Rightarrow$$
\be
\Rightarrow \left(x\partial_x+\sum q_iT_i\partial_i+N\mu\partial_{\mu}
\right)\lambda=\lambda
\label{11h}
\ee
Let us try now to derive the relation $(\spadesuit\spadesuit\spadesuit)$,
namely, $(\sum q_iT_i\partial_i+N\mu\partial_{\mu}-1)\lambda=0$.
Thus, $(\spadesuit\spadesuit\spadesuit)\equiv$ (\ref{11W}) and thence
(\ref{11X}) which means
\be
\left(\sum q_iT_i\partial_i-1\right)\lambda\sim -\frac{Whdx}{\sqrt{W^2-4\mu^2}}
\label{11i}
\ee
and this corresponds to (\ref{11P}) which we know to be true.  This shows
only however that an integrated form of $(\spadesuit\spadesuit
\spadesuit)$ is valid (i.e. $(\spadesuit\spadesuit\spadesuit)$ is
valid modulo $hd(xW/\sqrt{W^2-4\mu^2})$.
\\[3mm]\indent
Another set of
differential equations for $\lambda_{SW}$ is obtained using (\ref{11L}), to wit
\be
\frac{\partial^2}{\partial T_i\partial T_j}\lambda_{SW}=\sum_kC^k_{ij}(T)
\frac{\partial^2}{\partial T_k\partial T_r}\lambda_{SW}
\label{11Z}
\ee
To see how this arises one writes from (\ref{11L}), (\ref{11M}), and
(\ref{11N}) (using (\ref{11I}))
\be
\frac{\partial^2\lambda}{\partial T_i\partial T_j}=-
\frac{W_{ij}dx}{\sqrt{W^2-4\mu^2}}+\frac{\phi_i\phi_jWdx}
{(W^2-4\mu^2)^{3/2}}=
\label{11j}
\ee
$$=-\frac{dQ_{ij}}{\sqrt{W^2-4\mu^2}}+\frac{\left(\sum C^k_{ij}\phi_k
+Q_{ij}W_x\right)Wdx}{(W^2-4\mu^2)^{3/2}}=$$
$$=\sum\frac{C^k_{ij}W_kWdx}{(W^2-4\mu^2)^{3/2}}-d\left(\frac{Q_{ij}}
{\sqrt{W^2-4\mu^2}}\right)$$
Then observe that from (\ref{11N})
\be
\frac{WW_kdx}{(W^2-4\mu^2)^{3/2}}\sim\frac{\partial^2\lambda}
{\partial T_k\partial T_r}
\label{11k}
\ee
resulting in (\ref{11Z}).
Then the PF equations (based on (\ref{11Y}) and (\ref{11Z})) for the SW
period integrals $\Pi=\oint\lambda_{SW}$ are nothing but the Gauss-Manin
differential equations for period integrals expressed in the flat
coordinates of topological LG models.  These can be converted into $u_k$
parameters (where $\partial u_k/\partial T_r=-\delta_{kr}$) as
\be
{\cal L}_0\Pi\equiv\left(\sum_1^rq_iu_i\frac{\partial}{\partial u_i}-1
\right)^2\Pi-4\mu^2h^2\frac{\partial^2\Pi}{\partial u_r^2}=0;
\label{11a}
\ee
$${\cal L}_{ij}\Pi\equiv\frac{\partial^2\Pi}{\partial u_i\partial u_j}+\sum_1^r
A_{ijk}(u)\frac{\partial^2\Pi}{\partial u_k\partial u_r}+\sum_1^r
B_{ijk}(u)\frac{\partial\Pi}{\partial u_k}=0$$
where
\be
A_{ijk}(u)=\sum_1^r\frac{\partial T_m}{\partial u_i}\frac{\partial T_n}
{\partial u_j}\frac{\partial u_k}{\partial T_{\ell}}C^{\ell}_{mn}(u);\,\,
B_{ijk}(u)=-\sum_1^r\frac{\partial^2T_n}{\partial u_i\partial u_j}\frac
{\partial u_k}{\partial T_n}
\label{11b}
\ee
which are all polynomials in $u_i$.  One can emphasize that the PF equations
in 4-D $N=2$ YM are then essentially governed by the data in 2-D topological
LG models.

\subsection{Frobenius algebras and manifolds}

We sketch here very briefly and somewhat incompletely 
some basic material on Frobenius
algebras (FA) and Frobenius manifolds (FM) following \cite{dc,dd}
(cf. also \cite{kw,mf,mi}) with special emphasis on LG models and TFT.
This is not meant to be complete in any sense but will lead to a better
understanding of Section 10.1 and serve simultaneously as a prelude to
WDVV.  Generally one is looking for a function $F(t_1,\cdots,t_n)$ such
that $c_{\alpha\beta\gamma}=\partial^3F/\partial t^{\alpha}
\partial t^{\beta}\partial t^{\gamma}$ satisfying ${\bf (C})\,\,\eta_
{\alpha\beta}=c_{1\alpha\beta}(t)$ is a constant nondegenerate matrix
with $\eta^{\alpha\beta}=(\eta_{\alpha\beta})^{-1}$, ${\bf (D)}\,\,
c^{\gamma}_{\alpha\beta}=\eta^{\gamma\epsilon}c_{\epsilon\alpha\beta}(t)$
determines a structure of associative algebra $A_t:\,\,e_{\alpha}\cdot
e_{\beta}=c^{\gamma}_{\alpha\beta}e_{\gamma}$ where $e_1,\cdots,e_n$
is a basis of ${\bf R}^n$ with $e_1\sim$ unity via $c^{\beta}_{1\alpha}=\delta_
{\alpha}^{\beta}$, and ${\bf (E)}\,\,F(c^{d_1}t^1,\cdots,c^{d_n}t^n)=
c^{d_F}F(t^1,\cdots,t^n)$ which corresponds to ${\cal L}_EF=E^{\alpha}
\partial_{\alpha}F=d_F\cdot F$ for $E=E^{\alpha}\partial_{\alpha}$
with $E^{\alpha}=d_{\alpha}t^{\alpha}$ here (we use $t^k\sim t_k$ and 
later, for a certain LG model as in Remark 10.2, $t^k\sim T_k$ - cf.
(\ref{E108})). 
In \cite{dc} one looks at
$d_1=1$ and physics notation involves $d_{\alpha}=1-q_{\alpha},\,\,d_F=3-d,
\,\,q_n=d$ and $q_{\alpha}+q_{n-\alpha+1}=d$.  The associativity condition
in {\bf (D)} reads as (WDVV equations)
\be
\frac{\partial^3F}{\partial t^{\alpha}\partial t^{\beta}\partial 
t^{\lambda}}\eta^{\lambda\mu}\frac{\partial^3F}{\partial t^{\gamma}
\partial t^{\delta}\partial^{\mu}}=\frac{\partial^3F}{\partial t^{\gamma}
\partial t^{\beta}\partial t^{\lambda}}\eta^{\lambda\mu}
\frac{\partial^3F}{\partial t^{\alpha}\partial t^{\delta}
\partial t^{\,u}}
\label{11l}
\ee
\indent
Next one defines a (commutative) FA with identity $e$ via a
multiplication ${\bf (F)}\,\,(a,b)\to <a,b>$ with $<ab,c>=<a,bc>$.
Here if $\omega\in A^*$ is defined by $\omega(a)=<e,a>$ then we have
$<a,b>=\omega(ab)$.  The algebra $A$ is called semisimple (ss) if it
contains no nilpotent $a\,\,(a^m=0$).  Given a family $A_t$ of FA one
often identifies $A_t$ with the tangent space $T_tM$ at $t$ to a manifold
$M\,\,(t\in M$).  $M$ is a Frobenius manifold
(FM) if there is a FA structure on $T_tM$ such
that ${\bf (G)}\,\,<\,\,,\,\,>$ determines a flat metric on $M$, 
${\bf (H)}\,\,e$ is covariantly constant for the Levi-Civita (LC)
connection $\nabla$ based on $<\,\,,\,\,>$, i.e. $\nabla e=0$,
{\bf (I)} If $c(u,v,w)=<u\cdot v,w>$ then $(\nabla_zc)(u,v,w)$ should
be symmetric in the vector fields $(u,v,w,z)$, and {\bf (J)} There is 
an Euler vector field $E$ such that $\nabla(\nabla E)=0$ and the 
corresponding one parameter group of diffeomorphisms acts by conformal
transformation on $<\,\,,\,\,>$ and by rescaling on $T_tM$.  One shows
that solutions of WDVV with $d_1\not= 0$ are characterized by the FM
structure ($\partial_{\alpha}\sim\partial/\partial t^{\alpha},\,\,{\cal L}_E$
as in {\bf (E)})
\be
\partial_{\alpha}\cdot\partial_{\beta}=c^{\gamma}_{\alpha\beta}(t)\partial_
{\gamma};\,\,<\partial_{\alpha},\partial_{\beta}>=\eta_{\alpha\beta};\,\,
e_1=\partial_1
\label{11m}
\ee
and ${\cal L}_EF=d_FF+A_{\alpha\beta}t^{\alpha}t^{\beta}+B_{\alpha}
t^{\alpha}+c$ (where the extra terms in ${\cal L}_EF$
can be killed when $d_F\not= 0,\,\,d_F-d_{\alpha}
\not= 0$, and $d_F-d_{\alpha}-d_{\beta}\not= 0$).
\\[3mm]\indent
{\bf REMARK 10.2.}$\,\,$  The case of interest here is based on
$M=\{W(p)=p^{n+1}+a_np^{n-1}+\cdots +a_1;\,\,a_i\in {\bf C}\}$ where
$T_WM\sim$ all polynomials of degree less than $n$ and $A_W$ on
$T_WM$ is $A_W={\bf C}/W'(p)$ where $'\sim d/dp$ and $<f,g>_W=
Res_{\infty}[f(p)g(p)/W'(p)]$.  Then $e\sim\partial/\partial a_1$ and
$E=(1/(n+1))\sum(n-i+1)a_i(\partial/\partial a_i)$.  This should be
compared to Section 11.1 where $W=x^{r+1}-u_1x^{r-1}-\cdots -u_r$
so $a_k\sim -u_{r-k+1}$ and $e\sim -\partial/\partial u_r$ with $n=r$.  Note
the indexing leads to 
\be
W(tp,t^{n-k+2}c_k)=t^{n+1}W(p,a_k)\Rightarrow \left(p\partial_p
+\sum (n-k+2)a_k\frac{\partial}{\partial a_k}\right)W=(n+1)W
\label{11n}
\ee
and $(n-k+2)a_k\sim-(n-k+2)u_{n-k+1}\sim-(m+1)u_m$ (cf. (\ref{11d})).
One checks the vanishing of the curvature for the metric
$<f,g>_W=Res_{\infty}[fg/W']$ as follows.  Consider
$p=p(W)$ inverse to $W=W(p)$ obtained via Puiseaux series
\be
p=p(k)=k+\frac{1}{n+1}\left(\frac{t^n}{k}+\frac{t^{n-1}}{k^2}+\cdots
+\frac{t^1}{k^n}\right)+O\left(\frac{1}{k^{n+1}}\right)
\label{E106}
\ee
where $k^{n+1}=W$; this determines the coefficients $t^i(a_k)$ where
\be
p(k)^{n+1}+a_np(k)^{n-1}+\cdots+a_1=k^{n+1}=W
\label{E107}
\ee
will determine the expansion (\ref{E106}).  There is then a triangular
change of coordinates ${\bf (L)}\,\,a_i=-t^i+f_i(t^{i+1},\cdots,t^n)$
for $i=1,\cdots,n$.  Evidently
\be
t^{\alpha}=-\frac{n+1}{n-\alpha+1}Res_{\infty}\,W^{\frac{n-\alpha+1}{n+1}}(p)
dp
\label{E108}
\ee
(cf. (\ref{11H})) which suggests that $e_i=n-i+1$ and $c_i=-(n+1)/(n-i+1)$
in Section 10.1 and we
identify $t^k$ and $T_k$).  One can verify (\ref{E108}) by looking at 
$dp=dk-(1/(n+1))[(t^n/k^2)+\cdots+(nt^1/k^{n+1})]dk+O(dk/k^{n+2})$ with
$W=k^{n+1}$.  To prove that the $t^{\alpha}$ are flat coordinates one uses
the thermodynamic identity
\be
\partial_{\alpha}(Wdp)|_{p=c}=-\partial_{\alpha}(pdW)|_{W=c}
\label{E109}
\ee
which follows from $W(p(W,t),t)=W$ via $\partial_{\alpha}W|_{p=c}+(\partial_
pW)\partial_{\alpha}p|_{W=c}=0$ which says $\partial_{\alpha}(Wdp)|_{p=c}
+\partial_{\alpha}(pdW)|_{W=c}=0$ since $\partial_pW\sim dW/dp$.  Next
we have for $1\leq \alpha\leq n$
\be
\partial_{\alpha}(Wdp)|_{p=c}=-\left[k^{\alpha-1}dk\right]_{+}
\label{E110}
\ee
where $[fdk]_{+}=[f(dk/dp]_{+}dp$.  To see this note $k=W^{1/(n+1)}=
p+O(1/p)$ via (\ref{E107}) and from (\ref{E107}) and (\ref{E109}) we have
\be
-\partial_{\alpha}(Wdp)|_{p=c}=\partial_{\alpha}(pdW)|_{W=c}=
(\partial_{\alpha}p)dk^{n+1}=
\label{E111}
\ee
$$=\left[\frac{1}{n+1}\frac{1}{k^{n-\alpha+1}}+O\left(\frac{1}{k^{n+1}}
\right)\right]dk^{n+1}=k^{\alpha-1}dk+O\left(\frac{dk}{k}\right)$$
The left side is polynomial in $p$ and $[O(1/k)dk]_{+}=0$ since
$dk=dp+O(1/p^2)dp$ while $(1/k)=O(1/p)$ so (\ref{E110}) is proved.
Now use (\ref{E111}) in a standard formula (cf. \cite{dc})
\be
<\partial_{\alpha},\partial_{\beta}>=Res_{\infty}\frac
{\partial_{\alpha}(W(p)dp)\partial_{\beta}(W(p)dp)}{dW(p)};
\label{Era}
\ee
$$c_{\alpha\beta\gamma}=Res_{\infty}\frac
{\partial_{\alpha}(Wdp)\partial_{\beta}(Wdp)
\partial_{\gamma}(Wdp)}{dpdW(p)}$$
which yields
\be
<\partial_{\alpha},\partial_{\beta}>=Res_{p=\infty}\frac{k^{\alpha-1}
k^{\beta-1}dk}{dk^{n+1}}=\frac{1}{n+1}\delta_{\alpha+\beta,n+1}
\label{E112}
\ee
so the $t^{\alpha}$ are flat coordinates since $\eta_{\alpha\beta}$ is 
constant.  The corresponding $F$ arises via
\be
\partial_{\alpha}F=\frac{1}{(\alpha+1)(n+\alpha+2)}Res_{p=\infty}
W^{\frac{n+\alpha+2}{n+1}}dp
\label{E114}
\ee
(based on \cite{dn,dc}).
Thus one will have some WDVV equations based on the LG model
corresponding to those specified by $W^R_G$ in Section 10.1.
Strictly speaking there is no Whitham theory here - only a LG model
based perhaps on a dispersionless KdV hierarchy (see Remark 10.3).

\subsection{Witten-Dijkgraaf-Verlinde-Verlinde (WDVV) equations}

We go first to \cite{ib} and continue the context of Section 10.1.  The
main point is to show that the WDVV equations for SW theory 
(involving variables $a_i$) arise from
those of the corresponding LG model (this is quite cute).  
Thus one takes $z+(\mu^2/z)
=W_G(x,T_1,\cdots,T_r)$ as in (\ref{11E}) - (\ref{11F}) but $W_G$ is
now expressed in terms of flat coordinates as in (\ref{11H}) (again
only $A_n$ is considered).  We tentatively identify $T_{\alpha}$ and 
$t^{\alpha}$ in (\ref{11H}) and (\ref{E108}), where $n=r$.  One writes
$\phi_i\sim\phi_i^R$ as in (\ref{11I}) with $\phi_r=1$ and flatness
of $\eta_{ij}$ in (\ref{11K}) (where $P\sim\phi_r=1$) implies
(\ref{11M}), namely, $\partial_xQ_{ij}(x)=\partial_i\partial_jW(x)$ with
$\eta_{ij}=<\phi_i\phi_j\phi_r>=\delta_{e_i+e_j,h}$.  Associativity
of the chiral ring (cf. (\ref{11c})) $\phi_i(\phi_j\phi_k)=(\phi_i\phi_j)
\phi_k$ implies that $C^{\ell}_{ij}C^m_{\ell k}=C^{\ell}_{jk}C^m_{\ell i}$
or ${\bf (M)}\,\,[C_i,C_j]=0$ where $(C_i)^k_j=C^k_{ij}$.  From 
$F_{ijk}=C^{\ell}_{ij}\eta_{\ell k}$ one obtains then WDVV in the form
\be
F_i\eta^{-1}F_j=F_j\eta^{-1}F_i;\,\,(F_i)_{jk}=F_{ijk}
\label{E115}
\ee
This is all based on TFT for the LG model with $W\sim P$ in 
(\ref{11E}) - (\ref{E58}) or in Remark 10.2.
\\[3mm]\indent
Now look at the SW theory based on $W$ with ${\bf (N)}\,\,\mu^2=
\Lambda^{2N}/4$ where $N\sim h\sim r+1$.  We used $\mu^2=\Lambda^{2N}$
in Section 10.1 (as in \cite{ib} (9712018)) but switch now to {\bf (N)} plus
$\lambda=\lambda_{SW}=(1/2\pi i)(xdz/z)$ (instead of $\lambda=xdz/z$ 
in Section 10.1) in order to conform to the notation of \cite{ib}
(9803126).  The
PF equations (\ref{11a}) - (\ref{11b}) can be written in flat coordinates
as (cf. \cite{ib} or simply integrate $\oint\lambda_{SW}$ in 
(\ref{11Y}) and (\ref{11Z}))
\be
{\cal L}_0\Pi=\left(\sum_1^rq_iT_i\partial_i-1\right)^2\Pi-4\mu^2h^2
\frac{\partial^2\Pi}{\partial T_r^2}=0;
\label{E116}
\ee
$${\cal L}_{ij}\Pi=\partial_i\partial_j\Pi-\sum_1^rC^k_{ij}\partial_k\partial_r
\Pi=0\,\,(\partial_i\sim\frac{\partial}{\partial T_i})$$
Now one makes a change of variables $T_i\to a_i=\oint_{A_i}\lambda$
(philosophy below) so that ($\partial_i=\partial/\partial T_i$)
\be
\left(\partial_ia_I\partial_ja_J-\sum_1^rC^k_{ij}\partial_ka_I
\partial_ra_J\right)\frac{\partial^2\Pi}{\partial a_I\partial a_J}+
P^I_{ij}\frac{\partial\Pi}{\partial a_I}=0
\label{E117}
\ee
where $P^I_{ij}=\partial_i\partial_ja_I-\sum_1^rC^k_{ij}\partial_k\partial_r
a_I$.  Since $a_I$ satisfies ${\cal L}_{ij}a_I=0$ one knows that
$\Pi=a_I$ satisfies (\ref{E117}) and 
$P^I_{ij}=0$. 
Next take 
$\Pi=a^D_I=\partial{\cal F}/\partial a_I$ to get the third order
equation for ${\cal F}\sim F^{SW}$ ($P^I_{ij}=0$)
\be
\tilde{{\cal F}}_{ijk}=\sum_1^rC^{\ell}_{ij}\tilde{{\cal F}}_{\ell rk};\,\,
\tilde{{\cal F}}_{ijk}=\partial_ia_I\partial_ja_J\partial_ka_K{\cal F}_
{IJK};\,\,{\cal F}_{IJK}=\frac{\partial^3{\cal F}(a)}{\partial a_I\partial
a_J\partial a_K}
\label{E118}
\ee
Defining a metric by ${\cal G}_{ij}=\tilde{{\cal F}}_{ijr}$ one has
${\bf (O)}\,\,\tilde{{\cal F}}_i=C_i{\cal G}$ for $\tilde{{\cal F}}_i
=(\tilde{{\cal F}})_{jk}=\tilde{{\cal F}}_{ijk}$ and from commutativity
of the $C_i$ there results
\be
\tilde{{\cal F}}_i{\cal G}^{-1}\tilde{{\cal F}}_j=\tilde{{\cal F}}_j
{\cal G}^{-1}\tilde{{\cal F}}_i
\label{E119}
\ee
Hence the ${\cal G}^{-1}\tilde{{\cal F}}_i$ commute and consequently
the matrices ${\bf (P)}\,\,\tilde{{\cal F}}_k^{-1}\tilde{{\cal F}}_i
=({\cal G}^{-1}\tilde{{\cal F}}_k)^{-1}
{\cal G}^{-1}\tilde{{\cal F}}_i$
also commute for fixed $k$.  Therefore we obtain ${\bf (Q)}\,\,
\tilde{{\cal F}}_i\tilde{{\cal F}}_k^{-1}\tilde{{\cal F}}_j=
\tilde{{\cal F}}_j\tilde{{\cal F}}_k^{-1}\tilde{{\cal F}}_i$ and
removing the Jacobians $\partial a_I/\partial T_i$ from {\bf (Q)} implies
the general WDVV equations
\be
{\cal F}_I{\cal F}_K^{-1}{\cal F}_J={\cal F}_J{\cal F}_K^{-1}
{\cal F}_I
\label{E120}
\ee
as in \cite{bw,bz,mr,mw,mp,mz} where an association $a_I\sim d\omega_I\sim$
holomorphic differential is used in the constructions and proof
(cf. also \cite{cb,cy,dc,iz,ka,kc,ko,mf,mi} for WDVV).  
Here
${\cal F}\sim F^{SW}$ and the corresponding WDVV equations are a direct
consequence of the associativity of the chiral ring in the $A_n$ LG model 
(hence of WDVV equations of the form (\ref{11l})).
\\[3mm]\indent 
Regarding philosophy note we have assumed no a priori connection
between $F$ and ${\cal F}$.  $F$ comes from the TFT for the LG model 
while ${\cal F}$ is defined via $a^D_I=\partial{\cal F}/\partial a_I
=\oint_{B_i}\lambda_{SW}$ and 
(general) WDVV for ${\cal F}$ follows from the PF equations.  In
particular $F$ is not related to $F^W\sim$ Whitham prepotential
for the SW curve.  There is also a WDVV theory for a Whitham 
hierarchy on a RS involving the LG type $T_k$ for
$1\leq k\leq n$ plus other variables
including the $a_i$ (see Remark 10.3 below) and we note that the metric is
quite 
different from ${\cal G}$ insofar as the $a_i$ are concerned.
\\[3mm]\indent
{\bf REMARK 10.3.}$\,\,$  In this direction one can make the following
comments.
In (\ref{11H}) or (\ref{E108}) we have a formula ${\bf (AS)}
\,\,T_m=-[(n+1)/(n+1-m)]
Res\,W^{1-[m/(n+1)]}(p)dp$ for LG times based on a superpotential
${\bf (AT)}\,\,W=p^{n+1}-q_1p^{n-1}-\cdots -q_n$ 
($q_k\sim u_k$ and $a_k=-q_{n-k+1}$).  This
expresses the $T_m$ as functions of the $q_k$ and one requires $m=1,\cdots,n$
so there are $n$ primary times and $n$ coefficients $q_k$.  
Such objects $W$ as above often arise
from the dispersionless form of n-KdV situations for example and one
can refer to them as dispersionless times for a TFT of LG type.  Note
that diagonal coordinates (Riemann invariants) are obtained via
$u_i=W(p_i)$ where $W'(p_i)=0\,\,(i=1,\cdots,n)$.  Further in \cite{dc}
for example one shows how this extends, in the context of Hurwitz spaces, to 
a new class of Whitham times as follows
(note this differs from Section 7 and see also \cite{ka,kc,ko}).
Consider $g$-gap solutions of KdV problems based on $L=\partial^{n+1}
-q_1\partial^{n-1}-\cdots - q_n$ and let $M_{g,n+1}$ be the space of
such $g$-gap solutions.  This
moduli space 
${\cal M}=M_{g,n+1}$ is the moduli space of algebraic curves $\Sigma_g$
(with fixed homology) whose ramification is determined by the meromorphic
function {\bf (AT)}.  We recall by Riemann-Roch that $[\#(zeros)-
\#(poles)](dW)=2g-2$ and for $p\sim 1/z$ a pole $p^ndp$ corresponds to
$-z^{-n-2}dz$ so this has order $n+2$ (not $n$).  Hence $dW$ will
have $N=2g-2+(n+2)=2g+n$ zeros $p_i$ where $dW(p_i)=0$ and one defines
$u_i=W(p_i)\,\,(i=1,\cdots,N)$ as local coordinates at the branch points
$p_i$.  The averaged KdV hierarchy (or Whitham hierarchy) then has the form
${\bf (AU)}\,\,
\partial_md\Omega_1=\partial_Xd\Omega_m$ (or more generally $\partial_m
d\Omega_s=\partial_sd\Omega_m$) where the $d\Omega_m$ corresond
to $dW^{m/(n+1)}+\,\,regular\,\,terms$ as $p\to\infty$ with 
$\oint_{A_s}d\Omega_m=0$ (see below - note for $z\sim 1/p$ near $\infty$
we want to deal with $d\Omega_m\sim -mz^{-m-1}dz+\,\,regular\,\,terms$.
Further one identifies $p$ here with $\Omega_1$, i.e.
$dp=d\Omega_1$, and defines a flat metric via ${\bf (AV)}\,\,
g_{ii}=Res_{p_i}(dp^2/dW)$ as $ds^2=\sum_1^Ng_{ii}(u)du_i^2$ and the
corresponding flat times $t_i\,\,(1\leq i\leq N
=2g+n)$ for $ds^2$ have the form
\be
t_i=-(n+1)Res_{\infty}\frac{W(p)^{1-i/(n+1)}}{n+1-i}dp\,\,(i=1,\cdots,n);
\label{200}
\ee
$$t_{n+\alpha}=\frac{1}{2\pi i}\oint_{A_{\alpha}}pdW;\,\,t_{g+n+\alpha}=
\oint_{B_{\alpha}}dp\,\,(\alpha=1,\cdots,g)$$
(as indicated earlier $t^i\sim t_i\sim T_i$ for $1\leq i\leq n$).
Then (\ref{E112}) holds,
$<\partial_{n+\alpha},\partial_{g+n+\beta}>=\delta_{\alpha,\beta}$,
and otherwise
$<\,\,,\,\,>$ is 
zero.  Thus the set of TFT times {\bf (AS)} is enlarged to contain
$2g$ additional primary times and this is a basic sort of Whitham
theory (cf. also \cite{cb,dc,ka,kc,ko} for further enlargments).
The associated Whitham-LG theory for $A_n$ now involves
a coupling space ${\cal M}=M_{g,n+1}$ with flat coordinates $T_1,\cdots,
T_N$ as above and primary fields $\phi_1\sim -dp,\,\,\phi_2,\cdots,\phi_N$
of the form ${\bf (AW)}\,\,\phi_i=-(n+1)d\Omega_i\,\,(i=1\cdots,n)$
with $\phi_{n+\alpha}=d\omega_{\alpha}$ and $\phi_{g+n+\alpha}=
d\sigma_{\alpha}\,\,(\alpha=1,\cdots,g)$ where the $d\omega_{\alpha}$
are holomorphic differentials with $\oint_{A_j}d\omega_k=2\pi i\delta_{jk}$
and there are
multivalued holomorphic differentials $d\sigma_{\alpha}=d\sigma_{\alpha}^1$ 
satisfying $\Delta_{B_{\alpha}}d\sigma_{\alpha}=-dW$ and $\oint_{A_k}
d\sigma_{\alpha}=0$ (where $\Delta_{B_{\alpha}}d\sigma_{\alpha}=
d\sigma_{\alpha}(P+B_{\alpha})-d\sigma_{\alpha}(P)$).  Further the LG 
potential is $W=W(p)$ where $p\sim\int_{P_0}^Pdp\sim\int_{P_0}^Pd\Omega_1$
and 
\be
\left.\frac{\partial (Wdp)}{\partial t_{\alpha}}\right|_{p=c}=-\phi_{\alpha}=
-\left.\frac{\partial (pdW)}{\partial t_{\alpha}}\right|_{W=c}
\label{E2000}
\ee
with correlation functions (note the minus sign in $\phi_i$ above)
\be
<\phi_{\alpha}\phi_{\beta}>=\eta_{\alpha\beta}=\sum\,Res_{dW=0}
\frac{\phi_{\alpha}\phi_{\beta}}{dW};
\label{E201}
\ee
$$<\phi_{\alpha}\phi_{\beta}\phi_{\gamma}>=\sum\,Res_{dW=0}\frac
{\phi_{\alpha}\phi_{\beta}\phi_{\gamma}}{dWdp}=c_{\alpha\beta\gamma}(t)$$
The chiral algebra $c^{\gamma}_{\alpha\beta}(t)$ has the form ${\bf (AX)}\,\,
\phi_{\alpha}\phi_{\beta}=c^{\gamma}_{\alpha\beta}\phi_{\gamma}dp$
modulo $dW$ divisible differentials and one will have an expansion
${\bf (AY)}\,\,pdW=(n+1)d\Omega_{n+2}+\sum_1^Nt_{\alpha}\phi_{\alpha}$
for the multivalued differential $pdW$.  Writing $k^{n+1}=W$ one has also
\be
pdW=kdW-\sum_1^nt_{\alpha}k^{\alpha-1}+O(k^{-1})
\label{E202}
\ee
along with ($s=1,\cdots,g$)
\be
\oint_{A_s}pdW=2\pi it_{n+s};\,\,\Delta_{A_s}(pdW)=0;\,\,\Delta_{B_s}(pdW)
=t_{n+g+s}dW
\label{E203}
\ee
In keeping with $d\Omega_m\sim dW^{m/(n+1)}$ 
as indicated and {\bf (AW)} one has
$\phi_m=(-k^{m-1}+O(k^{-2}))dk$ with $\oint_{A_s}\phi_m=0$.  One can
also determine a partition function or prepotential $F$ such that
$c_{\alpha\beta\gamma}=\partial^3F/\partial t_{\alpha}\partial_{\beta}
\partial_{\gamma}$. 
This development is an obvious ancestor to the SW theories of more
recent vintage.  The analogous SW theory is best developed in a
two puncture Toda framework with two points $\infty_{\pm}$ corresponding
to $p\to\infty$ (cf. \cite{cb,gf,hb,ha,ka,kc,ko,na}); the analogous
$W$ term in not polynomial but logarithmic which severly limits 
the number of primary $T_i$ times.  
We will not develop this further here.

\end{document}